\documentclass[aip,graphicx,floatfix,groupedaddress]{revtex4-1}
\usepackage{graphicx, color, colortbl, soul}
\usepackage{amsmath, amssymb, amsfonts}
\usepackage{siunitx}

\RequirePackage{hyperref}
\hypersetup{
	colorlinks = true,
	linkcolor = black, 
	citecolor = black, 
	urlcolor = blue, 
	filecolor = black 
}
\usepackage{cleveref}

\usepackage[colorinlistoftodos]{todonotes}

\usepackage[normalem]{ulem}
\crefrangelabelformat{equation}{(#3#1#4--#5\crefstripprefix{#1}{#2}#6)}

\usepackage{lineno}

\usepackage{bm}

\DeclareMathAlphabet{\mathsfit}{\encodingdefault}{\sfdefault}{m}{sl}
\SetMathAlphabet{\mathsfit}{bold}{\encodingdefault}{\sfdefault}{bx}{sl}

\newcommand{\tens}[1]{\bm{\mathsfit{#1}}}

\usepackage[gen]{eurosym}

\newcommand{\unit}[1]{\ensuremath{\, \si{#1}}}

\newcommand{\fref}[1]{Figure~\ref{#1}}
\newcommand{\tref}[1]{Table~\ref{#1}}
\newcommand{\eref}[1]{Equation (\ref{#1})}
\newcommand{\sref}[1]{Section~\ref{#1}}

\newcommand{\usi}{\boldsymbol{u}_{s(i)}}

\newcommand{\xp}{\boldsymbol{x}_p}

\newcommand{\us}{\boldsymbol{u}_s}
\newcommand{\up}{\boldsymbol{u}_p}
\newcommand{\uf}{\boldsymbol{u}_f}
\newcommand{\ufAtP}{\boldsymbol{u}_{f@p}}

\newcommand{\upFilti}{\boldsymbol{\widetilde{u}}_{p(i)}}
\newcommand{\ufFilti}{\boldsymbol{\widetilde{u}}_{f@p(i)}}

\newcommand{\upFilt}{\widetilde{\boldsymbol{u}}_p}
\newcommand{\ufFilt}{\widetilde{\boldsymbol{u}}_f}

\newcommand{\upFiltAtP}{\widetilde{\boldsymbol{u}}_{p@p}}
\newcommand{\ufFiltAtP}{\widetilde{\boldsymbol{u}}_{f@p}}

\newcommand{\kfsgs}{k_{sgs}}

\newcommand{\Gbold}{\tens{G}}
\newcommand{\Bbold}{\tens{B}}

\def\code#1{\texttt{#1}}

\newcommand{\ReInit}{Re_\lambda^{t=0}}

\makeatletter
\let\@fnsymbol\@fnsymbol@latex
\@booleanfalse\altaffilletter@sw
\makeatother

\begin{document}

\title{Neural stochastic differential equations for particle dispersion in large-eddy simulations of homogeneous isotropic turbulence}

\author{J. Williams}
\author{U. Wolfram}
\thanks{AO and UW share last authorship. Address correspondence to a.ozel@hw.ac.uk}
\author{A. Ozel}
\thanks{AO and UW share last authorship. Address correspondence to a.ozel@hw.ac.uk}
\affiliation{School of Engineering and Physical Sciences, Heriot-Watt University, Edinburgh, UK}

\date{\today}

\begin{abstract}

In dilute turbulent particle-laden flows, such as atmospheric dispersion of pollutants or virus particles, the dynamics of tracer-like to low inertial particles are significantly altered by the fluctuating motion of the carrier fluid phase. Neglecting the effects of fluid velocity fluctuations on particle dynamics causes poor prediction of particle transport and dispersion. To account for the effects of fluid phase fluctuating velocity on the particle transport, stochastic differential equations coupled with large-eddy simulation are proposed to model the fluid velocity seen by the particle. The drift and diffusion terms in the stochastic differential equation are modelled using neural networks (`neural stochastic differential equations’). The neural networks are trained with direct numerical simulations (DNS) of decaying homogeneous isotropic turbulence at low and moderate Reynolds numbers. The predictability of the proposed models are assessed against DNS results through \textit{a priori} analyses and \textit{a posteriori} simulations of decaying homogeneous isotropic turbulence at low-to-high Reynolds numbers. Total particle fluctuating kinetic energy is under-predicted by 40\% with no model, compared to the DNS data. In contrast, the proposed model predictions match total particle fluctuating kinetic energy to within 5\% of the DNS data for low to high-inertia particles. For inertial particles, the model matches the variance of uncorrelated particle velocity to within 10\% of DNS results, compared to 60-70\% under-prediction with no model. It is concluded that the proposed model is applicable for flow configurations involving tracer and inertial particles, such as transport and dispersion of pollutants or virus particles.

\end{abstract}
\newcommand{\comment}[1]{}

\maketitle

\section{Introduction}
    Dilute turbulent particle-laden flows are common in various natural and industrial processes, such as aerosol drug inhalation \citep{williams2022effect,lambert11regional,koullapis18regional}, pollutant dispersion in urban environments \citep{wang2018quantifying}, and disease spread by cough plume dispersion \citep{calmet2021large, liu2021investigation, monroe2021pulsatility}. Due to low solid volume fraction and high density ratio between particles and carrier fluid (low particle mass loading), particle dynamics are mainly controlled by the carrier fluid. The fluid phase velocity fluctuations significantly alter particle transport and dispersion, and can fundamentally change how particles preferentially form in the carrier phase \citep{fede2006numerical}. The influence of the carrier fluid on particle dynamics is mainly characterised by the Stokes number, which is the ratio of particle drag timescale to the Kolmogorov timescale (timescale at which viscosity dominates) \citep{fessler1994preferential, dou2018effects, brandt2022particle}. Accurately capturing the influence of particle Stokes number on particle motion is a key challenge in dilute turbulent particle-laden flows \citep{moreau2010development, pozorski2009filtered} due to interactions between fluid and particles ranging over several length and timescales \citep{balachandar2010review}. Understanding of particle transport in turbulent flows is crucial for designing engineering solutions to mitigate pollutant and virus dispersion.
    
    Dilute turbulent particle-laden flows have been studied using computational modelling approaches such as particle-resolved direct numerical simulation (DNS) \citep{mehrabadi2018direct, vreman2018turbulent, costa2021near}, where all temporal-spatial structures of flow are resolved without any models \citep{sundaresan2018toward}. However, particle-resolved DNS is computationally infeasible for high Reynolds number flows or tracking many particles, meaning coarser modelling approaches are required for engineering applications such as modelling aerosol inhalation. \citet{mehrabadi2018direct} found point-particle DNS to agree well with particle-resolved DNS, particularly when mass loading was low and the particles have limited or no influence on the carrier fluid. In the point-particle approach, also called `unresolved' Eulerian-Lagrangian, particles are tracked in a Lagrangian fashion by solving Newton's equations of motion and an additional model is required for fluid-particle interactions \citep{and67fluimech, maxeyriley1983, capecelatro2013euler, sundaresan2018toward}. 
    When modelling flows in complex geometries or at high Reynolds number, it is also infeasible to fully resolve all temporal-spatial structures of carrier fluid motion. Therefore, it is necessary to use the Reynolds-averaged Navier-Stokes or large-eddy simulation (LES) approach \citep{popebook, sagaut06les}. In the Reynolds-averaged Navier-Stokes approach, the fluid fluctuating motion is modelled through additional transport equations for the evolution of variables such as turbulent kinetic energy and dissipation rate ($k$-$\varepsilon$) \citep{jones1972prediction, launder1983numerical, spalart1992one}. In the LES approach, the large scales of turbulence are directly resolved and the scales smaller than the characteristic mesh size are filtered out (see \citet{germano1992turbulence} for an overview of LES filtering). The effects of unresolved `subgrid' scales are accounted for with a subgrid model \citep{smagorinsky1963general, germano1992turbulence, sagaut06les}. Reynolds-averaged Navier-Stokes and LES are both widely used for modelling fluid transport in simulations of particle-laden turbulent flows where the ``resolved'' fluid flow field is used to calculate the fluid-particle interactions (e.g. drag force). However, in both approaches there is an unresolved fluctuating fluid velocity that can influence the drag force acting on a particle, thereby causing an under-prediction of particle fluctuating kinetic energy for low-moderate inertia particles \citep{fede2006numerical, marchioli2017review}. As LES provides predictions of fluid phase fluctuations with higher accuracy, in this study we choose to focus on LES modelling of dilute particle-laden flows. However, neglecting the effect of subgrid fluid fluctuations on particle dynamics can cause errors in macroscopic quantities of the particle phase such as total particle fluctuating kinetic energy and preferential concentration \citep{marchioli2008issues, marchioli2017review, fede2006numerical}. Therefore, we require an additional model to account for the effect of subgrid fluid fluctuations on the particle phase.
     
    To account for subgrid fluid fluctuations, a Langevin-type stochastic differential equation has been developed to predict the `fluid velocity seen by a particle' which is used to compute the drag force in dilute particle-laden turbulent flows \citep{minier2001pdf}. These models solve a stochastic differential equation (SDE) for the fluid velocity `seen' by a particle, where the fluctuating component represents the randomness of the unresolved turbulence \citep{haworth1986generalized, minier2015}. Langevin-type stochastic models have been well developed for large-eddy simulations of single-phase reactive flows \citep{pope1985pdf, pope1994relationship, gicquel2002velocity, jenny2001hybrid, pope2011simple, haworth1986generalized} and Reynolds-averaged Navier-Stokes models of particle-laden turbulent flows \citep{minier2001pdf, innocenti2021lagrangianpdf, innocenti2019lagrangian, peirano2006meanfield, pozorski1999probability, dreeben1998probability, chibbaro2008pipeflow, waclawczyk2004nearwall}. Recently, there have been studies on LES stochastic models for dilute particle-laden flows \citep{shotorban2006stochastic, fede2006numerical, fede2006stochastic, berrouk2007stochastic, pozorski2009filtered, minier2015, innocenti2016, knorps2021stochastic}. However, closures in LES-SDE models have not been widely developed \citep{fede2006numerical, innocenti2016, knorps2021stochastic}. Existing LES models are based on the simplified Langevin model for the fluid integral timescale \citep{knorps2021stochastic, innocenti2016, pozorski2009filtered, breuer2017influence, berrouk2007stochastic}, which has been used widely for Reynolds-averaged Navier-Stokes stochastic models of turbulent single-phase flow \citep{haworth1986generalized}. However, \citet{knorps2021stochastic} showed in \textit{a priori} analyses of turbulent channel flow that the integral timescale approaches infinity when the particle approaches the viscous sub-layer in LES-SDE models. As the timescale approaches infinity, particles entering the viscous sub-layer will become trapped as the time taken for a particle to respond to changes in the local flow is infinitely large. As the timescale approaches infinity, particles transitioning from turbulent to laminar flow regions will maintain their initial velocity for an infinite period of time and fail to adapt to the local fluid velocity. In decaying homogeneous isotropic turbulence, this would mean that the predicted decay of particle fluctuating kinetic energy is not consistent with the fluid as the particle velocity does not continue to decay. \citet{knorps2021stochastic} bypassed this by applying a near-wall damping for the timescale \citep{vandriest1956turbulent}, which is not applicable in unbounded cases such as homogeneous isotropic turbulence. Therefore, new models that are consistent with the underlying LES fluid solver are needed \citep{minier2021methodology}, in order to avoid unphysical predictions of kinetic energy in decaying turbulent flows \citep{minier2014}.
    
    Data-driven methods can be used as an alternative approach to developing closures for unknown terms in LES-SDE models. The merit of data-driven turbulence models has been demonstrated for (single-phase) fluid turbulence \citep{ling2016reynolds, fang2020neural, beetham2020formulating, park2021toward, zhou2019subgrid} and passive scalar dynamics \citep{milani2021turbulent, frezat2021physical}, where the data-driven approaches produced improved results compared to state-of-the-art closure models from theoretical derivation. By using data from highly resolved simulations (such as point-particle DNS) \citep{jiang2019neural, jiang2021development, dietrich2021learning, fang2020neural}, a neural network can learn a model for the evolution of a variable, such as the fluid velocity seen by a particle, in simulations of lower resolution. Recently, stochastic models for particle dynamics have been inferred with neural networks to create `neural SDEs' \citep{dietrich2021learning, yang2022generative, karniadakis2021solving, raissi2018stochastic, kidger2022neural}. Neural SDEs have been applied to learn drift and diffusion terms of a SDE model for Brownian motion \citep{yang2022generative} and epidemic dynamics \citep{dietrich2021learning} from particle observations at different time instants.
    Thus, neural SDEs provide a suitable framework to learn unknown physics required for model closures in LES-SDE models of dilute particle-laden flows. 
    
    Therefore, in this study we aim to develop data-driven closures for stochastic models of particle dispersion in decaying homogeneous isotropic turbulence. To do this, we first generate a \textit{ground truth} database, which is constructed on DNSs of decaying homogeneous isotropic turbulence. We then use DNS data to train neural networks to learn unknown terms in our stochastic model for the fluid velocity seen by particles. The developed model is assessed through \textit{a priori}, and \textit{a posteriori} analyses of decaying homogeneous isotropic turbulence at various Reynolds numbers, including a higher Reynolds number, which is not included in training. Our computational framework including OpenFOAM-based fluid-particle solvers and python scripts for neural network model training is publicly available online in a GitHub repository 
    (\url{https://github.com/jvwilliams23/turbulent-dispersion-neuralSDE}). Data for training is also available online through Kaggle \citep{williams2022filteredDNS}.

\section{Materials and methods}

\subsection{\textit{Ground truth: Direct numerical simulations of decaying homogeneous isotropic turbulence}} 
\label{sec:groundTruthGeneration}
To generate database for the neural SDE model training, we performed DNS in decaying homogeneous isotropic turbulence at $\ReInit=\{9, 33, 105\}$ with one-way coupling on an equispaced $N_{cell} = 256^3$ cell grid occupying a cubic domain of size $2\pi \unit{m}$ in each axis, where $N_{cell}$ is the number of cells. Simulations were performed in OpenFOAM v6 \citep{weller98tensorial}. The transport equations for the fluid phase are given as 
\begin{linenomath}
	\begin{eqnarray}  
    \boldsymbol \nabla \cdot \uf & = & 0, \label{eq:masscontDNS} \\
    \frac{\partial \uf }{\partial t} 
    + \uf  \cdot \boldsymbol \nabla \uf & = & 
	- \frac{1}{\rho_f} \boldsymbol{\nabla} p 
	+ \boldsymbol{\nabla} \cdot \tens{\tau}
	\label{eq:fluidmomDNS}
	\end{eqnarray}
\end{linenomath}
where $\uf$ is the fluid velocity, $p$ is the fluid pressure, $\rho_f$ is the fluid density and $\tens{\tau}$ is the viscous stress tensor, given by
\begin{equation}
    \tens{\tau} = \nu_f \left[\boldsymbol{\nabla} \uf + \boldsymbol{\nabla} \uf^T - \frac{2}{3}(\boldsymbol{\nabla}\cdot \uf) \tens{I} \right],
\end{equation}
where $\nu_f$ is the dynamic viscosity, and $\tens{I}$ is the identity tensor.

The dispersed phase is tracked in the Lagrangian reference frame by following Newton's equations of motion \citep{maxeyriley1983}. Considering only drag (gas-solid flows where $\rho_p>>\rho_f$), the evolution of a particle's position, $\xp$ and velocity $\up$ is
\begin{linenomath}
\begin{align}
    d \xp &= \up \, dt \label{eq:xp} \\
	d \up &= \frac{\boldsymbol{u}_{f @p} - \up}{\tau_p}dt \label{eq:up}
\end{align}
\end{linenomath}
where $\boldsymbol{u}_{f @p}$ is the fluid velocity at particle position and $\tau_p$ is the time for a particle to respond to changes in the velocity of the carrier fluid (called drag response time), which is defined as
\begin{equation} \label{eq:taup}
    \tau_p = \frac{\rho_p}{\rho_f}\frac{4\, d_p}{3\,C_D\,|\boldsymbol{u}_r|}
\end{equation}
where $\rho_p$ is the particle density, $d_p$ is the particle diameter and $\boldsymbol{u}_r$ is the relative velocity ($\boldsymbol{u}_{f@p}-\up$). The drag coefficient was given by \citet{schillernaumann} as $C_D = \frac{24}{Re_p} (1+0.15\,Re_p^{0.687})$, where $Re_p = |\boldsymbol{u}_r|\,d_p/\nu_f$.

\begin{figure}
    \centering
    \begin{tabular}{c c}
        (a) & (b) \\
        \includegraphics{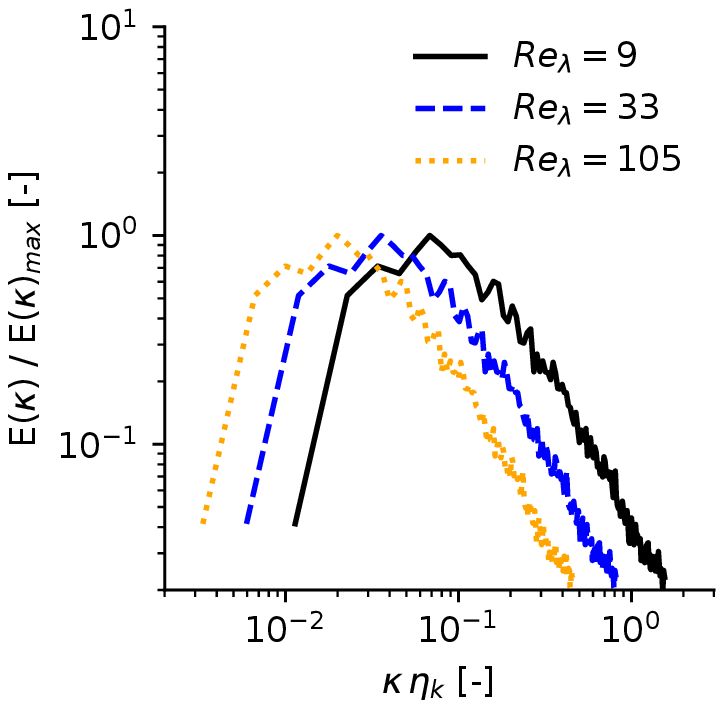} &
        \includegraphics{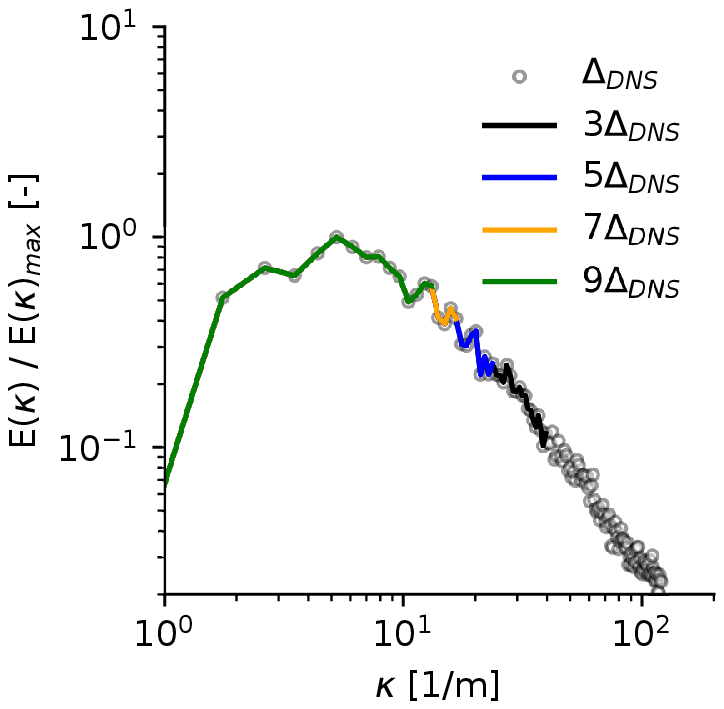}
    \end{tabular}
    \caption{Initial energy spectra in homogeneous isotropic turbulence. Panel (a) shows the effect of varying $\ReInit$ on the cutoff wavenumber relative to the Kolmogorov length in our DNS. Panel (b) shows effect of filter width on amount of large scale energy resolved in simulation.}
    \label{fig:spectra}
\end{figure}

The simulation timestep was based on the fluid Courant number, which was set be a maximum of 0.2 \citep{fureby1996comparative}. Fluid properties at the particle position were taken as the cell value, as we observed the trilinear interpolation in OpenFOAM to have considerable numerical error (Supplementary material). 
The number of particles in the domain, $N_p$, was $N_p=10^{5}$ and the particles were forced to be stationary in time to remove noise in fluid properties at particle position when the particle drifts into a neighbouring cell. To create the initial flowfield, we created an initial divergence-free velocity field with the OpenFOAM \code{createBoxTurb} function. The initial energy spectrum followed the relationship for energy with wavenumber, $E(\kappa)$, provided by \citet{kang2003decaying}, with 10,000 Fourier modes. We created the spectra using a Python tool for synthetic turbulence generation named \code{TurboGenPY} \citep{saad2017turbogenpy}. The flow statistics at the beginning of the simulation are presented in \tref{tab:simparams}. The resolution of the simulation can be assessed by multiplying the largest wave number $\kappa_{max}$ by the Kolmogorov length-scale, $\eta_k = {\nu^3 / \varepsilon}^{1/4}$ where $\varepsilon$ is the resolved scale kinetic energy dissipation rate \citep{yeung1989lagrangian}. For adequate numerical resolution of the scales of motion in DNS, $\kappa_{max} \eta_k > 1$ is required \citep{mansour1994decay, yeung1989lagrangian}. In \fref{fig:spectra}(a) we show the kinetic energy against $\kappa \eta_k$, which shows that $\ReInit=9$ is well-resolved ($\kappa_{max} \eta_k=1.5$) and $\ReInit=33$ is slightly under-resolved ($\kappa_{max} \eta_k=0.8$). The simulation is more under-resolved for $\ReInit=105$ ($\kappa_{max} \eta_k = 0.45$), where an estimated number of cells required to satisfy $\kappa_{max} \eta_k$ is $N_{cell} \approx 600^3$ ($\kappa_{max} = \sqrt{2} N_{ci} / 3$ where $N_{ci}$ is the number of cells in one axis \citep{yeung1989lagrangian}). We therefore did not include $\ReInit=105$ in training and reserve it for testing our model. In \fref{fig:spectra}(b), we show the kinetic energy spectra for DNS and filtered data (described below). We observe that for all mesh resolutions, the large scale features of the flow are retained, including a peak energy at $\kappa \approx 5$.

\begin{table}
    \centering
    \caption{Summary of flow parameters for decaying homogeneous isotropic turbulence test cases.}
    \begin{tabular}{l c}
        \hline
        \textbf{Parameter} & \textbf{Value} \\
        \hline
        Microscale Reynolds number, $\ReInit$ [-] & 9 / 33 / 105 \\
        Large-scale kinetic energy, $k^{t=0}$ [$\si{m^2 / s^2}$] & 0.36 / $4.7\times10^{-5}$ / $4.7\times10^{-4}$ \\
        Large-scale kinetic energy dissipation rate, $\varepsilon^{t=0}$ [$\si{m^2 / s^3}$] & $4.6$ / $1.86\times10^{-5}$ / $1.86\times10^{-5}$ \\
        Large-scale turbulent timescale, $\tau_L^{t=0}$ [s] & 0.079 / 25.1 / 25.1 \\
        Kolmogorov timescale, $\tau_k^{t=0}$ [s] & 0.033 / 2.9 / 0.9 \\
        Kolmogorov length-scale, $\eta_k^{t=0}$ [m] & 0.013 / 0.0068 / 0.0038 \\
        Dynamic viscosity, $\nu_f$ [$\si{Pa \cdot s}$] & 0.005 / $1.57\times 10^{-5}$ / $1.57\times 10^{-5}$\\
        Particle diameter, $\rho_p$ [$\si{kg / m^3}$] & 1000 \\
        Particle Stokes number, $St^{t=0} = \tau_p / \tau_k^{t=0}$ [-] & 0.1 / 1 / 5 \\
        \hline
    \end{tabular}
    \label{tab:simparams}
\end{table}

\subsubsection{Filtering procedure for training data}
\begin{figure}
    \centering
    \includegraphics[width=\linewidth]{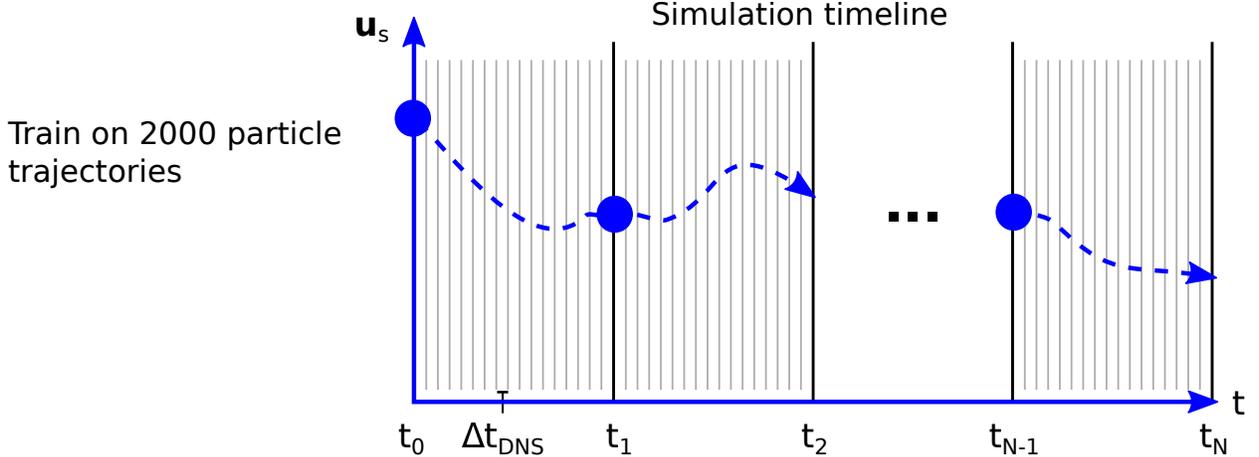}
    \caption{Schematic overview of timeline for data extraction from DNS. Particle properties ($\up$) and fluid properties at the particle position ($\ufAtP$, $k_{sgs@p}$, $\varepsilon_{sgs@p}$, $\ufFiltAtP$) were written from $t_0$ to $t_{N}$. At each saved timestep, we record properties at the start and end of timestep ($[t_n,t_n+\Delta t_{DNS}]$, where $n=1,...,N$).}
    \label{fig:timeline}
\end{figure}

The database for LES-SDE model training was deduced by filtering the results from DNS simulations. The filtered fluid velocity is defined as\citep{oze16fluipart} 
\begin{equation}
    \ufFilt = \int_{\mathbb{R}^3} g(\boldsymbol{r} - \boldsymbol{x}) \uf(\boldsymbol{r}, t) d\boldsymbol{r}.
\end{equation}
where $g(\mathbf{r}-\mathbf{x})$ is a weight function which satisfies $\int_{\mathbb{R}^3} g(\mathbf{r})d\mathbf{{r}}=1$. The box filter kernel is given by
\begin{equation}
  g(\mathbf{r}-\mathbf{x})=\left\{
                                                  \begin{array}{@{}ll@{}}
                                                  \frac{1}{\Delta_f^{3}}, & \text{if\,$|\mathbf{r}-\mathbf{x}|$} \leq \frac{\Delta_f}{2}  \\
                                                      0, & \text{otherwise.}
  \end{array}\right. \label{eq:tophat}
\end{equation}
We applied four widths of $\Delta / \Delta_{DNS} = \{3,5,7,9\}$. Training data consisted of using 2000 particles at each timestep from the $\ReInit=33$ cases (30 timesteps) and 1000 particles from the $\ReInit=9$ case (20 timesteps). This means that the total number of particles included in training is only 1-2\% of the total number of particles available from the DNS data. All filter widths were included in training. This gave 320,000 data points for training, of which, 70\% was used for training and 30\% for testing. 
For \textit{a priori} testing, we used data from filtered DNS as model inputs and advanced the transport equation from our neural SDE one timestep ($t_n \rightarrow t_n + \Delta t_{DNS}$) to eliminate error from time integration over a large timestep. To train the neural SDE, as discussed later, we minimise the difference between neural SDE transport equation integrated over $t_n \rightarrow t_{n+1}$ (write-time interval) and the ground truth at $t_{n+1}$ (\fref{fig:timeline}). This is due to the practical limitation, that the change in velocity over the interval $\Delta t_{DNS}$ is small and impractical for training (discussed below).

\subsection{LES modelling and particle dynamics}
To assess particle dispersion \textit{a posteriori}, we performed LES on mesh resolutions $N_{cell} = \{85^3, 51^3, 36^3, 28^3\}$, where the cell size is approximately the same as the filter width in our \textit{a priori} filtering ($\Delta / \Delta_{DNS}=\{3,5,7,9\}$). The particles are transported following a system of stochastic differential equations (SDEs). To focus on dilute particle-laden flows, we assume that there is no effect of particles on carrier phase dynamics (one-way coupling) \citep{balachandar2010review}. Therefore, the volume-filtered equations of motion for the carrier phase are described as \citep{and67fluimech, capecelatro2013euler}
\begin{linenomath}
	\begin{eqnarray}  
    \boldsymbol \nabla \cdot \ufFilt & = & 0, \label{eq:masscont} \\
    \frac{\partial \ufFilt }{\partial t} 
    + \ufFilt  \cdot \boldsymbol \nabla \ufFilt & = & 
	- \frac{1}{\rho_f} \boldsymbol{\nabla} \widetilde{p} 
	+ \boldsymbol{\nabla} \cdot (\widetilde{\tens{\tau}} - \tens{\tau}_{sgs}),
	\label{eq:fluidmom}
	\end{eqnarray}
\end{linenomath}
where bold lower-case symbols represent a vector, bold upper-case symbols represent a second-order tensor (unless stated otherwise). $(\,\widetilde{\cdot}\,)$ represents a volume-filtered field, $\ufFilt$ is the filtered fluid velocity, $\widetilde{p}$ is the filtered fluid pressure, $\widetilde{\tens{\tau}}$ is the filtered viscous stress tensor, and $\tens{\tau}_{sgs}$ is the subgrid viscous stress from filtering. The volume-filtering is achieved by applying the top-hat filter following \eqref{eq:tophat}, where $\Delta=\sqrt[3]{V_c}$ where $V_c$ is the volume of a computational cell \citep{germano91dynamic, devilliersphd}.
We modelled $\tens{\tau}_{sgs}$ by an eddy-viscosity concept \citep{popebook} as:
\begin{equation}
    \tens{\tau}_{sgs}
    =
    \frac{2}{3} \kfsgs \tens{I} - 
    \, 2 \, \nu_{sgs} \, \widetilde{\tens{S}}, \label{eq:tausgs}
\end{equation}
where $\nu_{sgs}$ is the subgrid viscosity, $\widetilde{\tens{S}}=(\boldsymbol{\nabla} \ufFilt + \boldsymbol{\nabla} \ufFilt^T)/2$ is the filtered, or `resolved', strain rate tensor, and $\kfsgs$ is the subgrid kinetic energy. The subgrid viscosity and kinetic energy are computed using the WALE model proposed by \citet{nicoud1999subgrid}. This model is based on the filter width and invariants of the velocity gradient tensor, giving $k_{sgs}$ as
\begin{equation}
    k_{sgs} = \left(\frac{C_w^2 \Delta}{C_k}\right)^2 
    \frac{\left(|\widetilde{\boldsymbol{\mathcal{S}}}^d \cdot \widetilde{\boldsymbol{\mathcal{S}}}^d|\right)^3}
    {\left(\left(|\widetilde{\tens{S}} \cdot \widetilde{\tens{S}}|\right)^{5/2} + \left(|\widetilde{\boldsymbol{\mathcal{S}}}^d \cdot \widetilde{\boldsymbol{\mathcal{S}}}^d|\right)^{5/4} \right)^2},
\end{equation}
where $\widetilde{\boldsymbol{\mathcal{S}}}=(\boldsymbol{\nabla} \ufFilt^2 + [\boldsymbol{\nabla} \ufFilt^2]^T)/2$, $(\, \cdot \,)^d$ is the deviatoric part of a tensor, $| \cdot |$ is the magnitude which for a second-order tensor, $\tens{T}$, we find by $|\tens{T}|=\sqrt{\tens{T}:\tens{T}}$, wherein $:$ denotes the double scalar product. Modelling constants are $C_w=0.325$ and $C_k=0.094$ \citep{nicoud1999subgrid}. The subgrid kinetic energy dissipation rate, $\varepsilon_{sgs}$, was computed by 
\begin{equation}
    \varepsilon_{sgs} = C_\varepsilon \frac{ \kfsgs^{3/2} }{\Delta},
    \label{eq:epsilon}
\end{equation}
where $C_\varepsilon=1$ is a model parameter \citep{cernick2015particlesgs}.

The transport equation for particle velocity remains as \eref{eq:up}, where $\boldsymbol{u}_{f@p}$ is replaced by $\us$, which is the fluid velocity seen by a particle. For LES without our dispersion model, $\us=\widetilde{\boldsymbol{u}}_{f@p}$, where $\ufFiltAtP$ is the filtered fluid velocity at the particle position.

\subsection{SDE for the fluid velocity seen by a particle}
The fluid velocity seen by a particle, $\us$, is typically obtained by interpolating the `resolved' fluid phase velocity (from either LES or Reynolds-averaged Navier-Stokes modelling) to the particle position \citep{afkhami2015fully, marchioli2008issues, lambert11regional, koullapis18regional, wang1996LES}. However, this provides erroneous particle trajectories, particularly for low inertia particles whose dynamics are mainly controlled by turbulent fluctuations \citep{marchioli2008issues, innocenti2016, pozorski2009filtered, fede2006numerical}.
These fluctuations can be accounted for with a stochastic model with LES modelling given as \citep{innocenti2016}
\begin{equation}
\begin{split}
    d \us = &-  \frac{1}{\rho_f} \left( \boldsymbol{\nabla} \widetilde{p}\right)_{@p} \, dt
    + \nu_f \left(\boldsymbol{\nabla} \cdot \boldsymbol{\nabla} \ufFilt \right)_{@p} \, dt
    + (\upFiltAtP - \ufFiltAtP)\cdot \left(\boldsymbol{\nabla} \ufFilt\right)_{@p} \, dt \\
    &- \Gbold \cdot (\us - \ufFiltAtP) \,dt
    + \Bbold \cdot \, d\boldsymbol{w}
    \label{eq:Us}
\end{split}
\end{equation}
where $\upFiltAtP$ is the Eulerian particle velocity at the particle position obtained from ensemble averaging over all particles in a cell using the same filter kernel for LES filtering as shown in \eref{eq:tophat}. $d\boldsymbol{w}$ is a vector of independent random variables sampled from a Gaussian distribution ($\mathcal{N}(0,dt)$) named a Wiener process \citep{wiener1938homogeneous, ito1951multiple}. This stochastic process represents the disordered nature of the unresolved scales which occur at a timescale much smaller than the observed time interval, $dt$ \citep{peirano2006meanfield}.
$\Gbold$ and $\Bbold$ are drift and diffusion tensor coefficients, respectively.
As the flow configuration is homogeneous isotropic turbulence, the mean velocity is zero in all directions, and the drift and diffusion parameters are isotropic \citep{popebook}. Therefore, the model parameters can be treated as $\tens{G}=G \tens{I}$ and $\tens{B}=B \tens{I}$ \citep{popebook}. The drift coefficient is related to the fluid timescale, $T$ by $T = 1/G$ as described by \citet{popebook}. In this study, we determine the unclosed terms $G$ and $B$ with a neural network for each term, $G=G^{NN}$ and $B=B^{NN}$.

\subsubsection{Neural network model}
To construct a data-driven model, we first identify the flow features that should be used as model inputs. \citet{haworth1986generalized} stated that the model terms $\Gbold$ and $\Bbold$ depend on the turbulent kinetic energy, $k_{sgs}$, the turbulent dissipation rate, $\varepsilon_{sgs}$, and velocity gradients which we exclude here as the flow is homogeneous with no spatial gradients. Additionally, in the LES approach there is a dependence on the filter width, $\Delta$ (where $\Delta = \sqrt[3]{\mathcal{V}_c}$, $\mathcal{V}_c$ is the volume of a computational cell). Therefore, our model inputs are the subgrid time scale $\tau_{sgs}^* = \tau_{sgs} / \tau_k = k_{sgs@p}/(\varepsilon_{sgs@p} \tau_k)$, and subgrid length scale $\Delta^* = \Delta / \eta_{sgs}$. We have normalised our model inputs using the Kolmogorov length scale \citep{kolmogorov1941dissipation}, $\eta_{sgs}$ ($\eta_{sgs}=\sqrt[4]{\nu_f^3 / \varepsilon_{sgs@p}}$, where $\nu_f$ is the kinematic viscosity) and time scale, $\tau_k=\sqrt{\nu_f / \varepsilon_{sgs@p}}$. The Kolmogorov velocity scale is $u_k=\eta_{sgs} / \tau_k$. 

The model can be discretised in time using the numerical scheme presented by \citet{minier2003schemes}. This scheme is an adaptation of the explicit Euler-Maruyama scheme for SDEs which uses an exponential for the ratio between the timestep and fluid or particle timescale (such as $\exp(-\Delta t / \tau_p)$), which makes the scheme stable for any combination of $\Delta t / \tau_p$ or $\Delta / T$. The scheme is given as follows:
\begin{linenomath}
\begin{eqnarray}
    \usi^{n+1} &=& \usi^n\, \mathrm{exp} (-\Delta t / T^n) + \boldsymbol{c}_{(i)}^n\, T^n [1- \mathrm{exp}(-\Delta t/T^n)] + \boldsymbol{\gamma}_{(i)}^n \label{eq:analyticalSch1}, \\
    \boldsymbol{c}_{(i)}^n\, T^n &=& \left[- \boldsymbol{\nabla} \widetilde{p} 
                    + \nu_f\, \boldsymbol{\nabla} \cdot \boldsymbol{\nabla} \ufFilti
                    + (\upFilti - \ufFilti) \cdot \boldsymbol{\nabla} \ufFilti \right] \, T^n 
                    + \ufFilti, \\
    \boldsymbol{\gamma}_{(i)}^n &=& \sqrt{0.5\, (B^{NN})^2\, T^n [1 - \mathrm{exp}(-2\Delta t/T^n)]} \, \boldsymbol{\xi}_{(i)}, \label{eq:gamma}
\end{eqnarray}
\end{linenomath}
where the notation $(\cdot)_{(i)}$ represents the $i$th component of a vector, and the notation $(\cdot)^{n}$ represents the $n$th timestep. $\boldsymbol{\xi}_{(i)}$ is a random value sampled from a distribution $\mathcal{N}(0,1)$. $T$ is the fluid timescale found by $T=1/G^{NN}$, $\boldsymbol{c}$ is the drift vector. To account for the unknown subgrid contributions in the initial fluid velocity, we make the initial condition for $\up$ and $\us$ such that $\us^{t=0} = \up^{t=0} = \ufFilt^{t=0} + \sqrt{2 \, k_{sgs}^{t=0}/3} \, \boldsymbol{\xi}$.

\begin{figure}[ht]
    \centering
    \includegraphics[scale=1]{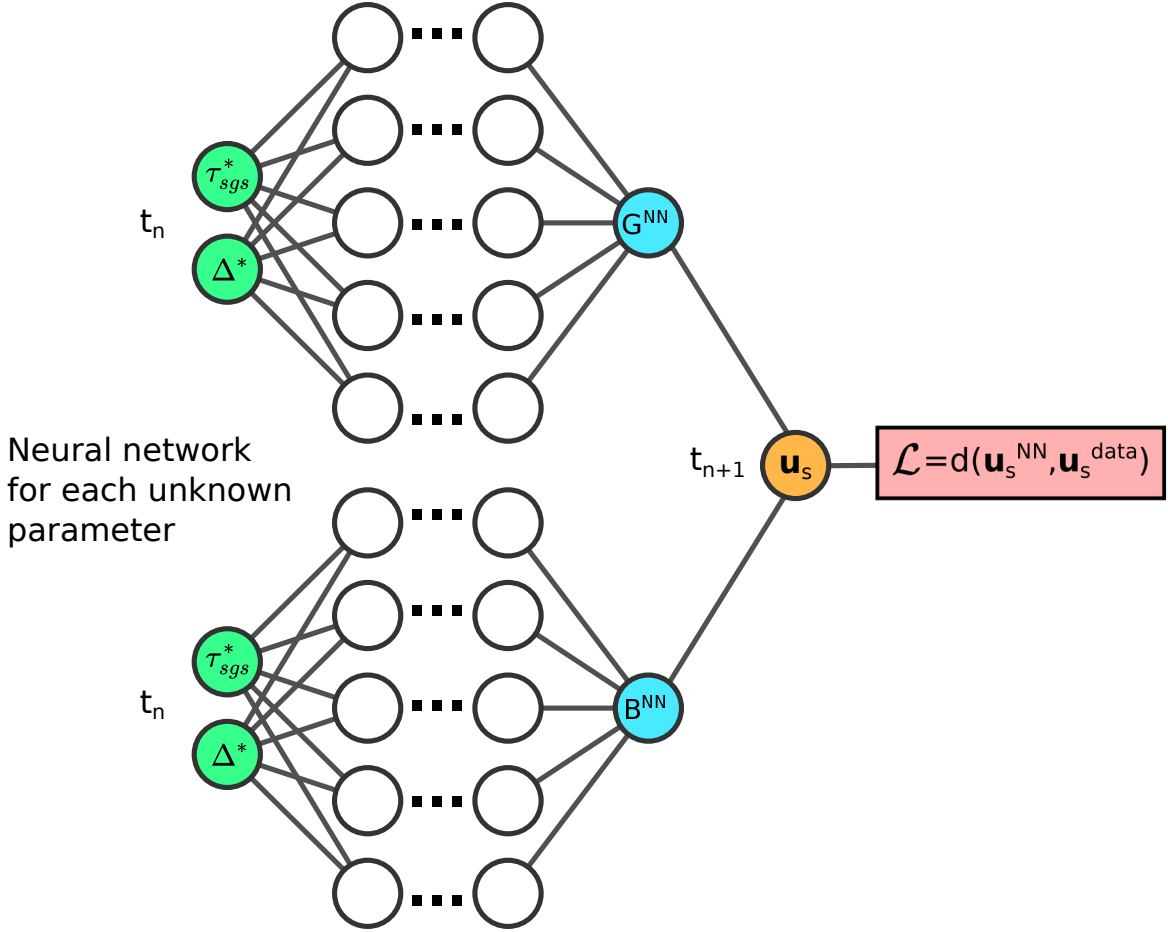}
    \caption{Schematic overview of neural SDE training and architecture. Simulation is trained on interval $t=n$ to $t=n+1$. Network inputs are given as green nodes, and outputs are blue nodes. The output parameters, $G^{NN}$ and $B^{NN}$ are then used to compute $\us^{NN}$ using filtered DNS data. Loss $\mathcal{L}$, \eref{eq:lossFinal}, is computed by comparing $\us^{NN}$ to DNS data $\us^{data}=\uf$.}
    \label{fig:schematicArch}
\end{figure}

Our target is to learn the closures for $G^{NN}$ and $B^{NN}$ such that the $\us$ predicted by our integration of \eref{eq:analyticalSch1} is the same as $\us$ obtained from DNS simulations (\fref{fig:schematicArch}). Therefore, we learn the increment of $d\us$ between one timestep where simulation data has been saved, $t=n$, and the next timestep where simulation data was saved $t=n+1$. For DNS, it is infeasible to save all timesteps due to storage limitations. Therefore, simulation data was saved at intervals of 40 DNS timesteps as shown in the timeline in \fref{fig:timeline}. We found it to be more effective to learn using the change in $\us$ over this large interval rather than a single timestep as the change in $\us$ over a single DNS timestep was very small. When training a neural network, this can cause a so-called vanishing gradients issue where the training does not converge, or converges to an incorrect solution \citep{hochreiter1998vanishing, hochreiter2001gradients}. Therefore, we minimise the following loss function based on the evolution of the SDE \eqref{eq:analyticalSch1} from timestep $t=n$ to $t=n+1$:
\begin{linenomath}
\begin{equation}
\begin{split}
    \mathcal{L} = \frac{1}{N_s}\sum^{N_s}_{i=1} &
    \Biggl[(\boldsymbol{u}_{s,(i)}^{n+1,NN} - \boldsymbol{u}_{s,(i)}^{n+1, data})^2 \\
    &+ \log \left(\frac{[\boldsymbol{u}_{s,(i)}^{t=n+1,data} - (\boldsymbol{u}_{s,(i)}^n\, \mathrm{exp} (-\Delta t / T^n) + \boldsymbol{c}_{(i)}^n\, T^n [1- \mathrm{exp}(-\Delta t/T^n)])]^2    }{ 0.5\, (B^{NN})^2\, T [1 - \mathrm{exp}(-2\Delta t/T)] }\right) \Biggl],
    \label{eq:lossFinal}
\end{split}
\end{equation}
\end{linenomath}
where $N_s$ is the number of samples used for training (one particle-timestep pair is one sample). The first term in the loss function compares the mean-squared error in the output of the NN with the data, which requires a sufficiently large batch size and number of samples to minimise the statistical error in the random term, $\boldsymbol{\gamma}$ \eqref{eq:gamma}.
Therefore, to minimise statistical error we took the average of the first term in \eqref{eq:lossFinal} over 50 independent realisations (50 separate values of $\boldsymbol{\xi}$). The second term in \eqref{eq:lossFinal} calculates the distance from $\boldsymbol{u}_{s,(i)}^{t=n+1,data}$ and the SDE distribution
\begin{equation}
    \mathcal{N}(\boldsymbol{u}_{s,(i)}^n\, \mathrm{exp} (-\Delta t / T^n) + \boldsymbol{c}_{(i)}^n\, T^n [1- \mathrm{exp}(-\Delta t/T^n)]), 0.5\, (B^{NN})^2\, T^n [1 - \mathrm{exp}(-2\Delta t/T^n)] ).
\end{equation}    
We take the logarithm of this to prevent huge fluctuations in the loss as the diffusion term approaches zero (exploding gradients issue \citep{hochreiter2001gradients}).

Results created from deep learning architectures such as neural networks are sensitive to parameters such as the number of layers, number of nodes per layer, the activation functions and the inital condition of the weights and biases. We determined these so-called hyperparameters through manually tuning. From this, we found a suitable architecture to be four layers with 120 nodes per layer.
The hidden layer biases were initialised to be zero, and weights were initialised as $\mathcal{N}(0, w_h^2)$ where $w_h = \sqrt{2 / n_{node,j}}$ and $n_{node,j}$ is the number of nodes in the input weight tensor for layer j \citep{he2015delving}. The drift term neural network used ReLu activation function for hidden layers, where the output of each node in the hidden layer, $y$, is $y=\max(0, w x + b)$ (for a simple 1 node layer, where $w$ is the node weight, $x$ is the node input, and $b$ is the node bias). The diffusion term neural network used hyperbolic tangent activation function for the hidden layers.
The output layer weights for $G^{NN}$ were initialised as $\mathcal{U}(-w_h, w_h)$. The output layer weights for $B^{NN}$ were initialised as $\mathcal{U}(-w_h, w_h)$. The output layer biases were fixed at zero in all networks trained. The output layers used a softplus activation function to ensure that $G^{NN}$ and $B^{NN}$ were positive-definite. The network was trained stochastic gradient-descent with a learning rate of $10^{-4}$. We implemented our neural network modelling framework using Keras \citep{chollet2015keras} and Tensorflow 2.4 \citep{tensorflow2015}. Our code was based on the code provided on GitHub by \citet{dietrich2021learning}.

Finally, as training a neural network is sensitive to local minima, some steps must be taken to optimise this process. The first step being to normalise our NN inputs to below one to minimise any large `exploding' gradients \citep{hochreiter2001gradients}. Therefore, we scale the normalised timescale as $\tau_{sgs}^{*} / C_\tau$ and the length-scale as $\Delta^{*} / C_\Delta$, where $C_\Delta$ and $C_\tau$ were chosen such that the maximum value of the normalised length-scale and timescale in the training set was 1. Additionally, optimisation algorithms such as stochastic gradient descent converge best if the initial output is near the expected, correct value \citep{sutskever2013importance}. This cannot be achieved by applying a constant scaling value, as when the flow becomes laminar, the drift term $G^{NN}$ should approach infinity ($T^{NN} \rightarrow 0$) and $B^{NN}$ should approach zero. Through trial and error, we found the following NN scalings to produce a close initial output which converged to a correct solution:
\begin{eqnarray}
    G^{NN} &=& \sqrt{\nu} \left[f^{NN}\left(\frac{\tau_{sgs}^{*}}{C_\tau}, \frac{\Delta^{*}}{C_\Delta}\right)\right]^{-1} \frac{C_G C_\tau}{\tau_{sgs}^{*}}, \label{eq:GNNscale} \\
    B^{NN} &=& \nu f^{NN}\left(\frac{\tau_{sgs}^{*}}{C_\tau}, \frac{\Delta^{*}}{C_\Delta}\right) \frac{C_B \Delta^{*}}{C_\Delta},
    \label{eq:BNNscale}
\end{eqnarray}
where $f^{NN}(\cdot)$ is a non-linear function output from a neural network, that depends on the parameters enclosed in brackets. $C_G$ and $C_B$ were chosen empirically as 5 and 1000. These constants and scaling were chosen to provide a NN that produced a kinetic energy decay close to the target value (DNS data) at initialisation (epoch 0). Scaling $G^{NN}$ with $1/\tau_{sgs}^{*}$ ensured that the integral scale $T^{NN}=1/G^{NN}\rightarrow0$ when the flow becomes laminar. 

\section{Results}
\subsection{\textit{A priori} analyses}

\begin{figure}[ht]
    \centering
    \begin{tabular}{c c}
        (a) $\Delta=3\Delta_{DNS}$ & (b) $\Delta=5\Delta_{DNS}$ \\
        \includegraphics{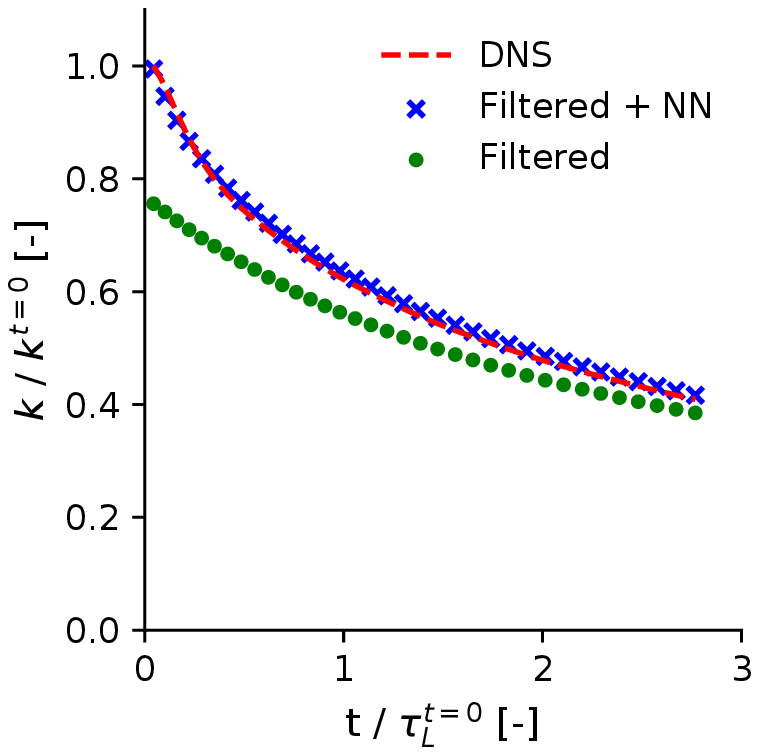} & 
        \includegraphics{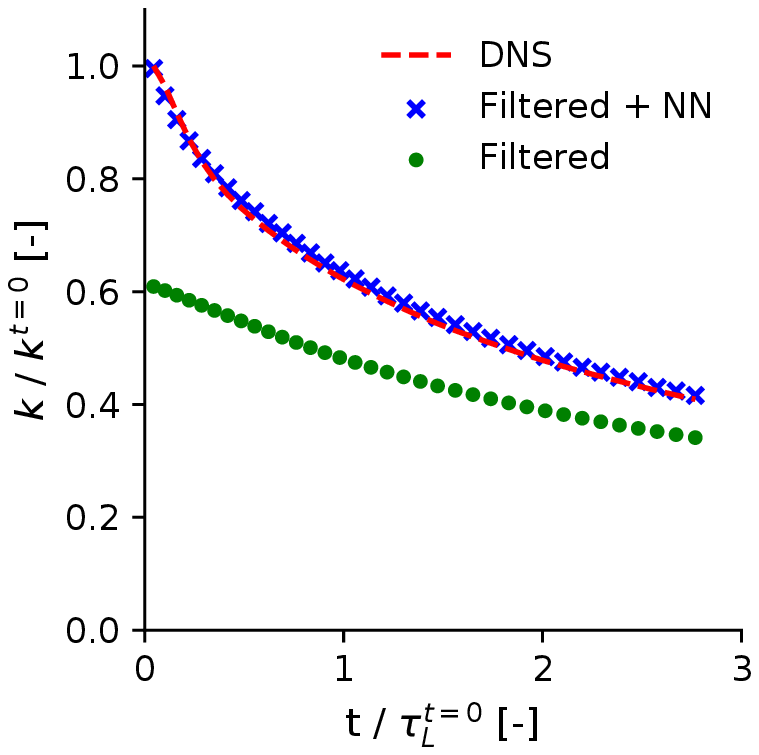} \\
        (c) $\Delta=7\Delta_{DNS}$ & (d) $\Delta=9\Delta_{DNS}$ \\
        \includegraphics{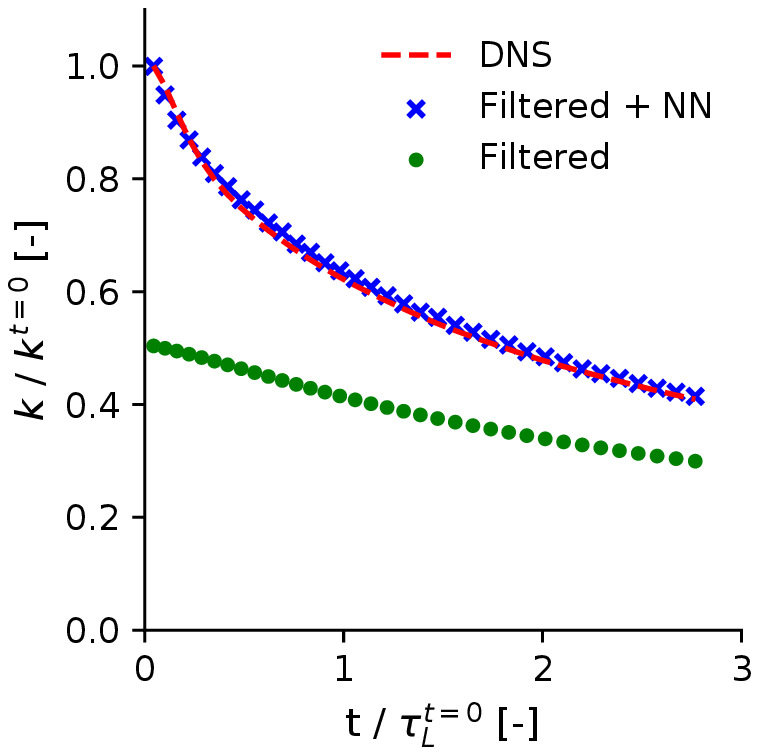} &
        \includegraphics{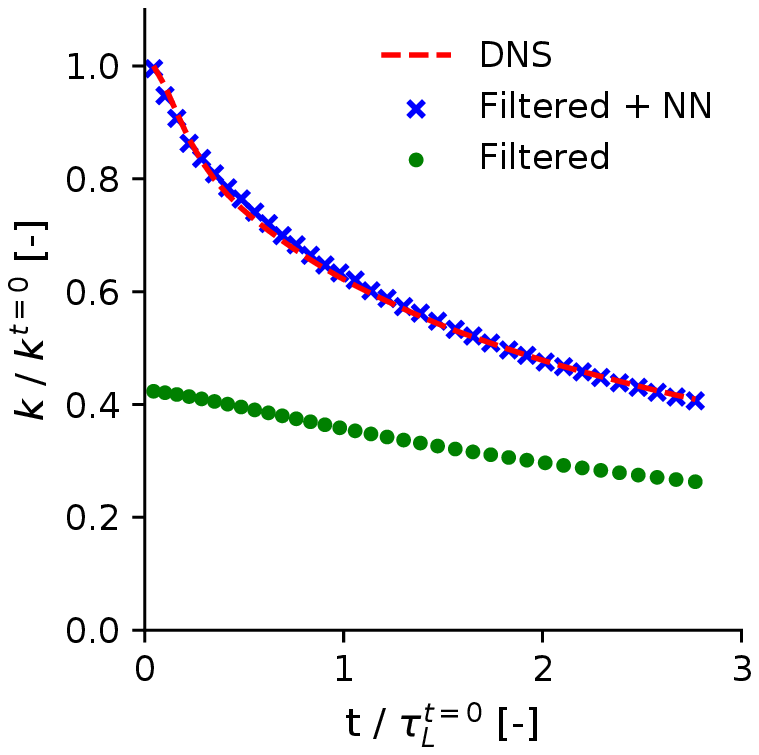} 
    \end{tabular}
    \caption{Decay of tracer particle fluctating kinetic energy in \textit{a priori} tests of homogeneous isotropic turbulence at $\ReInit=33$ for various filter sizes (a) $3\Delta_{DNS}$, (b) $5\Delta_{DNS}$, (c) $7\Delta_{DNS}$, (d) $9\Delta_{DNS}$. Flow configuration of homogeneous isotropic turbulence is given in Table \ref{tab:simparams}.  
    \label{fig:aprioriHITtracer}}%
\end{figure}

First we assessed the model by \textit{a priori} analysis at varying filter width in the $\ReInit=33$ case. We quantified model performance by comparing the fluctuating kinetic energy ($k=tr[(\boldsymbol{u}-\langle\boldsymbol{u}\rangle) \otimes (\boldsymbol{u}-\langle\boldsymbol{u}\rangle)]$ where $\langle\boldsymbol{u}\rangle$ is the domain-averaged velocity, which is equal to zero) for tracer particles ($\us=\up$) (\fref{fig:aprioriHITtracer}). This is compared to the fluctuating kinetic energy based on the DNS tracer particle velocity, $\ufAtP$, and based on the filtered DNS tracer particle velocity, $\ufFiltAtP$. Our neural SDE recovered the DNS kinetic energy at all filter widths shown. For larger filter widths, the effect of unresolved fluctuations on the kinetic energy became more apparent. In the largest filter width, the relative error in $k$ compared to the DNS data at the first time step was 0.5\% for $\us^{NN}$ and 41.3\% for $\ufFilt$ (without the model). For the smallest filter width, the relative error in $k$ in the first time step was 0.3\% for our neural SDE. The relative error in the filtered DNS data compared to the DNS data was 8\% in the first time-step. These results show that the dispersion model recovers filtered out scales in simulations with a range of filter widths.

\begin{figure}
    \begin{tabular}{c c}
        (a) $\ReInit=9$ & (b) $\ReInit=105$ \\
        \includegraphics{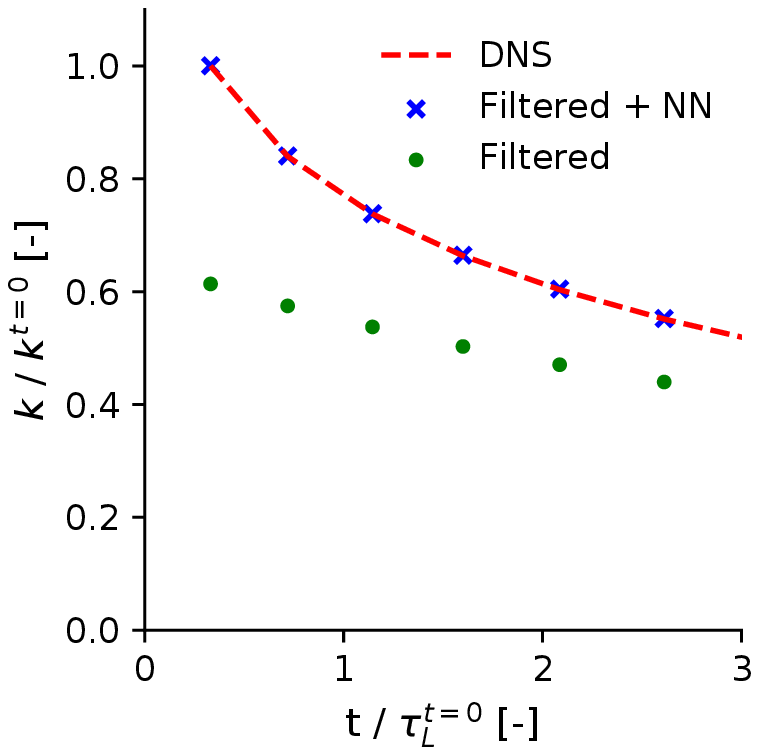} &
        \includegraphics{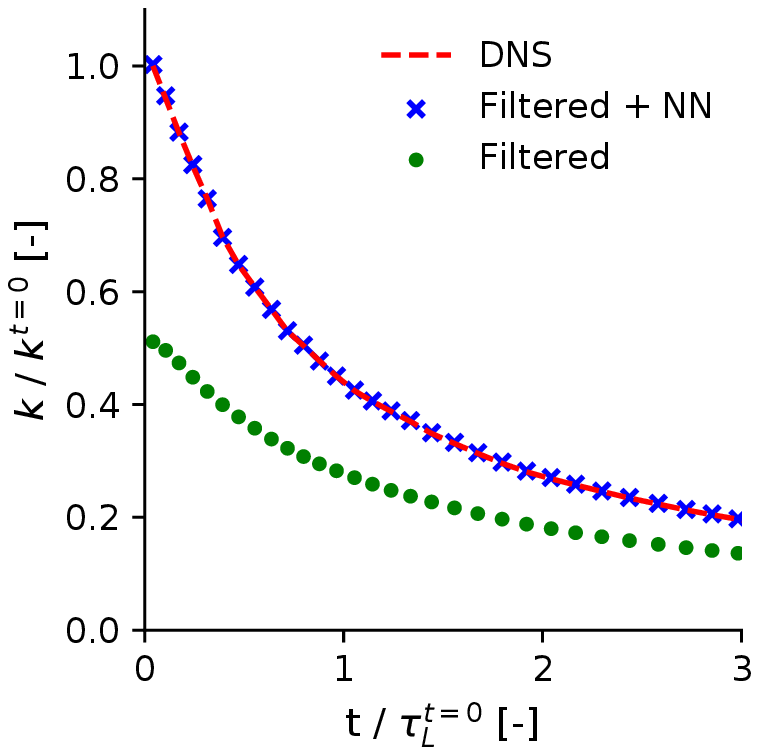}
    \end{tabular}
    \caption{Decay of tracer particle fluctuating kinetic energy in \textit{a priori} tests of homogeneous isotropic turbulence at (a) $\ReInit=9$ and (b) $\ReInit=105$ for a filter size, $\Delta=7\Delta_{DNS}$. Flow configuration of homogeneous isotropic turbulence is given in Table \ref{tab:simparams}.}
    \label{fig:aprioriHITtracerVaryRe}
\end{figure}

To assess generalisation to varying homogeneous isotropic turbulence flow configurations, we performed a priori tests at varying $\ReInit$ (Figures \ref{fig:aprioriHITtracer} and \ref{fig:aprioriHITtracerVaryRe}). For all three $\ReInit$ tested, our developed neural SDE gave good agreement with the DNS data. The maximum error was $<1\%$ for $\ReInit=9$ (\fref{fig:aprioriHITtracer}c), $7\%$ for $\ReInit=33$ (\fref{fig:aprioriHITtracerVaryRe}a), $4\%$ for $\ReInit=105$ (\fref{fig:aprioriHITtracerVaryRe}b). In contrast, the maximum error in the filtered dataset with no model was $31\%$ for $\ReInit=9$ and $47\%$ for $\ReInit=33$ and $\ReInit=105$. This \textit{a priori} analysis showed a significant improvement in the accuracy of fluctuating kinetic energy from using the proposed neural SDE dispersion model. Furthermore, the model is also consistent with the decay from turbulent to laminar flow suggesting that the neural SDE integral timescale is consistent with the underlying fluid solver \citep{minier2014, minier2021methodology}.

\begin{figure}
    \centering
    \begin{tabular}{c c}
        (a) $St^{t=0} = 0.1$ & (b) $St^{t=0} = 5$\\
        \includegraphics{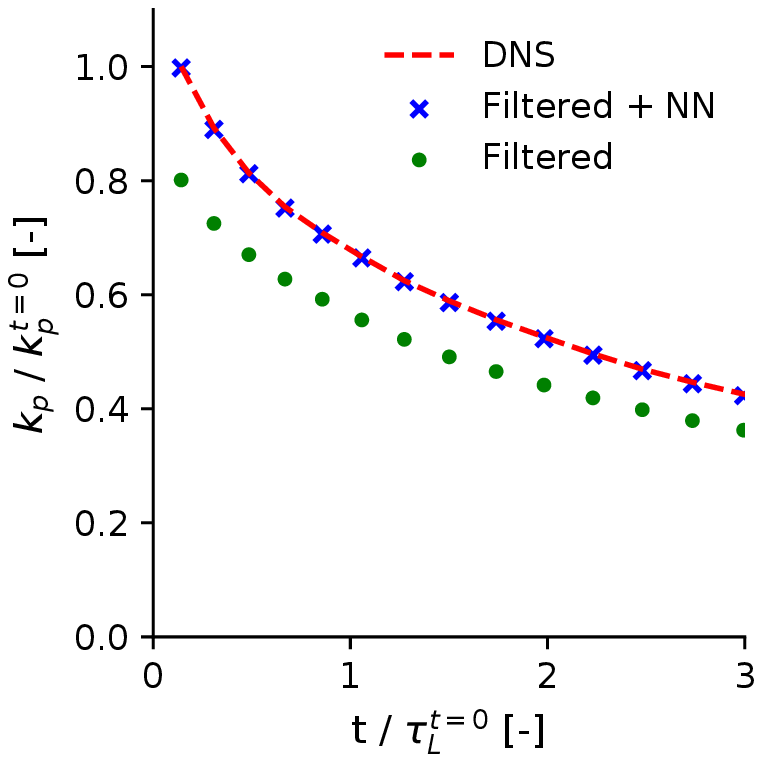} &
        \includegraphics{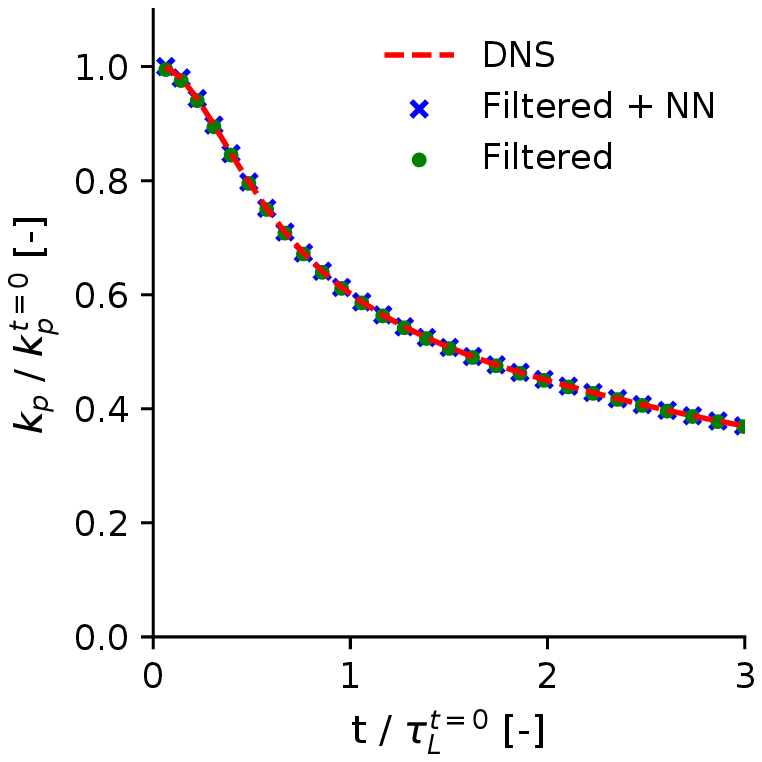}
    \end{tabular}
    \caption{Decay of inertial particle fluctuating kinetic energy in \textit{a priori} tests of homogeneous isotropic turbulence at (a) $St^{t=0}=0.1$ (low inertia), (b) $St^{t=0}=5$ (high inertia). Results show $\ReInit=33$ at filter size $\Delta = 7 \Delta_{DNS}$. For $St^{t=0}=5$, all markers are overlapping.}
    \label{fig:aprioriInertia}
\end{figure}

Particle fluctuating kinetic energy decay was assessed through \textit{a priori} analyses at varying Stokes number, $St^{t=0}=\tau_p / \tau_k^{t=0}$ (\fref{fig:aprioriInertia}). For low inertia particles ($St^{t=0}=0.1$, \fref{fig:aprioriInertia}a), the NN model predicted particle fluctuating kinetic energy to within 5\% error throughout the simulation and the filtered DNS data under-predicts particle fluctuating kinetic energy by a maximum of 19\%. For high inertia particles ($St^{t=0}=5$, \fref{fig:aprioriInertia}b), the results from both filtered tests (with and without our model) match the DNS data within 1\%. This agrees with literature as larger particles with large drag response times are less sensitive to fast fluctuations in the carrier fluid, meaning that turbulent dispersion is not influential \citep{balachandar2010review}.

From the presented \textit{a priori} analyses, we observe that our model is independent of filter width. In contrast, the filtered kinetic energy with no model showed high mesh dependence. Additionally, we have shown that our model works for a range of $\ReInit$, including high $\ReInit=105$ which is also not included in the training data. Finally, we see the effect of the model on inertial particles, which showed that low inertia particles are influenced by the dispersion model. The influence of filtering was found to be negligible for high inertia particles, which agrees with literature \citep{fede2006numerical}.

\subsection{\textit{A posteriori} analyses}
\begin{figure}
    \centering
    \begin{tabular}{c c}
        (a) $\Delta=3\Delta_{DNS}$ & (b) $\Delta=5\Delta_{DNS}$ \\
        \includegraphics{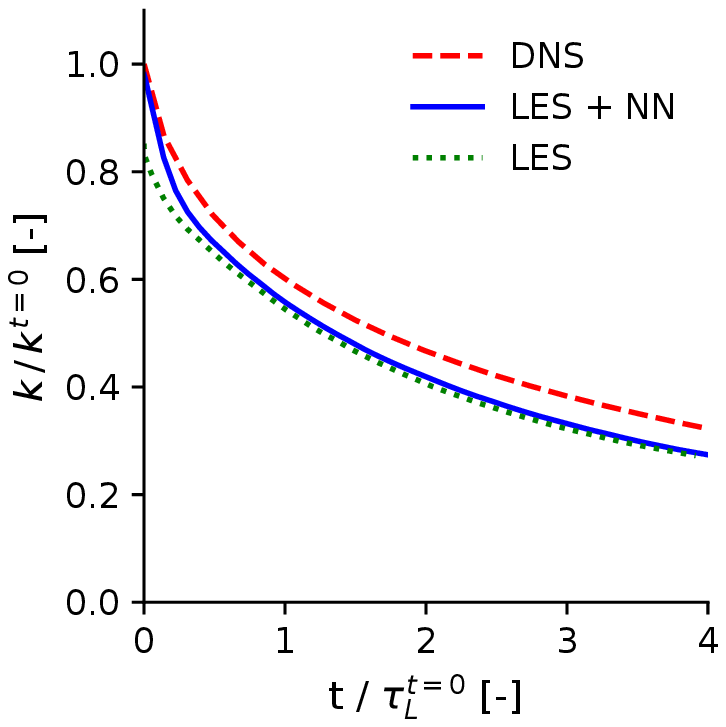} & 
        \includegraphics{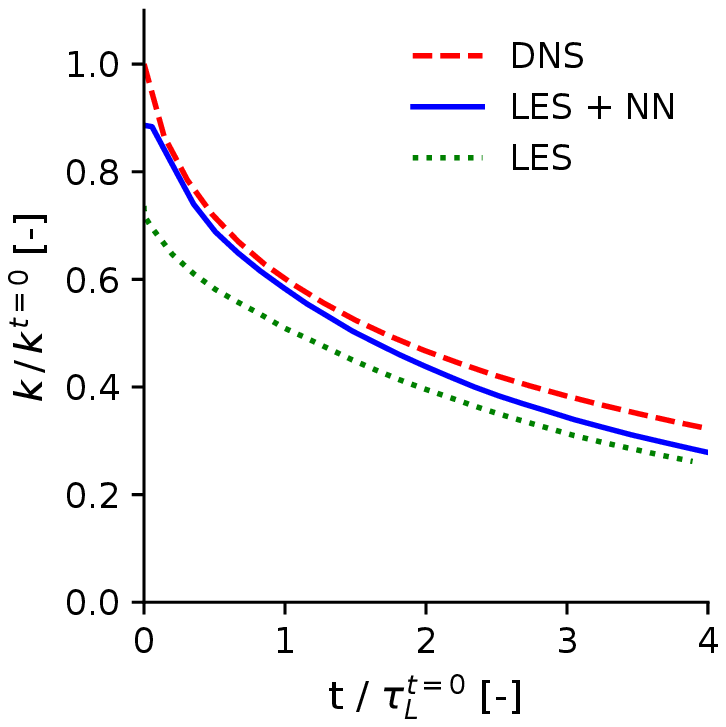} \\
        (c) $\Delta=7\Delta_{DNS}$ & (d) $\Delta=9\Delta_{DNS}$ \\
        \includegraphics{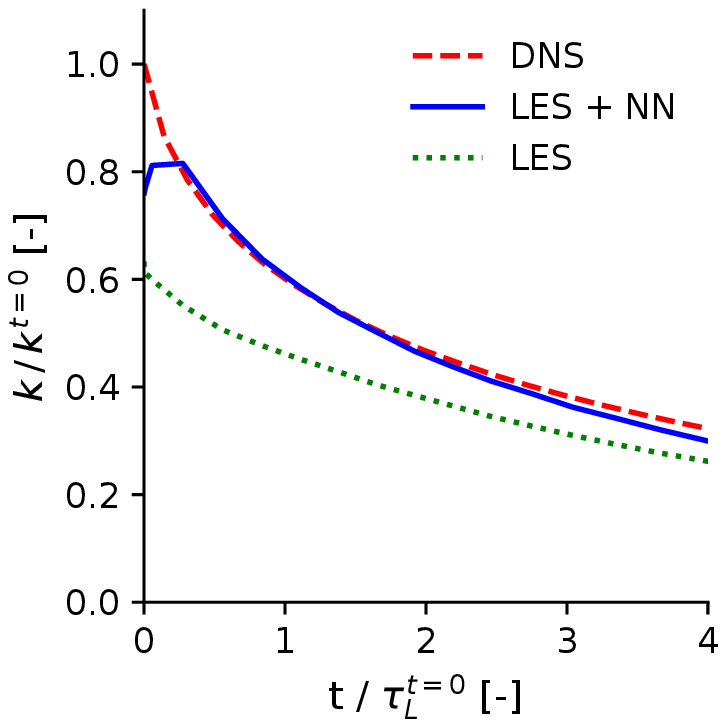} &
        \includegraphics{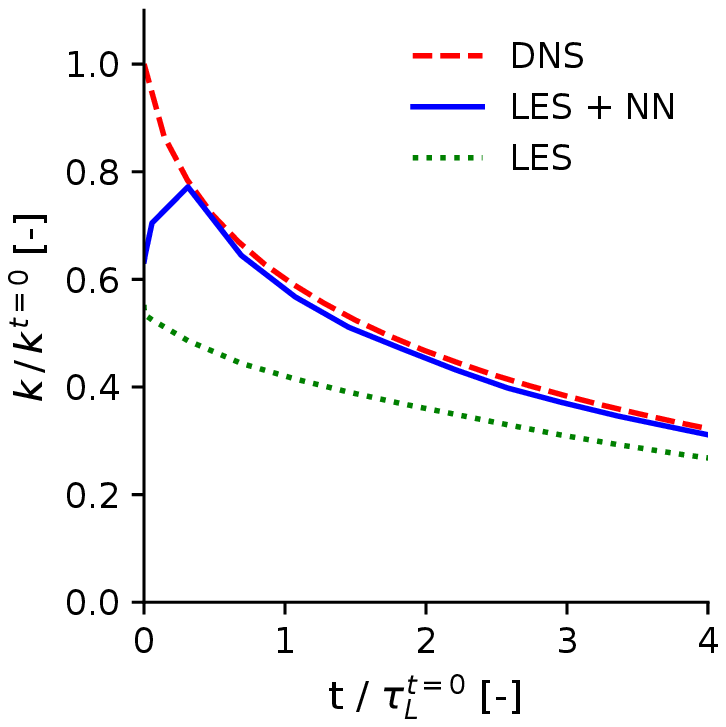} \\
    \end{tabular}
    \caption{Decay of tracer particle fluctuating kinetic energy in \textit{a posteriori} tests of homogeneous isotropic turbulence at $\ReInit=33$ for various mesh resolutions, $\Delta$, (a) $3\Delta_{DNS}$, (b) $5\Delta_{DNS}$, (c) $7\Delta_{DNS}$, (d) $9\Delta_{DNS}$. Flow configuration of homogeneous isotropic turbulence is given in Table \ref{tab:simparams}.}.
    \label{fig:apostHITtracer}
\end{figure}

To assess particle dispersion \textit{a posteriori}, we performed simulations of tracer dispersion to assess the performance of the neural SDE in the presence of additional errors such as subgrid modelling and numerical discretisation. This will alter the evolution of fluid properties used as network inputs and is therefore a necessary test to ensure robustness \citep{boivin2000prediction}. The domain had the same dimensions and initial parameters described in \sref{sec:groundTruthGeneration}. Similar to the \textit{a priori} tests, the \textit{a posteriori} simulations show good agreement with the DNS data for the LES with NN for all mesh resolutions, $N_{cell} = \{85^3, 51^3, 36^3, 28^3\}$ (\fref{fig:apostHITtracer}). The LES without the model cases also agree well with the DNS data for high mesh resolution, $N_{cell} = 85^3$ (\fref{fig:apostHITtracer}a). However, LES on the coarser grids show an under-prediction in $k$ (\fref{fig:apostHITtracer}b-d). For $9 \Delta_{DNS}$ (\fref{fig:apostHITtracer}d), our NN under-predicted $k$ by 14\% at the start of the simulation ($t/\tau_L^{t=0}=0.1$) compared to 40\% under-prediction for LES alone without the model. 

\begin{figure}
    \centering
    \begin{tabular}{c c}
        (a) $\ReInit=9$ & (b) $\ReInit=105$ \\
        \includegraphics{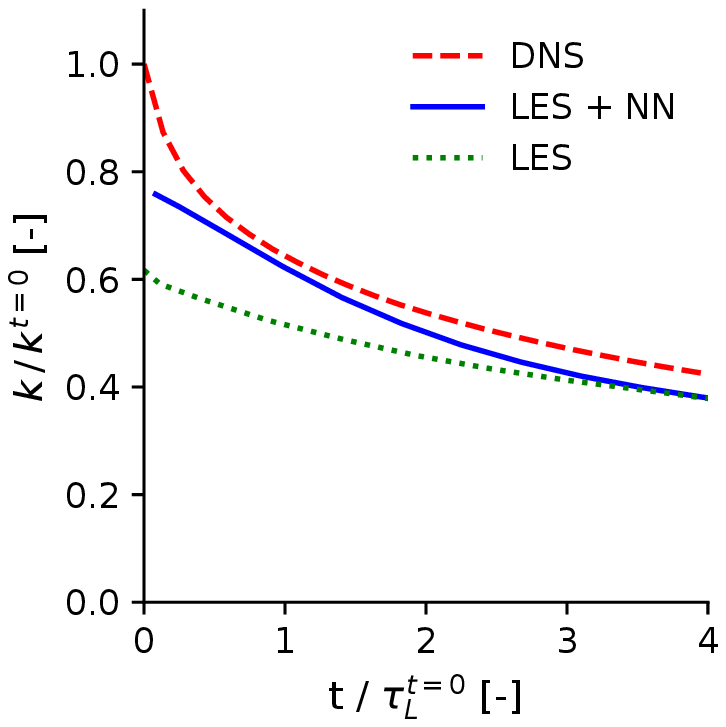} &
        \includegraphics{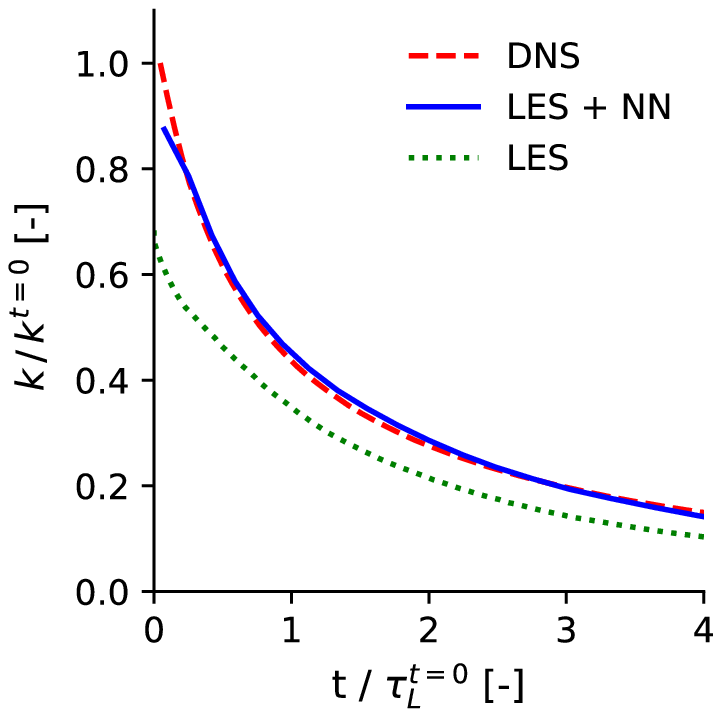} \\
    \end{tabular}
    \caption{Decay of tracer particle fluctuating kinetic energy in \textit{a posteriori} tests of homogeneous isotropic turbulence at (a) $\ReInit=9$ and (b) $\ReInit=105$. LES simulations performed with mesh resolution $\Delta = 7 \Delta_{DNS}$. Flow configuration of homogeneous isotropic turbulence is given in Table \ref{tab:simparams}.}
    \label{fig:apostHITvaryReTracerTKE}
\end{figure}

We have assessed whether the $\us$ model is mesh independent in \textit{a posteriori} analyses (see \fref{fig:apostHITtracer}). We performed simulations on a coarse mesh at larger $\ReInit$ (\fref{fig:apostHITvaryReTracerTKE}). For all $\ReInit$, the LES data under-estimates the $k$ predicted by DNS by around 40\%. For $\ReInit=9$ and 105, the LES with no model underestimates the DNS kinetic energy by 20\% at $t / \tau_L^{t=0} = 1$, whereas our dispersion model over-estimated the kinetic energy by 5\% (\fref{fig:apostHITvaryReTracerTKE}).

For our \textit{a posteriori} analysis of inertial particle dispersion, we decompose the particle velocity in two parts \citep{fevrier2005partitioning}. One is a component that is spatially correlated (the mean Eulerian particle velocity field), and the other component is uncorrelated and quasi-random in nature, caused by crossing-trajectories. The uncorrelated component is amplified as inertia increases and the heavy particles' trajectories will cross with the mean flow \citep{fevrier2005partitioning}. The correlated component is dominant for low inertia particles \citep{fevrier2005partitioning}.
\citet{moreau2010development} applied this velocity partitioning analysis to decaying homogeneous isotropic turbulence, with a single-flow realisation and a large number of particles ($N_p = 8\times 10^7$, $N_{ppc} \approx 300$, and $N_{cell}=64^3$). For our \textit{a posteriori} analyses of inertial particle dispersion, we used $N_p = 10^7$ ($N_{ppc} > 200$ when $N_{cell}=36^3$, where $N_{ppc}$ is the number of particles per cell) to minimise statistical error in calculation of second-order velocity moments \eqref{eq:uncorrelatedVel}.

By following these studies, we calculated the correlated velocity in each computational cell as
\begin{equation}
    \upFilt = \frac{1}{N_{ppc}} \frac{\sum^{N_{ppc}}_i \up^{(i)} g(\boldsymbol{x}^{(i)} - \boldsymbol{r})}
    {\sum^{N_{ppc}}_i g(\boldsymbol{x}^{(i)} - \boldsymbol{r})}
    \label{eq:correlatedVel}
\end{equation}
and the uncorrelated velocity of a particle, $i$, was found by
\begin{equation}
    \delta \up^{(i)} = \up^{(i)} - \upFiltAtP^{(i)}.
    \label{eq:uncorrelatedVel}
\end{equation}

With these definitions of velocity partitions, we can calculate the corresponding kinetic energy partitions. The domain-averaged particle total kinetic energy, $k_p$ can be found as
\begin{equation}
    k_p = \frac{1}{N_p} \sum^{N_p}_i tr(\up^{(i)} \otimes \up^{(i)}),
\end{equation}

The domain-averaged correlated and uncorrelated components of particle fluctuating kinetic energy are found as
\begin{eqnarray}
    \widetilde{k}_p &=& 
    \frac{\sum^{N_{cell}}_i 
        tr(\upFilt^{(i)} \otimes \upFilt^{(i)})\, \widetilde{n}_p^{(i)}}
        {\sum^{N_{cell}}_i \widetilde{n}_p^{(i)}}, \\
    \delta k_p &=& 
    \frac{1}{N_p}\sum^{N_p}_i tr(\delta \up^{(i)} \otimes \delta \up^{(i)})
\end{eqnarray}
where $\widetilde{n}_p$ is the local particle number density.

The correlated and uncorrelated velocity components are sensitive to the size of the filter, $r$, used for computing $\upFilt$ in \eref{eq:correlatedVel} \citep{capecelatro2014numerical, capecelatro2015fluid}. When calculating the velocity partitioning, we use the adaptive filter proposed by \citet{capecelatro2014numerical} to compute $\upFilt$, which is given as
\begin{equation}
    r(\alpha_p) = \sqrt[3]{\frac{N_{ppc} \, d_p^3}{\alpha_p} },
\end{equation}
where $\alpha_p$ is the local solid volume fraction and $N_{ppc}$ is the number of particles locally available for sampling. As an initial guess of filter width is required to calculate $\alpha_p$, we set the initial $r_0$ as the same size as the cell size of the LES mesh and perform six iterations to converge to a steady value of $r$. The same initial conditions for calculating $r$ were used for analysing both the LES and DNS datasets.

\begin{figure}
    \includegraphics[trim={0 0 0 1cm},clip]{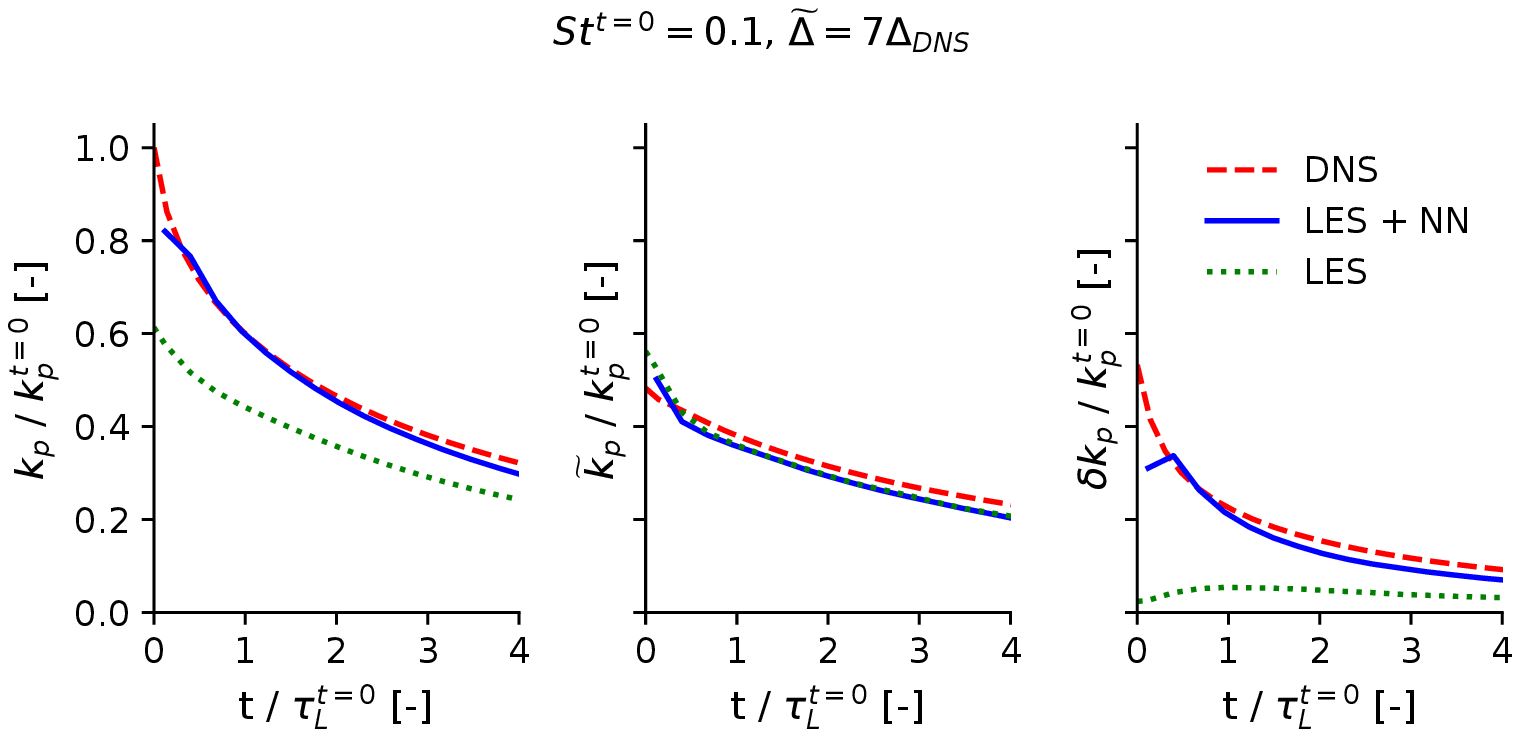} \\
    \includegraphics[trim={0 0 0 1cm},clip]{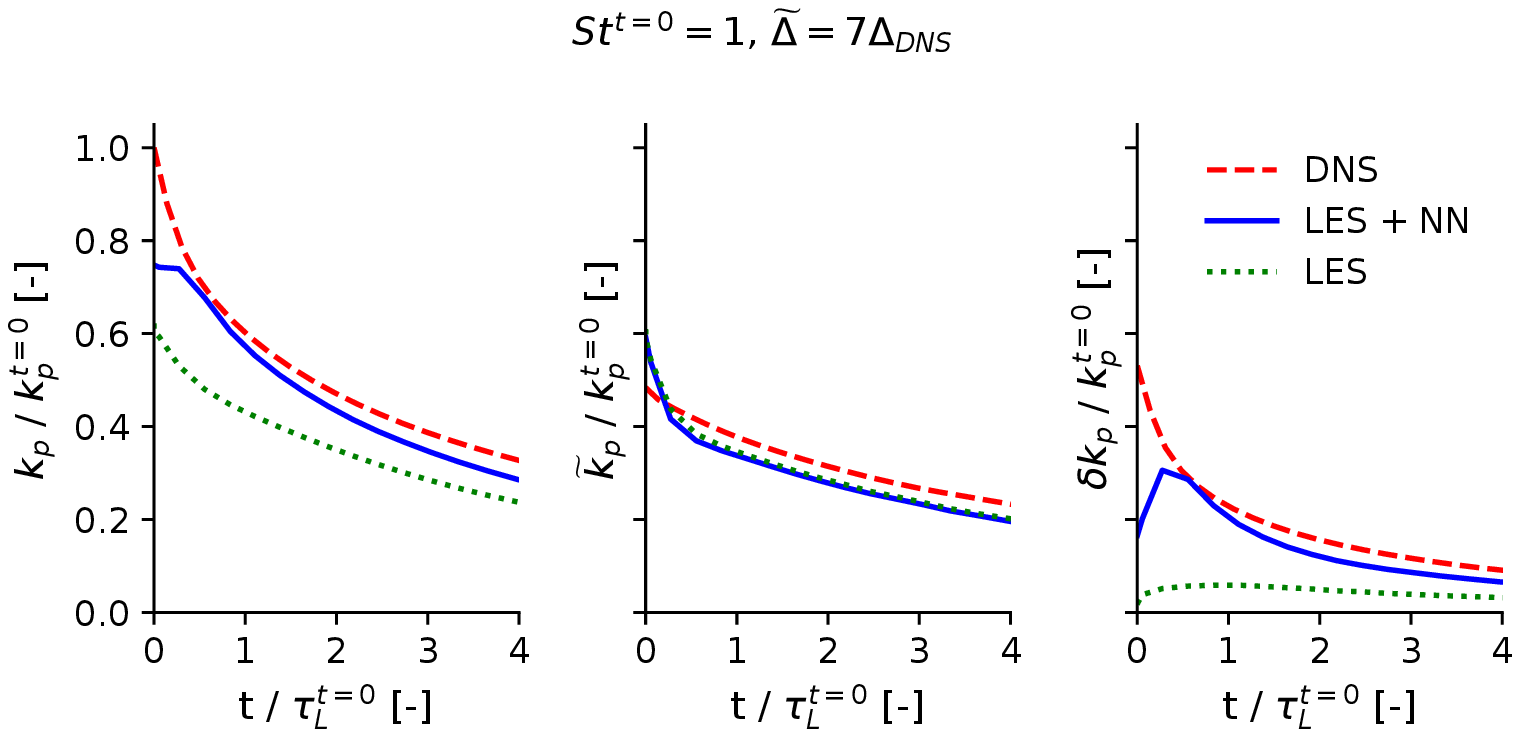} \\
    \includegraphics[trim={0 0 0 1cm},clip]{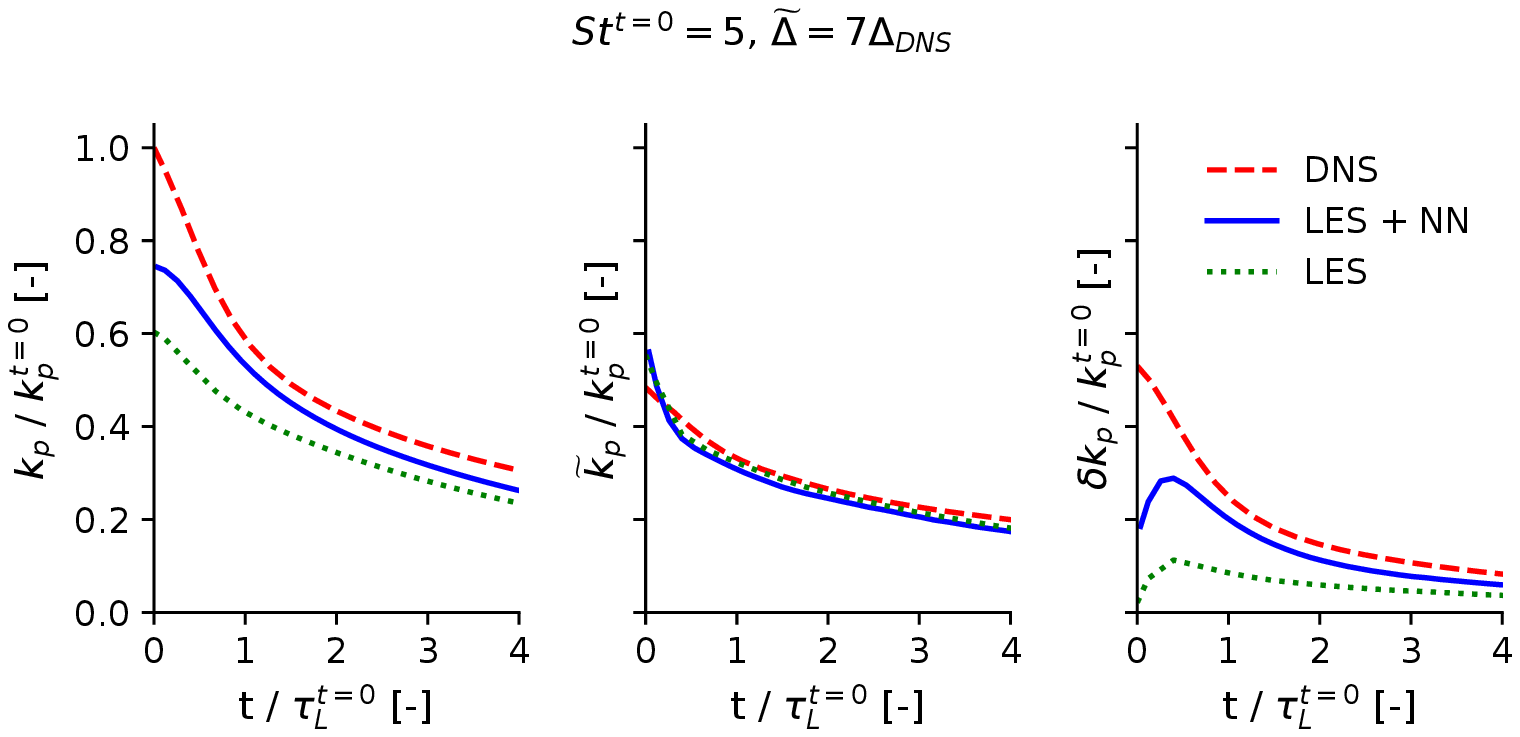} \\
    \caption{Time evolution of particle fluctuating kinetic energy components in \textit{a posteriori} tests of homogeneous isotropic turbulence, with varying $St^{t=0}$ number and filter size $\Delta=7\Delta_{DNS}$ at $\ReInit=33$. Rows are $St^{t=0}=0.1$, $St^{t=0}=1$, $St^{t=0}=5$ (top-to-bottom).
    }
    \label{fig:apostHITstTKE}
\end{figure}

In \fref{fig:apostHITstTKE}, we assess the total, correlated, and uncorrelated components of particle fluctuating kinetic energy from our dispersion model. For all $St^{t=0}$ number, the model with initial condition generator recovers the total particle fluctuating kinetic energy. We see that the model increases the uncorrelated (random) component, which is consistent as the stochastic term in our model follows a Gaussian distribution with zero mean. For all $St^{t=0}$ numbers, we can see that $\widetilde{k}_p$ is approximately same for DNS and both LES.
As particle inertia increases, the particle motion becomes increasingly uncorrelated with the mean fluid motion as shown by a larger $\delta k_p$ which agrees with previous studies \citep{fevrier2005partitioning, moreau2010development}.

\begin{figure}
    \includegraphics[trim={0 0 0 1cm},clip]{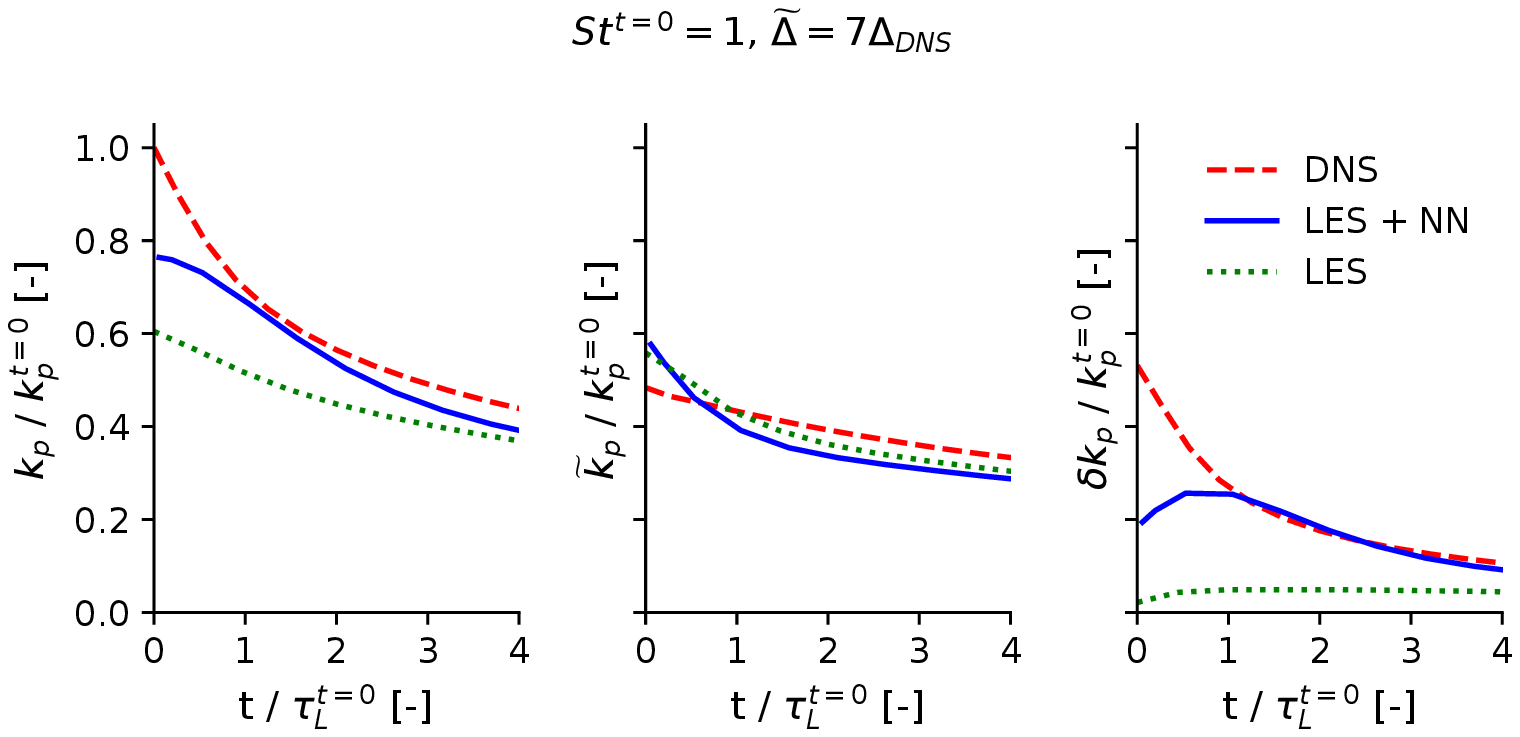}\\
    \includegraphics[trim={0 0 0 1cm},clip]{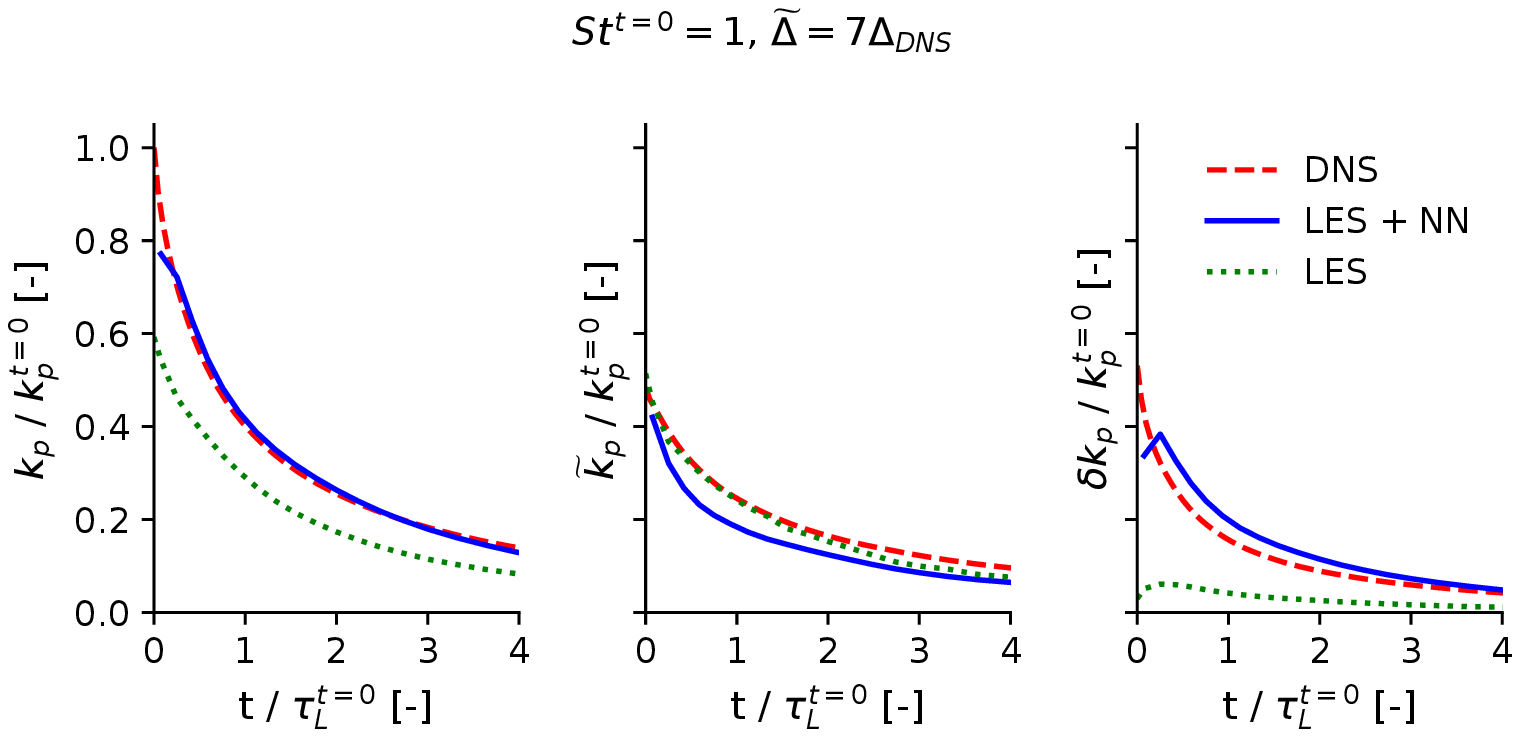} \\
    \caption{Time evolution of particle fluctuating kinetic energy components in \textit{a posteriori} tests of homogeneous isotropic turbulence, with varying $\ReInit$ at $St^{t=0}=1$. Filter size for LES was $\Delta=7\Delta_{DNS}$. Rows are (top) $\ReInit=9$, (bottom) $\ReInit=105$. 
    }
    \label{fig:apostHITinertialReTKE}
\end{figure}

In \fref{fig:apostHITinertialReTKE}, we assess the particle fluctuating kinetic energy decay at a low and high $\ReInit$. Similar to as shown for tracer particles (\fref{fig:apostHITvaryReTracerTKE}), our model has slightly underfit to the low $\ReInit$ case. For $\ReInit=105$ (not in training data), we achieve good agreement with the DNS predictions, particularly for $k_p$ and $\delta k_p$.

\begin{figure}
    \centering
    \begin{tabular}{c c}
        (a) $St^{t=0}=1$ & (b) $St^{t=0}=5$ \\
        \includegraphics{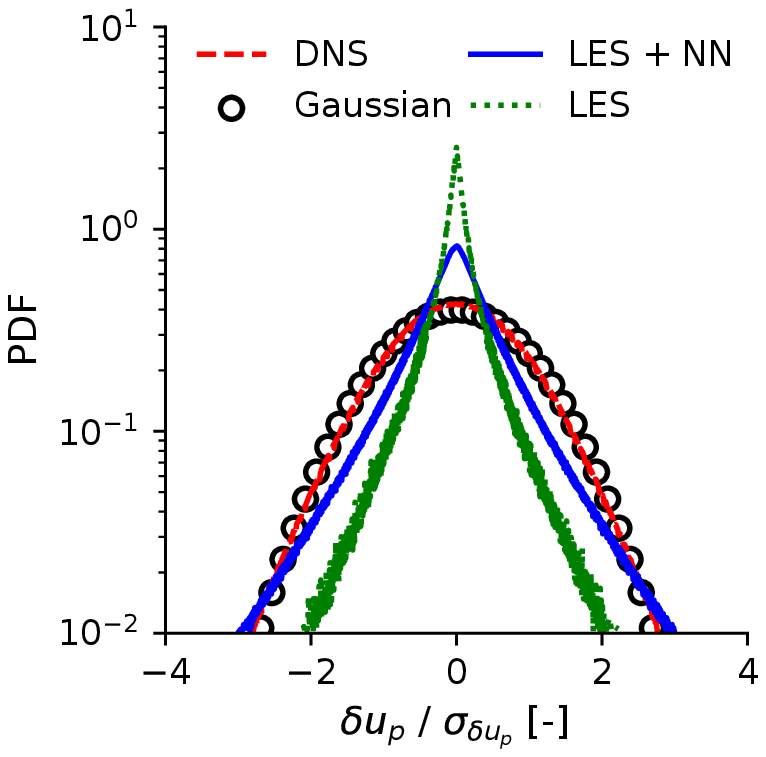} &
        \includegraphics{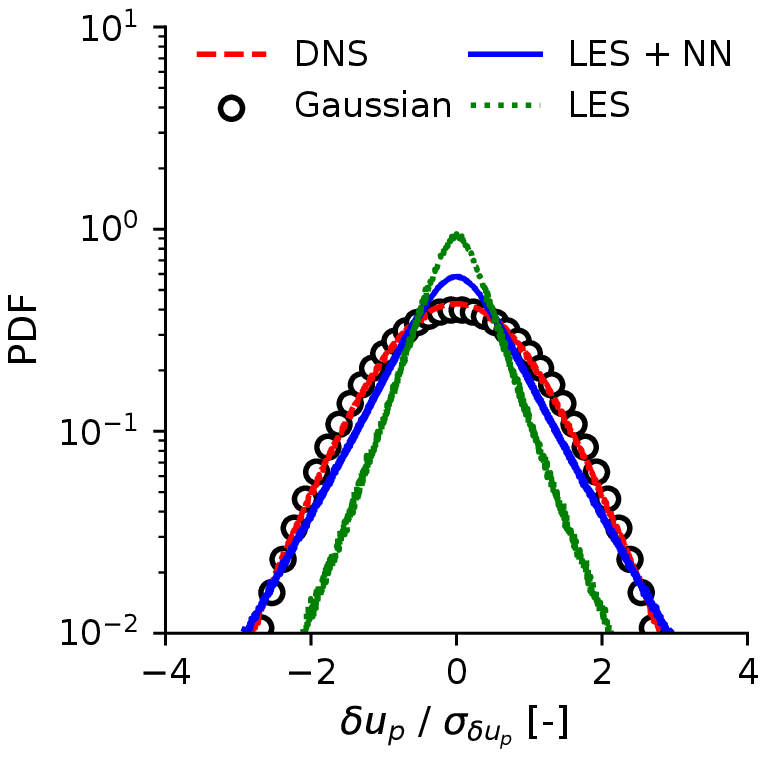} \\
        (c) & (d) \\
        \includegraphics{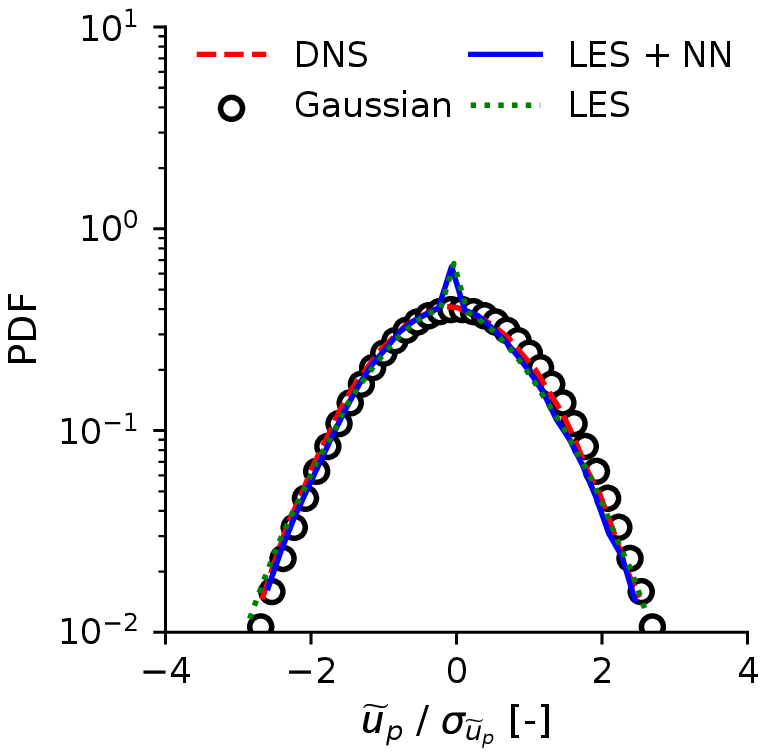} & 
        \includegraphics{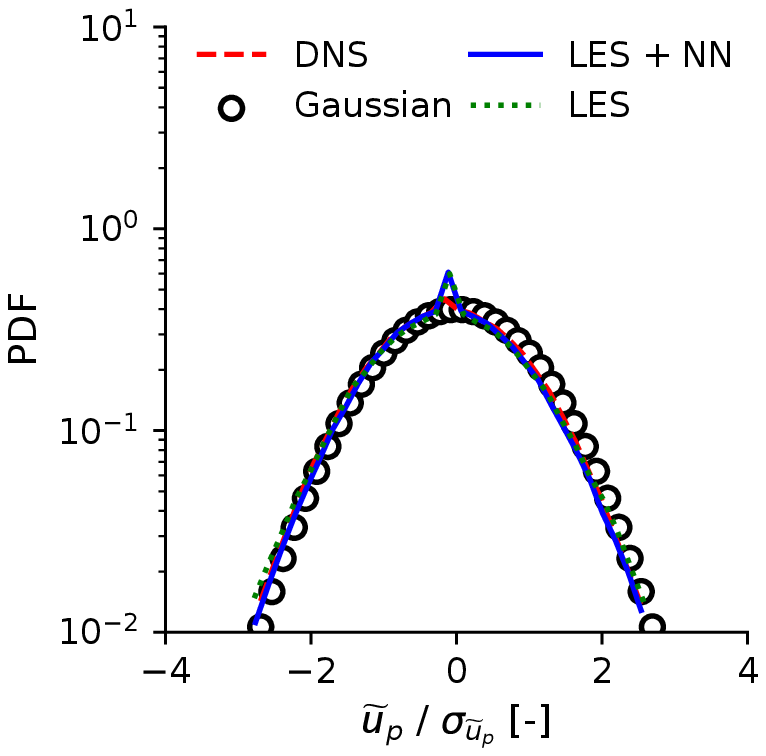} \\
    \end{tabular}
    \caption{Probability density function of instantaneous uncorrelated (top) and correlated (bottom) particle velocity at $t / \tau_L^{t=0} = 1$. Panels show (a,c) $St^{t=0}=1$ and (b,d) $St^{t=0}=5$. Velocities are normalised by the standard deviation of the DNS uncorrelated velocity. Simulations are performed at $\ReInit=33$ with $N_{cell}=36^3$ ($7\Delta_{DNS}$).}
    \label{fig:uncorPdf}
\end{figure}

Time evolution of particle fluctuating kinetic energies in Figures \ref{fig:apostHITstTKE} and \ref{fig:apostHITinertialReTKE} show good agreement between DNS and the dispersion model for predicted domain-averaged kinetic energy components. To gain a deeper insight into the velocity distribution, we evaluated the instantaneous PDF of correlated and uncorrelated particle velocity (\fref{fig:uncorPdf}). For $St^{t=0}=1$, the variance predicted by the neural SDE agrees well with the DNS data (within 3\% error), compared to the LES with no model, where the variance predicted is 70\% lower than the DNS (\fref{fig:uncorPdf}a). In contrast, the excess kurtosis (fourth central moment) for LES with our model was 6.4 compared to the DNS value of 0.6. The excess kurtosis for LES with no model was 6.7, which is a factor of 12 larger than the DNS. For $St^{t=0}=5$ (\fref{fig:uncorPdf}b), the variance predicted by the LES with our model is 6\% lower than the DNS data, compared to LES with no model under-predicted the DNS variance by 60\%. The excess kurtosis of $St^{t=0}=5$ particles for LES with no model is significantly lower than $St^{t=0}=1$ (3.1 compared to 6.7). The $\delta u_p$ distributions qualitatively agree with results of \citet{moreau2010development}, as the uncorrelated velocity for high inertia particles approaches a Gaussian distribution.

The instantaneous correlated particle velocity PDFs of both LES datasets agree well with the DNS for both $St^{t=0}$ (\fref{fig:uncorPdf}c,d). Only minor differences were observed, including that the variance of the PDF predicted by LES with our model under-predicted the DNS data by 9\% for $St^{t=0}=1$ and 6\% for $St^{t=0}=5$. The PDF variance predicted by LES with no model is within 3\% for both $St^{t=0}$. The excess kurtosis is less than 0.5 in all cases, showing the distribution is very close to Gaussian. Similar to Figures \ref{fig:apostHITstTKE} and \ref{fig:apostHITinertialReTKE}, these results confirm that LES filtering affects only the uncorrelated component of particle velocity. However, by examining the PDF (\fref{fig:uncorPdf}) our results show that a stochastic model can recover the same variance as DNS simulations, but not higher-order moments such as kurtosis.

\begin{figure}
    \centering
    \begin{tabular}{cc}
        (a) & (b)  \\
        \includegraphics{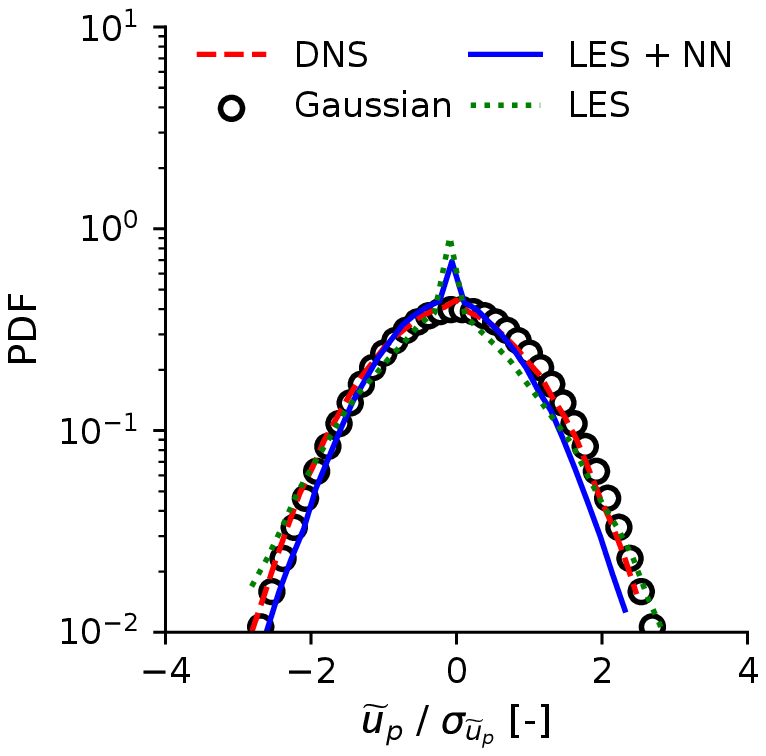} & 
        \includegraphics{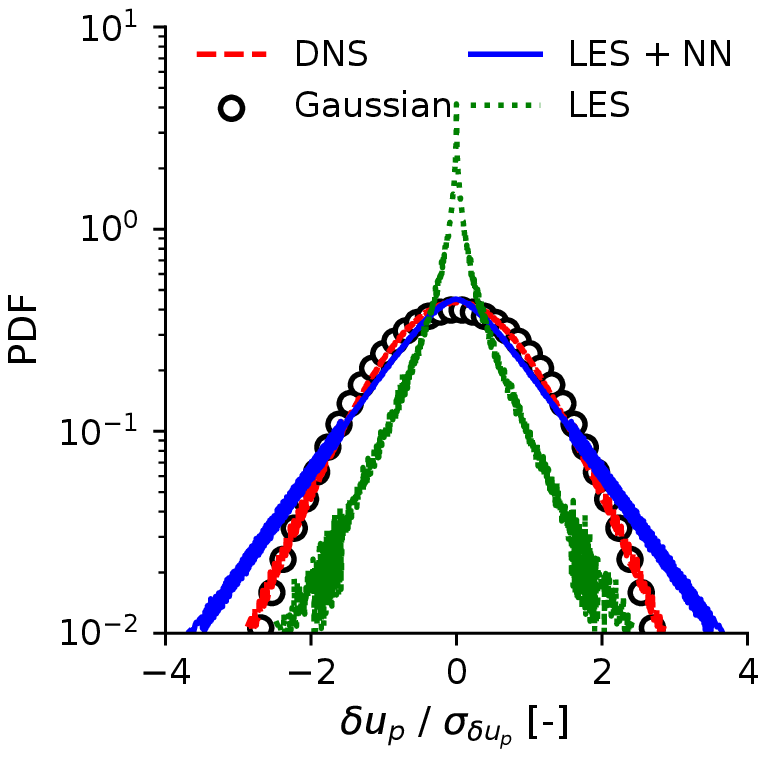}
    \end{tabular}
    \caption{Probability density function of instantaneous correlated (left) and uncorrelated (right) particle velocity at $\ReInit=105$ with $St^{t=0}=1$. Velocities are normalised by the standard deviation of the DNS velocity component. Simulations were performed with $N_{cell}=36^3$ ($7\Delta_{DNS}$). Distribution shown from time instant $t / \tau_L^{t=0} = 1$. }
    \label{fig:uncorPdfRe100}
\end{figure}

We have analysed the instantaneous PDF of particle velocity components for $\ReInit=105$ (\fref{fig:uncorPdfRe100}), to understand the behaviour of the particle fluctuating kinetic energy (\fref{fig:apostHITinertialReTKE} lower). The variance of the correlated velocity PDF predicted by LES with our model under-predicts the DNS variance by 25\%, compared to 1\% over-prediction by LES with no model (\fref{fig:uncorPdfRe100}a). The uncorrelated velocity variance is under-predicted by 60\% by LES with no model, and over-predicted by 60\% for LES with our model (\fref{fig:uncorPdfRe100}b). The excess kurtosis predicted by LES with no model was 8.6 times larger than the DNS predictions. The excess kurtosis predicted by our model was 3.4 times larger than the DNS predictions. These results show that although the domain-averaged particle fluctuating kinetic energy components are well-predicted by our model on out-of-distribution data (\fref{fig:apostHITinertialReTKE} lower), the moments of the uncorrelated velocity distribution show larger differences compared to the DNS data.

\section{Discussion and conclusions}
In this study, we aimed to develop a new approach for modelling turbulent dispersion in large-eddy simulation of dilute turbulent particle-laden flows. Our developed model is based on the classic Langevin-type equation for dispersion \citep{minier2015, innocenti2016, minier2021methodology}, where each unclosed term is inferred with a neural network (called a `neural SDE'). To create training data for the neural SDE, we performed DNS of decaying homogeneous isotropic turbulence. The DNS fluid velocity field was filtered in post-processing and this filtered data was used to extract subgrid turbulent properties ($k_{sgs}$ and $\varepsilon_{sgs}$), which were used for training and \textit{a priori} analysis. The trained network was then used for \textit{a posteriori} simulation in homogeneous isotropic turbulence to assess kinetic energy decay of tracer and inertial particles, which showed significant improvement over LES modelling with no dispersion model. The proposed neural SDE has been shown to provide improved predictions of particle velocity distributions and fluctuating kinetic energy decay in unbounded flows.

The proposed approach can be easily extended to include varying physics, such as shearing flow in the boundary layer of wall-bounded flows \citep{ling2016reynolds}. Furthermore, LES-SDE models for dispersion in moderate-dense flows may be derived by using two-way coupled simulations of particle transport for training data. Stochastic models with theoretically-derived closures have been developed for two-way and four-way coupled simulations with Reynolds-averaged Navier-Stokes fluid modelling \citep{innocenti2019lagrangian, innocenti2021lagrangianpdf}, but no such model exists for LES treatment of the fluid phase. 

Limitations of this study include the limited database used for training. We have trained the model on homogeneous isotropic turbulence at two Reynolds numbers, which obtained good agreement with DNS data in \textit{a priori} and \textit{a posteriori} analysis. Our model also showed good agreement with DNS data on a high Reynolds number case which was not included in training, showing that the model can generalise to unseen problems (\fref{fig:apostHITinertialReTKE}). To extend the model to more complex flow configurations, such as wall-bounded turbulence, one can train using filtered DNS data from particle-laden turbulent channel flow. Additionally, the current model could be adapted to include anisotropic velocity gradients that are essential in capturing near-wall flow physics \citep{ling2016reynolds}. The networks trained in this study can be used as a baseline, and additional networks may be added to each term, $G^{NN}$ and $B^{NN}$, to represent additional physics while satisfying the fundamental criterion that the model produces a correct decay of inertial and tracer particle fluctuating kinetic energy in homogeneous isotropic turbulence \citep{minier2014}.

A numerical limitation of this study is the use of cell-value interpolation  (Supplementary material). We have found that the trilinear interpolation scheme in OpenFOAM 6.0 does not conserve particle momentum and produces nonphysical results. We are aware that using cell values without interpolation to the particle position is generally less accurate than high-order Lagrange polynomials, cubic spline or Hermite interpolation \citep{yeungpope1988algorithm, balachandarmaxey1989, rovelstad1994interpolation, garg2007accurate}. However, OpenFOAM is a viable framework for Eulerian-Lagrangian modelling of gas-solid flows because of its open-source aspects with applicability for wide range of applications in complex geometries. Accurate and robust interpolation schemes in OpenFOAM would be a complementary development for improved prediction of particle velocity in turbulent particle-laden flows.

The proposed model has been shown to improve predictions of particle fluctuating kinetic energy and particle velocity distributions at varying $\ReInit$ and $St^{t=0}$ numbers in decaying homogeneous isotropic turbulence, compared to using no dispersion model. Applications of the model include modelling cough plume dispersion with LES for understanding disease spread \citep{liu2021investigation, monroe2021pulsatility} or pollutant dispersion in atmospheric flows. Future efforts should extend the existing neural SDE to account for the influence of velocity gradients and validate this against DNS of turbulent channel flow \citep{marchioli2008benchmark, soldati2009physics}. This will allow the neural SDE to be used for modelling particle dispersion in wall-bounded flows such as drug transport in lung airways.

\section*{Supplementary material}
See supplementary material for comparison of errors in OpenFOAM interpolation schemes.

\begin{acknowledgments}
JW was funded by a 2019 PhD Scholarship from the Carnegie-Trust for the Universities of Scotland.
\end{acknowledgments}

\section*{Author declarations}
\subsection*{Conflict of interest}
The authors have no conflicts to disclose.

\subsection*{Author contributions}
\textbf{Josh Williams:} Conceptualization, Methodology, Software, Validation, Formal analysis, Investigation, Data curation, Writing – Original draft, Writing – Review and editing, Visualisation, Funding acquisition. \textbf{Uwe Wolfram:} Conceptualization, Methodology, Resources, Supervision, Writing – Review and editing, Funding Acquisition,  Project administration. \textbf{Ali Ozel:} Conceptualization, Methodology, Software, Resources, Supervision, Writing – Review and editing, Funding Acquisition, Project administration. 

\section*{Data availability statement}
The data that support the findings of this study are openly available in Kaggle at http://doi.org/10.34740/KAGGLE/DSV/3998403, reference number 3998403.

\bibliography{refs_appraisal}

\begin{thebibliography}{103}%
\makeatletter
\providecommand \@ifxundefined [1]{%
 \@ifx{#1\undefined}
}%
\providecommand \@ifnum [1]{%
 \ifnum #1\expandafter \@firstoftwo
 \else \expandafter \@secondoftwo
 \fi
}%
\providecommand \@ifx [1]{%
 \ifx #1\expandafter \@firstoftwo
 \else \expandafter \@secondoftwo
 \fi
}%
\providecommand \natexlab [1]{#1}%
\providecommand \enquote  [1]{``#1''}%
\providecommand \bibnamefont  [1]{#1}%
\providecommand \bibfnamefont [1]{#1}%
\providecommand \citenamefont [1]{#1}%
\providecommand \href@noop [0]{\@secondoftwo}%
\providecommand \href [0]{\begingroup \@sanitize@url \@href}%
\providecommand \@href[1]{\@@startlink{#1}\@@href}%
\providecommand \@@href[1]{\endgroup#1\@@endlink}%
\providecommand \@sanitize@url [0]{\catcode `\\12\catcode `\$12\catcode
  `\&12\catcode `\#12\catcode `\^12\catcode `\_12\catcode `\%12\relax}%
\providecommand \@@startlink[1]{}%
\providecommand \@@endlink[0]{}%
\providecommand \url  [0]{\begingroup\@sanitize@url \@url }%
\providecommand \@url [1]{\endgroup\@href {#1}{\urlprefix }}%
\providecommand \urlprefix  [0]{URL }%
\providecommand \Eprint [0]{\href }%
\providecommand \doibase [0]{http://dx.doi.org/}%
\providecommand \selectlanguage [0]{\@gobble}%
\providecommand \bibinfo  [0]{\@secondoftwo}%
\providecommand \bibfield  [0]{\@secondoftwo}%
\providecommand \translation [1]{[#1]}%
\providecommand \BibitemOpen [0]{}%
\providecommand \bibitemStop [0]{}%
\providecommand \bibitemNoStop [0]{.\EOS\space}%
\providecommand \EOS [0]{\spacefactor3000\relax}%
\providecommand \BibitemShut  [1]{\csname bibitem#1\endcsname}%
\let\auto@bib@innerbib\@empty
\bibitem [{\citenamefont {Williams}\ \emph {et~al.}(2022)\citenamefont
  {Williams}, \citenamefont {Kolehmainen}, \citenamefont {Cunningham},
  \citenamefont {Ozel},\ and\ \citenamefont {Wolfram}}]{williams2022effect}%
  \BibitemOpen
  \bibfield  {author} {\bibinfo {author} {\bibfnamefont {J.}~\bibnamefont
  {Williams}}, \bibinfo {author} {\bibfnamefont {J.}~\bibnamefont
  {Kolehmainen}}, \bibinfo {author} {\bibfnamefont {S.}~\bibnamefont
  {Cunningham}}, \bibinfo {author} {\bibfnamefont {A.}~\bibnamefont {Ozel}}, \
  and\ \bibinfo {author} {\bibfnamefont {U.}~\bibnamefont {Wolfram}},\
  }\bibfield  {title} {\enquote {\bibinfo {title} {Effect of patient inhalation
  profile and airway structure on drug deposition in image-based models with
  particle-particle interactions},}\ }\href@noop {} {\bibfield  {journal}
  {\bibinfo  {journal} {International Journal of Pharmaceutics}\ }\textbf
  {\bibinfo {volume} {612}},\ \bibinfo {pages} {121321} (\bibinfo {year}
  {2022})}\BibitemShut {NoStop}%
\bibitem [{\citenamefont {Lambert}\ \emph {et~al.}(2011)\citenamefont
  {Lambert}, \citenamefont {O'shaughnessy}, \citenamefont {Tawhai},
  \citenamefont {Hoffman},\ and\ \citenamefont {Lin}}]{lambert11regional}%
  \BibitemOpen
  \bibfield  {author} {\bibinfo {author} {\bibfnamefont {A.~R.}\ \bibnamefont
  {Lambert}}, \bibinfo {author} {\bibfnamefont {P.~T.}\ \bibnamefont
  {O'shaughnessy}}, \bibinfo {author} {\bibfnamefont {M.~H.}\ \bibnamefont
  {Tawhai}}, \bibinfo {author} {\bibfnamefont {E.~A.}\ \bibnamefont {Hoffman}},
  \ and\ \bibinfo {author} {\bibfnamefont {C.-L.}\ \bibnamefont {Lin}},\
  }\bibfield  {title} {\enquote {\bibinfo {title} {Regional deposition of
  particles in an image-based airway model: large-eddy simulation and
  left-right lung ventilation asymmetry},}\ }\href@noop {} {\bibfield
  {journal} {\bibinfo  {journal} {Aerosol Science and Technology}\ }\textbf
  {\bibinfo {volume} {45}},\ \bibinfo {pages} {11--25} (\bibinfo {year}
  {2011})}\BibitemShut {NoStop}%
\bibitem [{\citenamefont {Koullapis}\ \emph {et~al.}(2018)\citenamefont
  {Koullapis}, \citenamefont {Kassinos}, \citenamefont {Muela}, \citenamefont
  {Perez-Segarra}, \citenamefont {Rigola}, \citenamefont {Lehmkuhl},
  \citenamefont {Cui}, \citenamefont {Sommerfeld}, \citenamefont {Elcner},
  \citenamefont {Jicha} \emph {et~al.}}]{koullapis18regional}%
  \BibitemOpen
  \bibfield  {author} {\bibinfo {author} {\bibfnamefont {P.}~\bibnamefont
  {Koullapis}}, \bibinfo {author} {\bibfnamefont {S.~C.}\ \bibnamefont
  {Kassinos}}, \bibinfo {author} {\bibfnamefont {J.}~\bibnamefont {Muela}},
  \bibinfo {author} {\bibfnamefont {C.}~\bibnamefont {Perez-Segarra}}, \bibinfo
  {author} {\bibfnamefont {J.}~\bibnamefont {Rigola}}, \bibinfo {author}
  {\bibfnamefont {O.}~\bibnamefont {Lehmkuhl}}, \bibinfo {author}
  {\bibfnamefont {Y.}~\bibnamefont {Cui}}, \bibinfo {author} {\bibfnamefont
  {M.}~\bibnamefont {Sommerfeld}}, \bibinfo {author} {\bibfnamefont
  {J.}~\bibnamefont {Elcner}}, \bibinfo {author} {\bibfnamefont
  {M.}~\bibnamefont {Jicha}},  \emph {et~al.},\ }\bibfield  {title} {\enquote
  {\bibinfo {title} {Regional aerosol deposition in the human airways: The
  siminhale benchmark case and a critical assessment of in silico methods},}\
  }\href@noop {} {\bibfield  {journal} {\bibinfo  {journal} {European Journal
  of Pharmaceutical Sciences}\ }\textbf {\bibinfo {volume} {113}},\ \bibinfo
  {pages} {77--94} (\bibinfo {year} {2018})}\BibitemShut {NoStop}%
\bibitem [{\citenamefont {Wang}, \citenamefont {Li},\ and\ \citenamefont
  {Wang}(2018)}]{wang2018quantifying}%
  \BibitemOpen
  \bibfield  {author} {\bibinfo {author} {\bibfnamefont {C.}~\bibnamefont
  {Wang}}, \bibinfo {author} {\bibfnamefont {Q.}~\bibnamefont {Li}}, \ and\
  \bibinfo {author} {\bibfnamefont {Z.-H.}\ \bibnamefont {Wang}},\ }\bibfield
  {title} {\enquote {\bibinfo {title} {Quantifying the impact of urban trees on
  passive pollutant dispersion using a coupled large-eddy
  simulation--lagrangian stochastic model},}\ }\href@noop {} {\bibfield
  {journal} {\bibinfo  {journal} {Building and Environment}\ }\textbf {\bibinfo
  {volume} {145}},\ \bibinfo {pages} {33--49} (\bibinfo {year}
  {2018})}\BibitemShut {NoStop}%
\bibitem [{\citenamefont {Calmet}\ \emph {et~al.}(2021)\citenamefont {Calmet},
  \citenamefont {Inthavong}, \citenamefont {Both}, \citenamefont {Surapaneni},
  \citenamefont {Mira}, \citenamefont {Egukitza},\ and\ \citenamefont
  {Houzeaux}}]{calmet2021large}%
  \BibitemOpen
  \bibfield  {author} {\bibinfo {author} {\bibfnamefont {H.}~\bibnamefont
  {Calmet}}, \bibinfo {author} {\bibfnamefont {K.}~\bibnamefont {Inthavong}},
  \bibinfo {author} {\bibfnamefont {A.}~\bibnamefont {Both}}, \bibinfo {author}
  {\bibfnamefont {A.}~\bibnamefont {Surapaneni}}, \bibinfo {author}
  {\bibfnamefont {D.}~\bibnamefont {Mira}}, \bibinfo {author} {\bibfnamefont
  {B.}~\bibnamefont {Egukitza}}, \ and\ \bibinfo {author} {\bibfnamefont
  {G.}~\bibnamefont {Houzeaux}},\ }\bibfield  {title} {\enquote {\bibinfo
  {title} {Large eddy simulation of cough jet dynamics, droplet transport, and
  inhalability over a ten minute exposure},}\ }\href@noop {} {\bibfield
  {journal} {\bibinfo  {journal} {Physics of Fluids}\ }\textbf {\bibinfo
  {volume} {33}},\ \bibinfo {pages} {125122} (\bibinfo {year}
  {2021})}\BibitemShut {NoStop}%
\bibitem [{\citenamefont {Liu}\ \emph {et~al.}(2021)\citenamefont {Liu},
  \citenamefont {Allahyari}, \citenamefont {Salinas}, \citenamefont {Zgheib},\
  and\ \citenamefont {Balachandar}}]{liu2021investigation}%
  \BibitemOpen
  \bibfield  {author} {\bibinfo {author} {\bibfnamefont {K.}~\bibnamefont
  {Liu}}, \bibinfo {author} {\bibfnamefont {M.}~\bibnamefont {Allahyari}},
  \bibinfo {author} {\bibfnamefont {J.}~\bibnamefont {Salinas}}, \bibinfo
  {author} {\bibfnamefont {N.}~\bibnamefont {Zgheib}}, \ and\ \bibinfo {author}
  {\bibfnamefont {S.}~\bibnamefont {Balachandar}},\ }\bibfield  {title}
  {\enquote {\bibinfo {title} {Investigation of theoretical scaling laws using
  large eddy simulations for airborne spreading of viral contagion from
  sneezing and coughing},}\ }\href@noop {} {\bibfield  {journal} {\bibinfo
  {journal} {Physics of Fluids}\ }\textbf {\bibinfo {volume} {33}},\ \bibinfo
  {pages} {063318} (\bibinfo {year} {2021})}\BibitemShut {NoStop}%
\bibitem [{\citenamefont {Monroe}\ \emph {et~al.}(2021)\citenamefont {Monroe},
  \citenamefont {Yao}, \citenamefont {Lattanzi}, \citenamefont {Raghav},\ and\
  \citenamefont {Capecelatro}}]{monroe2021pulsatility}%
  \BibitemOpen
  \bibfield  {author} {\bibinfo {author} {\bibfnamefont {K.}~\bibnamefont
  {Monroe}}, \bibinfo {author} {\bibfnamefont {Y.}~\bibnamefont {Yao}},
  \bibinfo {author} {\bibfnamefont {A.}~\bibnamefont {Lattanzi}}, \bibinfo
  {author} {\bibfnamefont {V.}~\bibnamefont {Raghav}}, \ and\ \bibinfo {author}
  {\bibfnamefont {J.}~\bibnamefont {Capecelatro}},\ }\bibfield  {title}
  {\enquote {\bibinfo {title} {Role of pulsatility on particle dispersion in
  expiratory flows},}\ }\href@noop {} {\bibfield  {journal} {\bibinfo
  {journal} {Physics of Fluids}\ }\textbf {\bibinfo {volume} {33}},\ \bibinfo
  {pages} {043311} (\bibinfo {year} {2021})}\BibitemShut {NoStop}%
\bibitem [{\citenamefont {Fede}\ and\ \citenamefont
  {Simonin}(2006)}]{fede2006numerical}%
  \BibitemOpen
  \bibfield  {author} {\bibinfo {author} {\bibfnamefont {P.}~\bibnamefont
  {Fede}}\ and\ \bibinfo {author} {\bibfnamefont {O.}~\bibnamefont {Simonin}},\
  }\bibfield  {title} {\enquote {\bibinfo {title} {Numerical study of the
  subgrid fluid turbulence effects on the statistics of heavy colliding
  particles},}\ }\href@noop {} {\bibfield  {journal} {\bibinfo  {journal}
  {Physics of Fluids}\ }\textbf {\bibinfo {volume} {18}},\ \bibinfo {pages}
  {045103} (\bibinfo {year} {2006})}\BibitemShut {NoStop}%
\bibitem [{\citenamefont {Fessler}, \citenamefont {Kulick},\ and\ \citenamefont
  {Eaton}(1994)}]{fessler1994preferential}%
  \BibitemOpen
  \bibfield  {author} {\bibinfo {author} {\bibfnamefont {J.~R.}\ \bibnamefont
  {Fessler}}, \bibinfo {author} {\bibfnamefont {J.~D.}\ \bibnamefont {Kulick}},
  \ and\ \bibinfo {author} {\bibfnamefont {J.~K.}\ \bibnamefont {Eaton}},\
  }\bibfield  {title} {\enquote {\bibinfo {title} {Preferential concentration
  of heavy particles in a turbulent channel flow},}\ }\href@noop {} {\bibfield
  {journal} {\bibinfo  {journal} {Physics of Fluids}\ }\textbf {\bibinfo
  {volume} {6}},\ \bibinfo {pages} {3742--3749} (\bibinfo {year}
  {1994})}\BibitemShut {NoStop}%
\bibitem [{\citenamefont {Dou}\ \emph {et~al.}(2018)\citenamefont {Dou},
  \citenamefont {Bragg}, \citenamefont {Hammond}, \citenamefont {Liang},
  \citenamefont {Collins},\ and\ \citenamefont {Meng}}]{dou2018effects}%
  \BibitemOpen
  \bibfield  {author} {\bibinfo {author} {\bibfnamefont {Z.}~\bibnamefont
  {Dou}}, \bibinfo {author} {\bibfnamefont {A.~D.}\ \bibnamefont {Bragg}},
  \bibinfo {author} {\bibfnamefont {A.~L.}\ \bibnamefont {Hammond}}, \bibinfo
  {author} {\bibfnamefont {Z.}~\bibnamefont {Liang}}, \bibinfo {author}
  {\bibfnamefont {L.~R.}\ \bibnamefont {Collins}}, \ and\ \bibinfo {author}
  {\bibfnamefont {H.}~\bibnamefont {Meng}},\ }\bibfield  {title} {\enquote
  {\bibinfo {title} {Effects of reynolds number and stokes number on
  particle-pair relative velocity in isotropic turbulence: a systematic
  experimental study},}\ }\href@noop {} {\bibfield  {journal} {\bibinfo
  {journal} {Journal of Fluid Mechanics}\ }\textbf {\bibinfo {volume} {839}},\
  \bibinfo {pages} {271--292} (\bibinfo {year} {2018})}\BibitemShut {NoStop}%
\bibitem [{\citenamefont {Brandt}\ and\ \citenamefont
  {Coletti}(2022)}]{brandt2022particle}%
  \BibitemOpen
  \bibfield  {author} {\bibinfo {author} {\bibfnamefont {L.}~\bibnamefont
  {Brandt}}\ and\ \bibinfo {author} {\bibfnamefont {F.}~\bibnamefont
  {Coletti}},\ }\bibfield  {title} {\enquote {\bibinfo {title} {Particle-laden
  turbulence: progress and perspectives},}\ }\href@noop {} {\bibfield
  {journal} {\bibinfo  {journal} {Annual Review of Fluid Mechanics}\ }\textbf
  {\bibinfo {volume} {54}},\ \bibinfo {pages} {159--189} (\bibinfo {year}
  {2022})}\BibitemShut {NoStop}%
\bibitem [{\citenamefont {Moreau}, \citenamefont {Simonin},\ and\ \citenamefont
  {B{\'e}dat}(2010)}]{moreau2010development}%
  \BibitemOpen
  \bibfield  {author} {\bibinfo {author} {\bibfnamefont {M.}~\bibnamefont
  {Moreau}}, \bibinfo {author} {\bibfnamefont {O.}~\bibnamefont {Simonin}}, \
  and\ \bibinfo {author} {\bibfnamefont {B.}~\bibnamefont {B{\'e}dat}},\
  }\bibfield  {title} {\enquote {\bibinfo {title} {Development of gas-particle
  euler-euler les approach: a priori analysis of particle sub-grid models in
  homogeneous isotropic turbulence},}\ }\href@noop {} {\bibfield  {journal}
  {\bibinfo  {journal} {Flow, turbulence and combustion}\ }\textbf {\bibinfo
  {volume} {84}},\ \bibinfo {pages} {295--324} (\bibinfo {year}
  {2010})}\BibitemShut {NoStop}%
\bibitem [{\citenamefont {Pozorski}\ and\ \citenamefont
  {Apte}(2009)}]{pozorski2009filtered}%
  \BibitemOpen
  \bibfield  {author} {\bibinfo {author} {\bibfnamefont {J.}~\bibnamefont
  {Pozorski}}\ and\ \bibinfo {author} {\bibfnamefont {S.~V.}\ \bibnamefont
  {Apte}},\ }\bibfield  {title} {\enquote {\bibinfo {title} {Filtered particle
  tracking in isotropic turbulence and stochastic modeling of subgrid-scale
  dispersion},}\ }\href@noop {} {\bibfield  {journal} {\bibinfo  {journal}
  {International Journal of Multiphase Flow}\ }\textbf {\bibinfo {volume}
  {35}},\ \bibinfo {pages} {118--128} (\bibinfo {year} {2009})}\BibitemShut
  {NoStop}%
\bibitem [{\citenamefont {Balachandar}\ and\ \citenamefont
  {Eaton}(2010)}]{balachandar2010review}%
  \BibitemOpen
  \bibfield  {author} {\bibinfo {author} {\bibfnamefont {S.}~\bibnamefont
  {Balachandar}}\ and\ \bibinfo {author} {\bibfnamefont {J.~K.}\ \bibnamefont
  {Eaton}},\ }\bibfield  {title} {\enquote {\bibinfo {title} {Turbulent
  dispersed multiphase flow},}\ }\href@noop {} {\bibfield  {journal} {\bibinfo
  {journal} {Annual review of fluid mechanics}\ }\textbf {\bibinfo {volume}
  {42}},\ \bibinfo {pages} {111--133} (\bibinfo {year} {2010})}\BibitemShut
  {NoStop}%
\bibitem [{\citenamefont {Mehrabadi}\ \emph {et~al.}(2018)\citenamefont
  {Mehrabadi}, \citenamefont {Horwitz}, \citenamefont {Subramaniam},\ and\
  \citenamefont {Mani}}]{mehrabadi2018direct}%
  \BibitemOpen
  \bibfield  {author} {\bibinfo {author} {\bibfnamefont {M.}~\bibnamefont
  {Mehrabadi}}, \bibinfo {author} {\bibfnamefont {J.}~\bibnamefont {Horwitz}},
  \bibinfo {author} {\bibfnamefont {S.}~\bibnamefont {Subramaniam}}, \ and\
  \bibinfo {author} {\bibfnamefont {A.}~\bibnamefont {Mani}},\ }\bibfield
  {title} {\enquote {\bibinfo {title} {A direct comparison of particle-resolved
  and point-particle methods in decaying turbulence},}\ }\href@noop {}
  {\bibfield  {journal} {\bibinfo  {journal} {Journal of Fluid Mechanics}\
  }\textbf {\bibinfo {volume} {850}},\ \bibinfo {pages} {336--369} (\bibinfo
  {year} {2018})}\BibitemShut {NoStop}%
\bibitem [{\citenamefont {Vreman}\ and\ \citenamefont
  {Kuerten}(2018)}]{vreman2018turbulent}%
  \BibitemOpen
  \bibfield  {author} {\bibinfo {author} {\bibfnamefont {A.}~\bibnamefont
  {Vreman}}\ and\ \bibinfo {author} {\bibfnamefont {J.}~\bibnamefont
  {Kuerten}},\ }\bibfield  {title} {\enquote {\bibinfo {title} {Turbulent
  channel flow past a moving array of spheres},}\ }\href@noop {} {\bibfield
  {journal} {\bibinfo  {journal} {Journal of Fluid Mechanics}\ }\textbf
  {\bibinfo {volume} {856}},\ \bibinfo {pages} {580--632} (\bibinfo {year}
  {2018})}\BibitemShut {NoStop}%
\bibitem [{\citenamefont {Costa}, \citenamefont {Brandt},\ and\ \citenamefont
  {Picano}(2021)}]{costa2021near}%
  \BibitemOpen
  \bibfield  {author} {\bibinfo {author} {\bibfnamefont {P.}~\bibnamefont
  {Costa}}, \bibinfo {author} {\bibfnamefont {L.}~\bibnamefont {Brandt}}, \
  and\ \bibinfo {author} {\bibfnamefont {F.}~\bibnamefont {Picano}},\
  }\bibfield  {title} {\enquote {\bibinfo {title} {Near-wall turbulence
  modulation by small inertial particles},}\ }\href@noop {} {\bibfield
  {journal} {\bibinfo  {journal} {Journal of Fluid Mechanics}\ }\textbf
  {\bibinfo {volume} {922}} (\bibinfo {year} {2021})}\BibitemShut {NoStop}%
\bibitem [{\citenamefont {Sundaresan}, \citenamefont {Ozel},\ and\
  \citenamefont {Kolehmainen}(2018)}]{sundaresan2018toward}%
  \BibitemOpen
  \bibfield  {author} {\bibinfo {author} {\bibfnamefont {S.}~\bibnamefont
  {Sundaresan}}, \bibinfo {author} {\bibfnamefont {A.}~\bibnamefont {Ozel}}, \
  and\ \bibinfo {author} {\bibfnamefont {J.}~\bibnamefont {Kolehmainen}},\
  }\bibfield  {title} {\enquote {\bibinfo {title} {Toward constitutive models
  for momentum, species, and energy transport in gas--particle flows},}\
  }\href@noop {} {\bibfield  {journal} {\bibinfo  {journal} {Annual review of
  chemical and biomolecular engineering}\ }\textbf {\bibinfo {volume} {9}},\
  \bibinfo {pages} {61--81} (\bibinfo {year} {2018})}\BibitemShut {NoStop}%
\bibitem [{\citenamefont {Anderson}\ and\ \citenamefont
  {Jackson}(1967)}]{and67fluimech}%
  \BibitemOpen
  \bibfield  {author} {\bibinfo {author} {\bibfnamefont {T.~B.}\ \bibnamefont
  {Anderson}}\ and\ \bibinfo {author} {\bibfnamefont {R.}~\bibnamefont
  {Jackson}},\ }\bibfield  {title} {\enquote {\bibinfo {title} {Fluid
  mechanical description of fluidized beds. equations of motion},}\ }\href@noop
  {} {\bibfield  {journal} {\bibinfo  {journal} {Industrial \& Engineering
  Chemistry Fundamentals}\ }\textbf {\bibinfo {volume} {6}},\ \bibinfo {pages}
  {527--539} (\bibinfo {year} {1967})}\BibitemShut {NoStop}%
\bibitem [{\citenamefont {Maxey}\ and\ \citenamefont
  {Riley}(1983)}]{maxeyriley1983}%
  \BibitemOpen
  \bibfield  {author} {\bibinfo {author} {\bibfnamefont {M.~R.}\ \bibnamefont
  {Maxey}}\ and\ \bibinfo {author} {\bibfnamefont {J.~J.}\ \bibnamefont
  {Riley}},\ }\bibfield  {title} {\enquote {\bibinfo {title} {Equation of
  motion for a small rigid sphere in a nonuniform flow},}\ }\href@noop {}
  {\bibfield  {journal} {\bibinfo  {journal} {The Physics of Fluids}\ }\textbf
  {\bibinfo {volume} {26}},\ \bibinfo {pages} {883--889} (\bibinfo {year}
  {1983})}\BibitemShut {NoStop}%
\bibitem [{\citenamefont {Capecelatro}\ and\ \citenamefont
  {Desjardins}(2013)}]{capecelatro2013euler}%
  \BibitemOpen
  \bibfield  {author} {\bibinfo {author} {\bibfnamefont {J.}~\bibnamefont
  {Capecelatro}}\ and\ \bibinfo {author} {\bibfnamefont {O.}~\bibnamefont
  {Desjardins}},\ }\bibfield  {title} {\enquote {\bibinfo {title} {An
  euler--lagrange strategy for simulating particle-laden flows},}\ }\href@noop
  {} {\bibfield  {journal} {\bibinfo  {journal} {Journal of Computational
  Physics}\ }\textbf {\bibinfo {volume} {238}},\ \bibinfo {pages} {1--31}
  (\bibinfo {year} {2013})}\BibitemShut {NoStop}%
\bibitem [{\citenamefont {Pope}(2000)}]{popebook}%
  \BibitemOpen
  \bibfield  {author} {\bibinfo {author} {\bibfnamefont {S.~B.}\ \bibnamefont
  {Pope}},\ }\href@noop {} {\emph {\bibinfo {title} {Turbulent Flows}}}\
  (\bibinfo  {publisher} {Cambridge University Press},\ \bibinfo {year}
  {2000})\BibitemShut {NoStop}%
\bibitem [{\citenamefont {Sagaut}(2006)}]{sagaut06les}%
  \BibitemOpen
  \bibfield  {author} {\bibinfo {author} {\bibfnamefont {P.}~\bibnamefont
  {Sagaut}},\ }\href@noop {} {\emph {\bibinfo {title} {Large eddy simulation
  for incompressible flows: an introduction}}}\ (\bibinfo  {publisher}
  {Springer Science \& Business Media},\ \bibinfo {year} {2006})\BibitemShut
  {NoStop}%
\bibitem [{\citenamefont {Jones}\ and\ \citenamefont
  {Launder}(1972)}]{jones1972prediction}%
  \BibitemOpen
  \bibfield  {author} {\bibinfo {author} {\bibfnamefont {W.~P.}\ \bibnamefont
  {Jones}}\ and\ \bibinfo {author} {\bibfnamefont {B.~E.}\ \bibnamefont
  {Launder}},\ }\bibfield  {title} {\enquote {\bibinfo {title} {The prediction
  of laminarization with a two-equation model of turbulence},}\ }\href@noop {}
  {\bibfield  {journal} {\bibinfo  {journal} {International journal of heat and
  mass transfer}\ }\textbf {\bibinfo {volume} {15}},\ \bibinfo {pages}
  {301--314} (\bibinfo {year} {1972})}\BibitemShut {NoStop}%
\bibitem [{\citenamefont {Launder}\ and\ \citenamefont
  {Spalding}(1983)}]{launder1983numerical}%
  \BibitemOpen
  \bibfield  {author} {\bibinfo {author} {\bibfnamefont {B.~E.}\ \bibnamefont
  {Launder}}\ and\ \bibinfo {author} {\bibfnamefont {D.~B.}\ \bibnamefont
  {Spalding}},\ }\bibfield  {title} {\enquote {\bibinfo {title} {The numerical
  computation of turbulent flows},}\ }in\ \href@noop {} {\emph {\bibinfo
  {booktitle} {Numerical prediction of flow, heat transfer, turbulence and
  combustion}}}\ (\bibinfo  {publisher} {Elsevier},\ \bibinfo {year} {1983})\
  pp.\ \bibinfo {pages} {96--116}\BibitemShut {NoStop}%
\bibitem [{\citenamefont {Spalart}\ and\ \citenamefont
  {Allmaras}(1992)}]{spalart1992one}%
  \BibitemOpen
  \bibfield  {author} {\bibinfo {author} {\bibfnamefont {P.}~\bibnamefont
  {Spalart}}\ and\ \bibinfo {author} {\bibfnamefont {S.}~\bibnamefont
  {Allmaras}},\ }\bibfield  {title} {\enquote {\bibinfo {title} {A one-equation
  turbulence model for aerodynamic flows},}\ }in\ \href@noop {} {\emph
  {\bibinfo {booktitle} {30th aerospace sciences meeting and exhibit}}}\
  (\bibinfo {year} {1992})\ p.\ \bibinfo {pages} {439}\BibitemShut {NoStop}%
\bibitem [{\citenamefont {Germano}(1992)}]{germano1992turbulence}%
  \BibitemOpen
  \bibfield  {author} {\bibinfo {author} {\bibfnamefont {M.}~\bibnamefont
  {Germano}},\ }\bibfield  {title} {\enquote {\bibinfo {title} {Turbulence: the
  filtering approach},}\ }\href@noop {} {\bibfield  {journal} {\bibinfo
  {journal} {Journal of Fluid Mechanics}\ }\textbf {\bibinfo {volume} {238}},\
  \bibinfo {pages} {325--336} (\bibinfo {year} {1992})}\BibitemShut {NoStop}%
\bibitem [{\citenamefont {Smagorinsky}(1963)}]{smagorinsky1963general}%
  \BibitemOpen
  \bibfield  {author} {\bibinfo {author} {\bibfnamefont {J.}~\bibnamefont
  {Smagorinsky}},\ }\bibfield  {title} {\enquote {\bibinfo {title} {General
  circulation experiments with the primitive equations: I. the basic
  experiment},}\ }\href@noop {} {\bibfield  {journal} {\bibinfo  {journal}
  {Monthly weather review}\ }\textbf {\bibinfo {volume} {91}},\ \bibinfo
  {pages} {99--164} (\bibinfo {year} {1963})}\BibitemShut {NoStop}%
\bibitem [{\citenamefont {Marchioli}(2017)}]{marchioli2017review}%
  \BibitemOpen
  \bibfield  {author} {\bibinfo {author} {\bibfnamefont {C.}~\bibnamefont
  {Marchioli}},\ }\bibfield  {title} {\enquote {\bibinfo {title} {Large-eddy
  simulation of turbulent dispersed flows: a review of modelling approaches},}\
  }\href@noop {} {\bibfield  {journal} {\bibinfo  {journal} {Acta Mechanica}\
  }\textbf {\bibinfo {volume} {228}},\ \bibinfo {pages} {741--771} (\bibinfo
  {year} {2017})}\BibitemShut {NoStop}%
\bibitem [{\citenamefont {Marchioli}, \citenamefont {Salvetti},\ and\
  \citenamefont {Soldati}(2008)}]{marchioli2008issues}%
  \BibitemOpen
  \bibfield  {author} {\bibinfo {author} {\bibfnamefont {C.}~\bibnamefont
  {Marchioli}}, \bibinfo {author} {\bibfnamefont {M.~V.}\ \bibnamefont
  {Salvetti}}, \ and\ \bibinfo {author} {\bibfnamefont {A.}~\bibnamefont
  {Soldati}},\ }\bibfield  {title} {\enquote {\bibinfo {title} {Some issues
  concerning large-eddy simulation of inertial particle dispersion in turbulent
  bounded flows},}\ }\href@noop {} {\bibfield  {journal} {\bibinfo  {journal}
  {Physics of Fluids}\ }\textbf {\bibinfo {volume} {20}},\ \bibinfo {pages}
  {040603} (\bibinfo {year} {2008})}\BibitemShut {NoStop}%
\bibitem [{\citenamefont {Minier}\ and\ \citenamefont
  {Peirano}(2001)}]{minier2001pdf}%
  \BibitemOpen
  \bibfield  {author} {\bibinfo {author} {\bibfnamefont {J.-P.}\ \bibnamefont
  {Minier}}\ and\ \bibinfo {author} {\bibfnamefont {E.}~\bibnamefont
  {Peirano}},\ }\bibfield  {title} {\enquote {\bibinfo {title} {The pdf
  approach to turbulent polydispersed two-phase flows},}\ }\href@noop {}
  {\bibfield  {journal} {\bibinfo  {journal} {Physics reports}\ }\textbf
  {\bibinfo {volume} {352}},\ \bibinfo {pages} {1--214} (\bibinfo {year}
  {2001})}\BibitemShut {NoStop}%
\bibitem [{\citenamefont {Haworth}\ and\ \citenamefont
  {Pope}(1986)}]{haworth1986generalized}%
  \BibitemOpen
  \bibfield  {author} {\bibinfo {author} {\bibfnamefont {D.~C.}\ \bibnamefont
  {Haworth}}\ and\ \bibinfo {author} {\bibfnamefont {S.~B.}\ \bibnamefont
  {Pope}},\ }\bibfield  {title} {\enquote {\bibinfo {title} {A generalized
  langevin model for turbulent flows},}\ }\href@noop {} {\bibfield  {journal}
  {\bibinfo  {journal} {The Physics of fluids}\ }\textbf {\bibinfo {volume}
  {29}},\ \bibinfo {pages} {387--405} (\bibinfo {year} {1986})}\BibitemShut
  {NoStop}%
\bibitem [{\citenamefont {Minier}(2015)}]{minier2015}%
  \BibitemOpen
  \bibfield  {author} {\bibinfo {author} {\bibfnamefont {J.-P.}\ \bibnamefont
  {Minier}},\ }\bibfield  {title} {\enquote {\bibinfo {title} {{On Lagrangian
  stochastic methods for turbulent polydisperse two-phase reactive flows}},}\
  }\href@noop {} {\bibfield  {journal} {\bibinfo  {journal} {Progress in Energy
  and Combustion Science}\ }\textbf {\bibinfo {volume} {50}},\ \bibinfo {pages}
  {1--62} (\bibinfo {year} {2015})}\BibitemShut {NoStop}%
\bibitem [{\citenamefont {Pope}(1985)}]{pope1985pdf}%
  \BibitemOpen
  \bibfield  {author} {\bibinfo {author} {\bibfnamefont {S.~B.}\ \bibnamefont
  {Pope}},\ }\bibfield  {title} {\enquote {\bibinfo {title} {Pdf methods for
  turbulent reactive flows},}\ }\href@noop {} {\bibfield  {journal} {\bibinfo
  {journal} {Progress in energy and combustion science}\ }\textbf {\bibinfo
  {volume} {11}},\ \bibinfo {pages} {119--192} (\bibinfo {year}
  {1985})}\BibitemShut {NoStop}%
\bibitem [{\citenamefont {Pope}(1994)}]{pope1994relationship}%
  \BibitemOpen
  \bibfield  {author} {\bibinfo {author} {\bibfnamefont {S.}~\bibnamefont
  {Pope}},\ }\bibfield  {title} {\enquote {\bibinfo {title} {On the
  relationship between stochastic lagrangian models of turbulence and
  second-moment closures},}\ }\href@noop {} {\bibfield  {journal} {\bibinfo
  {journal} {Physics of Fluids}\ }\textbf {\bibinfo {volume} {6}},\ \bibinfo
  {pages} {973--985} (\bibinfo {year} {1994})}\BibitemShut {NoStop}%
\bibitem [{\citenamefont {Gicquel}\ \emph {et~al.}(2002)\citenamefont
  {Gicquel}, \citenamefont {Givi}, \citenamefont {Jaberi},\ and\ \citenamefont
  {Pope}}]{gicquel2002velocity}%
  \BibitemOpen
  \bibfield  {author} {\bibinfo {author} {\bibfnamefont {L.~Y.}\ \bibnamefont
  {Gicquel}}, \bibinfo {author} {\bibfnamefont {P.}~\bibnamefont {Givi}},
  \bibinfo {author} {\bibfnamefont {F.}~\bibnamefont {Jaberi}}, \ and\ \bibinfo
  {author} {\bibfnamefont {S.}~\bibnamefont {Pope}},\ }\bibfield  {title}
  {\enquote {\bibinfo {title} {Velocity filtered density function for large
  eddy simulation of turbulent flows},}\ }\href@noop {} {\bibfield  {journal}
  {\bibinfo  {journal} {Physics of Fluids}\ }\textbf {\bibinfo {volume} {14}},\
  \bibinfo {pages} {1196--1213} (\bibinfo {year} {2002})}\BibitemShut {NoStop}%
\bibitem [{\citenamefont {Jenny}\ \emph {et~al.}(2001)\citenamefont {Jenny},
  \citenamefont {Pope}, \citenamefont {Muradoglu},\ and\ \citenamefont
  {Caughey}}]{jenny2001hybrid}%
  \BibitemOpen
  \bibfield  {author} {\bibinfo {author} {\bibfnamefont {P.}~\bibnamefont
  {Jenny}}, \bibinfo {author} {\bibfnamefont {S.~B.}\ \bibnamefont {Pope}},
  \bibinfo {author} {\bibfnamefont {M.}~\bibnamefont {Muradoglu}}, \ and\
  \bibinfo {author} {\bibfnamefont {D.~A.}\ \bibnamefont {Caughey}},\
  }\bibfield  {title} {\enquote {\bibinfo {title} {A hybrid algorithm for the
  joint pdf equation of turbulent reactive flows},}\ }\href@noop {} {\bibfield
  {journal} {\bibinfo  {journal} {Journal of Computational Physics}\ }\textbf
  {\bibinfo {volume} {166}},\ \bibinfo {pages} {218--252} (\bibinfo {year}
  {2001})}\BibitemShut {NoStop}%
\bibitem [{\citenamefont {Pope}(2011)}]{pope2011simple}%
  \BibitemOpen
  \bibfield  {author} {\bibinfo {author} {\bibfnamefont {S.~B.}\ \bibnamefont
  {Pope}},\ }\bibfield  {title} {\enquote {\bibinfo {title} {Simple models of
  turbulent flows},}\ }\href@noop {} {\bibfield  {journal} {\bibinfo  {journal}
  {Physics of Fluids}\ }\textbf {\bibinfo {volume} {23}},\ \bibinfo {pages}
  {011301} (\bibinfo {year} {2011})}\BibitemShut {NoStop}%
\bibitem [{\citenamefont {Innocenti}, \citenamefont {Fox},\ and\ \citenamefont
  {Chibbaro}(2021)}]{innocenti2021lagrangianpdf}%
  \BibitemOpen
  \bibfield  {author} {\bibinfo {author} {\bibfnamefont {A.}~\bibnamefont
  {Innocenti}}, \bibinfo {author} {\bibfnamefont {R.~O.}\ \bibnamefont {Fox}},
  \ and\ \bibinfo {author} {\bibfnamefont {S.}~\bibnamefont {Chibbaro}},\
  }\bibfield  {title} {\enquote {\bibinfo {title} {A lagrangian
  probability-density-function model for turbulent particle-laden channel flow
  in the dense regime},}\ }\href@noop {} {\bibfield  {journal} {\bibinfo
  {journal} {Physics of Fluids}\ }\textbf {\bibinfo {volume} {33}},\ \bibinfo
  {pages} {053308} (\bibinfo {year} {2021})}\BibitemShut {NoStop}%
\bibitem [{\citenamefont {Innocenti}\ \emph {et~al.}(2019)\citenamefont
  {Innocenti}, \citenamefont {Fox}, \citenamefont {Salvetti},\ and\
  \citenamefont {Chibbaro}}]{innocenti2019lagrangian}%
  \BibitemOpen
  \bibfield  {author} {\bibinfo {author} {\bibfnamefont {A.}~\bibnamefont
  {Innocenti}}, \bibinfo {author} {\bibfnamefont {R.~O.}\ \bibnamefont {Fox}},
  \bibinfo {author} {\bibfnamefont {M.~V.}\ \bibnamefont {Salvetti}}, \ and\
  \bibinfo {author} {\bibfnamefont {S.}~\bibnamefont {Chibbaro}},\ }\bibfield
  {title} {\enquote {\bibinfo {title} {A lagrangian
  probability-density-function model for collisional turbulent fluid--particle
  flows},}\ }\href@noop {} {\bibfield  {journal} {\bibinfo  {journal} {Journal
  of Fluid Mechanics}\ }\textbf {\bibinfo {volume} {862}},\ \bibinfo {pages}
  {449--489} (\bibinfo {year} {2019})}\BibitemShut {NoStop}%
\bibitem [{\citenamefont {Peirano}\ \emph {et~al.}(2006)\citenamefont
  {Peirano}, \citenamefont {Chibbaro}, \citenamefont {Pozorski},\ and\
  \citenamefont {Minier}}]{peirano2006meanfield}%
  \BibitemOpen
  \bibfield  {author} {\bibinfo {author} {\bibfnamefont {E.}~\bibnamefont
  {Peirano}}, \bibinfo {author} {\bibfnamefont {S.}~\bibnamefont {Chibbaro}},
  \bibinfo {author} {\bibfnamefont {J.}~\bibnamefont {Pozorski}}, \ and\
  \bibinfo {author} {\bibfnamefont {J.-P.}\ \bibnamefont {Minier}},\ }\bibfield
   {title} {\enquote {\bibinfo {title} {Mean-field/pdf numerical approach for
  polydispersed turbulent two-phase flows},}\ }\href@noop {} {\bibfield
  {journal} {\bibinfo  {journal} {Progress in energy and combustion science}\
  }\textbf {\bibinfo {volume} {32}},\ \bibinfo {pages} {315--371} (\bibinfo
  {year} {2006})}\BibitemShut {NoStop}%
\bibitem [{\citenamefont {Pozorski}\ and\ \citenamefont
  {Minier}(1999)}]{pozorski1999probability}%
  \BibitemOpen
  \bibfield  {author} {\bibinfo {author} {\bibfnamefont {J.}~\bibnamefont
  {Pozorski}}\ and\ \bibinfo {author} {\bibfnamefont {J.-P.}\ \bibnamefont
  {Minier}},\ }\bibfield  {title} {\enquote {\bibinfo {title} {Probability
  density function modeling of dispersed two-phase turbulent flows},}\
  }\href@noop {} {\bibfield  {journal} {\bibinfo  {journal} {Physical Review
  E}\ }\textbf {\bibinfo {volume} {59}},\ \bibinfo {pages} {855} (\bibinfo
  {year} {1999})}\BibitemShut {NoStop}%
\bibitem [{\citenamefont {Dreeben}\ and\ \citenamefont
  {Pope}(1998)}]{dreeben1998probability}%
  \BibitemOpen
  \bibfield  {author} {\bibinfo {author} {\bibfnamefont {T.~D.}\ \bibnamefont
  {Dreeben}}\ and\ \bibinfo {author} {\bibfnamefont {S.~B.}\ \bibnamefont
  {Pope}},\ }\bibfield  {title} {\enquote {\bibinfo {title} {Probability
  density function/monte carlo simulation of near-wall turbulent flows},}\
  }\href@noop {} {\bibfield  {journal} {\bibinfo  {journal} {Journal of Fluid
  Mechanics}\ }\textbf {\bibinfo {volume} {357}},\ \bibinfo {pages} {141--166}
  (\bibinfo {year} {1998})}\BibitemShut {NoStop}%
\bibitem [{\citenamefont {Chibbaro}\ and\ \citenamefont
  {Minier}(2008)}]{chibbaro2008pipeflow}%
  \BibitemOpen
  \bibfield  {author} {\bibinfo {author} {\bibfnamefont {S.}~\bibnamefont
  {Chibbaro}}\ and\ \bibinfo {author} {\bibfnamefont {J.-P.}\ \bibnamefont
  {Minier}},\ }\bibfield  {title} {\enquote {\bibinfo {title} {Langevin pdf
  simulation of particle deposition in a turbulent pipe flow},}\ }\href@noop {}
  {\bibfield  {journal} {\bibinfo  {journal} {Journal of aerosol science}\
  }\textbf {\bibinfo {volume} {39}},\ \bibinfo {pages} {555--571} (\bibinfo
  {year} {2008})}\BibitemShut {NoStop}%
\bibitem [{\citenamefont {Wac{\l}awczyk}, \citenamefont {Pozorski},\ and\
  \citenamefont {Minier}(2004)}]{waclawczyk2004nearwall}%
  \BibitemOpen
  \bibfield  {author} {\bibinfo {author} {\bibfnamefont {M.}~\bibnamefont
  {Wac{\l}awczyk}}, \bibinfo {author} {\bibfnamefont {J.}~\bibnamefont
  {Pozorski}}, \ and\ \bibinfo {author} {\bibfnamefont {J.-P.}\ \bibnamefont
  {Minier}},\ }\bibfield  {title} {\enquote {\bibinfo {title} {Probability
  density function computation of turbulent flows with a new near-wall
  model},}\ }\href@noop {} {\bibfield  {journal} {\bibinfo  {journal} {Physics
  of Fluids}\ }\textbf {\bibinfo {volume} {16}},\ \bibinfo {pages} {1410--1422}
  (\bibinfo {year} {2004})}\BibitemShut {NoStop}%
\bibitem [{\citenamefont {Shotorban}\ and\ \citenamefont
  {Mashayek}(2006)}]{shotorban2006stochastic}%
  \BibitemOpen
  \bibfield  {author} {\bibinfo {author} {\bibfnamefont {B.}~\bibnamefont
  {Shotorban}}\ and\ \bibinfo {author} {\bibfnamefont {F.}~\bibnamefont
  {Mashayek}},\ }\bibfield  {title} {\enquote {\bibinfo {title} {A stochastic
  model for particle motion in large-eddy simulation},}\ }\href@noop {}
  {\bibfield  {journal} {\bibinfo  {journal} {Journal of Turbulence}\ ,\
  \bibinfo {pages} {N18}} (\bibinfo {year} {2006})}\BibitemShut {NoStop}%
\bibitem [{\citenamefont {Fede}\ \emph {et~al.}(2006)\citenamefont {Fede},
  \citenamefont {Simonin}, \citenamefont {Villedieu},\ and\ \citenamefont
  {Squires}}]{fede2006stochastic}%
  \BibitemOpen
  \bibfield  {author} {\bibinfo {author} {\bibfnamefont {P.}~\bibnamefont
  {Fede}}, \bibinfo {author} {\bibfnamefont {O.}~\bibnamefont {Simonin}},
  \bibinfo {author} {\bibfnamefont {P.}~\bibnamefont {Villedieu}}, \ and\
  \bibinfo {author} {\bibfnamefont {K.}~\bibnamefont {Squires}},\ }\bibfield
  {title} {\enquote {\bibinfo {title} {Stochastic modeling of the turbulent
  subgrid fluid velocity along inertial particle trajectories},}\ }in\
  \href@noop {} {\emph {\bibinfo {booktitle} {Proceedings of the Summer
  Program}}}\ (\bibinfo {organization} {Center for Turbulence Research},\
  \bibinfo {year} {2006})\ pp.\ \bibinfo {pages} {247--258}\BibitemShut
  {NoStop}%
\bibitem [{\citenamefont {Berrouk}\ \emph {et~al.}(2007)\citenamefont
  {Berrouk}, \citenamefont {Laurence}, \citenamefont {Riley},\ and\
  \citenamefont {Stock}}]{berrouk2007stochastic}%
  \BibitemOpen
  \bibfield  {author} {\bibinfo {author} {\bibfnamefont {A.}~\bibnamefont
  {Berrouk}}, \bibinfo {author} {\bibfnamefont {D.}~\bibnamefont {Laurence}},
  \bibinfo {author} {\bibfnamefont {J.}~\bibnamefont {Riley}}, \ and\ \bibinfo
  {author} {\bibfnamefont {D.}~\bibnamefont {Stock}},\ }\bibfield  {title}
  {\enquote {\bibinfo {title} {Stochastic modelling of inertial particle
  dispersion by subgrid motion for les of high reynolds number pipe flow},}\
  }\href {\doibase 10.1080/14685240701615952} {\bibfield  {journal} {\bibinfo
  {journal} {Journal of Turbulence}\ ,\ \bibinfo {pages} {N50}} (\bibinfo
  {year} {2007})}\BibitemShut {NoStop}%
\bibitem [{\citenamefont {Innocenti}, \citenamefont {Marchioli},\ and\
  \citenamefont {Chibbaro}(2016)}]{innocenti2016}%
  \BibitemOpen
  \bibfield  {author} {\bibinfo {author} {\bibfnamefont {A.}~\bibnamefont
  {Innocenti}}, \bibinfo {author} {\bibfnamefont {C.}~\bibnamefont
  {Marchioli}}, \ and\ \bibinfo {author} {\bibfnamefont {S.}~\bibnamefont
  {Chibbaro}},\ }\bibfield  {title} {\enquote {\bibinfo {title} {{Lagrangian
  filtered density function for LES-based stochastic modelling of turbulent
  particle-laden flows}},}\ }\href@noop {} {\bibfield  {journal} {\bibinfo
  {journal} {Physics of Fluids}\ }\textbf {\bibinfo {volume} {28}},\ \bibinfo
  {pages} {115106} (\bibinfo {year} {2016})}\BibitemShut {NoStop}%
\bibitem [{\citenamefont {Knorps}\ and\ \citenamefont
  {Pozorski}(2021)}]{knorps2021stochastic}%
  \BibitemOpen
  \bibfield  {author} {\bibinfo {author} {\bibfnamefont {M.}~\bibnamefont
  {Knorps}}\ and\ \bibinfo {author} {\bibfnamefont {J.}~\bibnamefont
  {Pozorski}},\ }\bibfield  {title} {\enquote {\bibinfo {title} {Stochastic
  modeling for subgrid-scale particle dispersion in large-eddy simulation of
  inhomogeneous turbulence},}\ }\href@noop {} {\bibfield  {journal} {\bibinfo
  {journal} {Physics of Fluids}\ }\textbf {\bibinfo {volume} {33}},\ \bibinfo
  {pages} {043323} (\bibinfo {year} {2021})}\BibitemShut {NoStop}%
\bibitem [{\citenamefont {Breuer}\ and\ \citenamefont
  {Hoppe}(2017)}]{breuer2017influence}%
  \BibitemOpen
  \bibfield  {author} {\bibinfo {author} {\bibfnamefont {M.}~\bibnamefont
  {Breuer}}\ and\ \bibinfo {author} {\bibfnamefont {F.}~\bibnamefont {Hoppe}},\
  }\bibfield  {title} {\enquote {\bibinfo {title} {Influence of a
  cost--efficient langevin subgrid-scale model on the dispersed phase of
  large--eddy simulations of turbulent bubble--laden and particle--laden
  flows},}\ }\href@noop {} {\bibfield  {journal} {\bibinfo  {journal}
  {International Journal of Multiphase Flow}\ }\textbf {\bibinfo {volume}
  {89}},\ \bibinfo {pages} {23--44} (\bibinfo {year} {2017})}\BibitemShut
  {NoStop}%
\bibitem [{\citenamefont {Van~Driest}(1956)}]{vandriest1956turbulent}%
  \BibitemOpen
  \bibfield  {author} {\bibinfo {author} {\bibfnamefont {E.~R.}\ \bibnamefont
  {Van~Driest}},\ }\bibfield  {title} {\enquote {\bibinfo {title} {On turbulent
  flow near a wall},}\ }\href@noop {} {\bibfield  {journal} {\bibinfo
  {journal} {Journal of the aeronautical sciences}\ }\textbf {\bibinfo {volume}
  {23}},\ \bibinfo {pages} {1007--1011} (\bibinfo {year} {1956})}\BibitemShut
  {NoStop}%
\bibitem [{\citenamefont {Minier}(2021)}]{minier2021methodology}%
  \BibitemOpen
  \bibfield  {author} {\bibinfo {author} {\bibfnamefont {J.-P.}\ \bibnamefont
  {Minier}},\ }\bibfield  {title} {\enquote {\bibinfo {title} {A methodology to
  devise consistent probability density function models for particle dynamics
  in turbulent dispersed two-phase flows},}\ }\href@noop {} {\bibfield
  {journal} {\bibinfo  {journal} {Physics of Fluids}\ }\textbf {\bibinfo
  {volume} {33}},\ \bibinfo {pages} {023312} (\bibinfo {year}
  {2021})}\BibitemShut {NoStop}%
\bibitem [{\citenamefont {Minier}, \citenamefont {Chibbaro},\ and\
  \citenamefont {Pope}(2014)}]{minier2014}%
  \BibitemOpen
  \bibfield  {author} {\bibinfo {author} {\bibfnamefont {J.-P.}\ \bibnamefont
  {Minier}}, \bibinfo {author} {\bibfnamefont {S.}~\bibnamefont {Chibbaro}}, \
  and\ \bibinfo {author} {\bibfnamefont {S.~B.}\ \bibnamefont {Pope}},\
  }\bibfield  {title} {\enquote {\bibinfo {title} {{Guidelines for the
  formulation of Lagrangian stochastic models for particle simulations of
  single-phase and dispersed two-phase turbulent flows}},}\ }\href@noop {}
  {\bibfield  {journal} {\bibinfo  {journal} {Physics of Fluids}\ }\textbf
  {\bibinfo {volume} {26}},\ \bibinfo {pages} {113303} (\bibinfo {year}
  {2014})}\BibitemShut {NoStop}%
\bibitem [{\citenamefont {Ling}, \citenamefont {Kurzawski},\ and\ \citenamefont
  {Templeton}(2016)}]{ling2016reynolds}%
  \BibitemOpen
  \bibfield  {author} {\bibinfo {author} {\bibfnamefont {J.}~\bibnamefont
  {Ling}}, \bibinfo {author} {\bibfnamefont {A.}~\bibnamefont {Kurzawski}}, \
  and\ \bibinfo {author} {\bibfnamefont {J.}~\bibnamefont {Templeton}},\
  }\bibfield  {title} {\enquote {\bibinfo {title} {Reynolds averaged turbulence
  modelling using deep neural networks with embedded invariance},}\ }\href@noop
  {} {\bibfield  {journal} {\bibinfo  {journal} {Journal of Fluid Mechanics}\
  }\textbf {\bibinfo {volume} {807}},\ \bibinfo {pages} {155--166} (\bibinfo
  {year} {2016})}\BibitemShut {NoStop}%
\bibitem [{\citenamefont {Fang}\ \emph {et~al.}(2020)\citenamefont {Fang},
  \citenamefont {Sondak}, \citenamefont {Protopapas},\ and\ \citenamefont
  {Succi}}]{fang2020neural}%
  \BibitemOpen
  \bibfield  {author} {\bibinfo {author} {\bibfnamefont {R.}~\bibnamefont
  {Fang}}, \bibinfo {author} {\bibfnamefont {D.}~\bibnamefont {Sondak}},
  \bibinfo {author} {\bibfnamefont {P.}~\bibnamefont {Protopapas}}, \ and\
  \bibinfo {author} {\bibfnamefont {S.}~\bibnamefont {Succi}},\ }\bibfield
  {title} {\enquote {\bibinfo {title} {Neural network models for the
  anisotropic reynolds stress tensor in turbulent channel flow},}\ }\href@noop
  {} {\bibfield  {journal} {\bibinfo  {journal} {Journal of Turbulence}\
  }\textbf {\bibinfo {volume} {21}},\ \bibinfo {pages} {525--543} (\bibinfo
  {year} {2020})}\BibitemShut {NoStop}%
\bibitem [{\citenamefont {Beetham}\ and\ \citenamefont
  {Capecelatro}(2020)}]{beetham2020formulating}%
  \BibitemOpen
  \bibfield  {author} {\bibinfo {author} {\bibfnamefont {S.}~\bibnamefont
  {Beetham}}\ and\ \bibinfo {author} {\bibfnamefont {J.}~\bibnamefont
  {Capecelatro}},\ }\bibfield  {title} {\enquote {\bibinfo {title} {Formulating
  turbulence closures using sparse regression with embedded form invariance},}\
  }\href@noop {} {\bibfield  {journal} {\bibinfo  {journal} {Physical Review
  Fluids}\ }\textbf {\bibinfo {volume} {5}},\ \bibinfo {pages} {084611}
  (\bibinfo {year} {2020})}\BibitemShut {NoStop}%
\bibitem [{\citenamefont {Park}\ and\ \citenamefont
  {Choi}(2021)}]{park2021toward}%
  \BibitemOpen
  \bibfield  {author} {\bibinfo {author} {\bibfnamefont {J.}~\bibnamefont
  {Park}}\ and\ \bibinfo {author} {\bibfnamefont {H.}~\bibnamefont {Choi}},\
  }\bibfield  {title} {\enquote {\bibinfo {title} {Toward neural-network-based
  large eddy simulation: Application to turbulent channel flow},}\ }\href@noop
  {} {\bibfield  {journal} {\bibinfo  {journal} {Journal of Fluid Mechanics}\
  }\textbf {\bibinfo {volume} {914}} (\bibinfo {year} {2021})}\BibitemShut
  {NoStop}%
\bibitem [{\citenamefont {Zhou}\ \emph {et~al.}(2019)\citenamefont {Zhou},
  \citenamefont {He}, \citenamefont {Wang},\ and\ \citenamefont
  {Jin}}]{zhou2019subgrid}%
  \BibitemOpen
  \bibfield  {author} {\bibinfo {author} {\bibfnamefont {Z.}~\bibnamefont
  {Zhou}}, \bibinfo {author} {\bibfnamefont {G.}~\bibnamefont {He}}, \bibinfo
  {author} {\bibfnamefont {S.}~\bibnamefont {Wang}}, \ and\ \bibinfo {author}
  {\bibfnamefont {G.}~\bibnamefont {Jin}},\ }\bibfield  {title} {\enquote
  {\bibinfo {title} {Subgrid-scale model for large-eddy simulation of isotropic
  turbulent flows using an artificial neural network},}\ }\href@noop {}
  {\bibfield  {journal} {\bibinfo  {journal} {Computers \& Fluids}\ }\textbf
  {\bibinfo {volume} {195}},\ \bibinfo {pages} {104319} (\bibinfo {year}
  {2019})}\BibitemShut {NoStop}%
\bibitem [{\citenamefont {Milani}, \citenamefont {Ling},\ and\ \citenamefont
  {Eaton}(2021)}]{milani2021turbulent}%
  \BibitemOpen
  \bibfield  {author} {\bibinfo {author} {\bibfnamefont {P.~M.}\ \bibnamefont
  {Milani}}, \bibinfo {author} {\bibfnamefont {J.}~\bibnamefont {Ling}}, \ and\
  \bibinfo {author} {\bibfnamefont {J.~K.}\ \bibnamefont {Eaton}},\ }\bibfield
  {title} {\enquote {\bibinfo {title} {Turbulent scalar flux in inclined jets
  in crossflow: counter gradient transport and deep learning modelling},}\
  }\href@noop {} {\bibfield  {journal} {\bibinfo  {journal} {Journal of Fluid
  Mechanics}\ }\textbf {\bibinfo {volume} {906}} (\bibinfo {year}
  {2021})}\BibitemShut {NoStop}%
\bibitem [{\citenamefont {Frezat}\ \emph {et~al.}(2021)\citenamefont {Frezat},
  \citenamefont {Balarac}, \citenamefont {Le~Sommer}, \citenamefont {Fablet},\
  and\ \citenamefont {Lguensat}}]{frezat2021physical}%
  \BibitemOpen
  \bibfield  {author} {\bibinfo {author} {\bibfnamefont {H.}~\bibnamefont
  {Frezat}}, \bibinfo {author} {\bibfnamefont {G.}~\bibnamefont {Balarac}},
  \bibinfo {author} {\bibfnamefont {J.}~\bibnamefont {Le~Sommer}}, \bibinfo
  {author} {\bibfnamefont {R.}~\bibnamefont {Fablet}}, \ and\ \bibinfo {author}
  {\bibfnamefont {R.}~\bibnamefont {Lguensat}},\ }\bibfield  {title} {\enquote
  {\bibinfo {title} {Physical invariance in neural networks for subgrid-scale
  scalar flux modeling},}\ }\href@noop {} {\bibfield  {journal} {\bibinfo
  {journal} {Physical Review Fluids}\ }\textbf {\bibinfo {volume} {6}},\
  \bibinfo {pages} {024607} (\bibinfo {year} {2021})}\BibitemShut {NoStop}%
\bibitem [{\citenamefont {Jiang}\ \emph {et~al.}(2019)\citenamefont {Jiang},
  \citenamefont {Kolehmainen}, \citenamefont {Gu}, \citenamefont {Kevrekidis},
  \citenamefont {Ozel},\ and\ \citenamefont {Sundaresan}}]{jiang2019neural}%
  \BibitemOpen
  \bibfield  {author} {\bibinfo {author} {\bibfnamefont {Y.}~\bibnamefont
  {Jiang}}, \bibinfo {author} {\bibfnamefont {J.}~\bibnamefont {Kolehmainen}},
  \bibinfo {author} {\bibfnamefont {Y.}~\bibnamefont {Gu}}, \bibinfo {author}
  {\bibfnamefont {Y.~G.}\ \bibnamefont {Kevrekidis}}, \bibinfo {author}
  {\bibfnamefont {A.}~\bibnamefont {Ozel}}, \ and\ \bibinfo {author}
  {\bibfnamefont {S.}~\bibnamefont {Sundaresan}},\ }\bibfield  {title}
  {\enquote {\bibinfo {title} {Neural-network-based filtered drag model for
  gas-particle flows},}\ }\href@noop {} {\bibfield  {journal} {\bibinfo
  {journal} {Powder Technology}\ }\textbf {\bibinfo {volume} {346}},\ \bibinfo
  {pages} {403--413} (\bibinfo {year} {2019})}\BibitemShut {NoStop}%
\bibitem [{\citenamefont {Jiang}\ \emph {et~al.}(2021)\citenamefont {Jiang},
  \citenamefont {Chen}, \citenamefont {Kolehmainen}, \citenamefont
  {Kevrekidis}, \citenamefont {Ozel},\ and\ \citenamefont
  {Sundaresan}}]{jiang2021development}%
  \BibitemOpen
  \bibfield  {author} {\bibinfo {author} {\bibfnamefont {Y.}~\bibnamefont
  {Jiang}}, \bibinfo {author} {\bibfnamefont {X.}~\bibnamefont {Chen}},
  \bibinfo {author} {\bibfnamefont {J.}~\bibnamefont {Kolehmainen}}, \bibinfo
  {author} {\bibfnamefont {I.~G.}\ \bibnamefont {Kevrekidis}}, \bibinfo
  {author} {\bibfnamefont {A.}~\bibnamefont {Ozel}}, \ and\ \bibinfo {author}
  {\bibfnamefont {S.}~\bibnamefont {Sundaresan}},\ }\bibfield  {title}
  {\enquote {\bibinfo {title} {Development of data-driven filtered drag model
  for industrial-scale fluidized beds},}\ }\href@noop {} {\bibfield  {journal}
  {\bibinfo  {journal} {Chemical Engineering Science}\ }\textbf {\bibinfo
  {volume} {230}},\ \bibinfo {pages} {116235} (\bibinfo {year}
  {2021})}\BibitemShut {NoStop}%
\bibitem [{\citenamefont {Dietrich}\ \emph {et~al.}(2021)\citenamefont
  {Dietrich}, \citenamefont {Makeev}, \citenamefont {Kevrekidis}, \citenamefont
  {Evangelou}, \citenamefont {Bertalan}, \citenamefont {Reich},\ and\
  \citenamefont {Kevrekidis}}]{dietrich2021learning}%
  \BibitemOpen
  \bibfield  {author} {\bibinfo {author} {\bibfnamefont {F.}~\bibnamefont
  {Dietrich}}, \bibinfo {author} {\bibfnamefont {A.}~\bibnamefont {Makeev}},
  \bibinfo {author} {\bibfnamefont {G.}~\bibnamefont {Kevrekidis}}, \bibinfo
  {author} {\bibfnamefont {N.}~\bibnamefont {Evangelou}}, \bibinfo {author}
  {\bibfnamefont {T.}~\bibnamefont {Bertalan}}, \bibinfo {author}
  {\bibfnamefont {S.}~\bibnamefont {Reich}}, \ and\ \bibinfo {author}
  {\bibfnamefont {I.~G.}\ \bibnamefont {Kevrekidis}},\ }\bibfield  {title}
  {\enquote {\bibinfo {title} {Learning effective stochastic differential
  equations from microscopic simulations: combining stochastic numerics and
  deep learning},}\ }\href@noop {} {\bibfield  {journal} {\bibinfo  {journal}
  {arXiv preprint arXiv:2106.09004}\ } (\bibinfo {year} {2021})}\BibitemShut
  {NoStop}%
\bibitem [{\citenamefont {Yang}, \citenamefont {Daskalakis},\ and\
  \citenamefont {Karniadakis}(2022)}]{yang2022generative}%
  \BibitemOpen
  \bibfield  {author} {\bibinfo {author} {\bibfnamefont {L.}~\bibnamefont
  {Yang}}, \bibinfo {author} {\bibfnamefont {C.}~\bibnamefont {Daskalakis}}, \
  and\ \bibinfo {author} {\bibfnamefont {G.~E.}\ \bibnamefont {Karniadakis}},\
  }\bibfield  {title} {\enquote {\bibinfo {title} {Generative ensemble
  regression: Learning particle dynamics from observations of ensembles with
  physics-informed deep generative models},}\ }\href@noop {} {\bibfield
  {journal} {\bibinfo  {journal} {SIAM Journal on Scientific Computing}\
  }\textbf {\bibinfo {volume} {44}},\ \bibinfo {pages} {B80--B99} (\bibinfo
  {year} {2022})}\BibitemShut {NoStop}%
\bibitem [{\citenamefont {Chen}\ \emph {et~al.}(2021)\citenamefont {Chen},
  \citenamefont {Yang}, \citenamefont {Duan},\ and\ \citenamefont
  {Karniadakis}}]{karniadakis2021solving}%
  \BibitemOpen
  \bibfield  {author} {\bibinfo {author} {\bibfnamefont {X.}~\bibnamefont
  {Chen}}, \bibinfo {author} {\bibfnamefont {L.}~\bibnamefont {Yang}}, \bibinfo
  {author} {\bibfnamefont {J.}~\bibnamefont {Duan}}, \ and\ \bibinfo {author}
  {\bibfnamefont {G.~E.}\ \bibnamefont {Karniadakis}},\ }\bibfield  {title}
  {\enquote {\bibinfo {title} {Solving inverse stochastic problems from
  discrete particle observations using the fokker--planck equation and
  physics-informed neural networks},}\ }\href@noop {} {\bibfield  {journal}
  {\bibinfo  {journal} {SIAM Journal on Scientific Computing}\ }\textbf
  {\bibinfo {volume} {43}},\ \bibinfo {pages} {B811--B830} (\bibinfo {year}
  {2021})}\BibitemShut {NoStop}%
\bibitem [{\citenamefont {Raissi}(2018)}]{raissi2018stochastic}%
  \BibitemOpen
  \bibfield  {author} {\bibinfo {author} {\bibfnamefont {M.}~\bibnamefont
  {Raissi}},\ }\bibfield  {title} {\enquote {\bibinfo {title} {Forward-backward
  stochastic neural networks: Deep learning of high-dimensional partial
  differential equations},}\ }\href@noop {} {\bibfield  {journal} {\bibinfo
  {journal} {arXiv preprint arXiv:1804.07010}\ } (\bibinfo {year}
  {2018})}\BibitemShut {NoStop}%
\bibitem [{\citenamefont {Kidger}(2022)}]{kidger2022neural}%
  \BibitemOpen
  \bibfield  {author} {\bibinfo {author} {\bibfnamefont {P.}~\bibnamefont
  {Kidger}},\ }\bibfield  {title} {\enquote {\bibinfo {title} {On neural
  differential equations},}\ }\href@noop {} {\bibfield  {journal} {\bibinfo
  {journal} {arXiv preprint arXiv:2202.02435}\ } (\bibinfo {year}
  {2022})}\BibitemShut {NoStop}%
\bibitem [{\citenamefont {Williams}, \citenamefont {Wolfram},\ and\
  \citenamefont {Ozel}(2022)}]{williams2022filteredDNS}%
  \BibitemOpen
  \bibfield  {author} {\bibinfo {author} {\bibfnamefont {J.}~\bibnamefont
  {Williams}}, \bibinfo {author} {\bibfnamefont {U.}~\bibnamefont {Wolfram}}, \
  and\ \bibinfo {author} {\bibfnamefont {A.}~\bibnamefont {Ozel}},\ }\href
  {\doibase 10.34740/KAGGLE/DSV/3998403} {\enquote {\bibinfo {title} {Filtered
  direct numerical simulation dataset},}\ } (\bibinfo {year} {2022}),\ \bibinfo
  {note} {doi: \url{10.34740/KAGGLE/DSV/3998403}}\BibitemShut {NoStop}%
\bibitem [{\citenamefont {Weller}\ \emph {et~al.}(1998)\citenamefont {Weller},
  \citenamefont {Tabor}, \citenamefont {Jasak},\ and\ \citenamefont
  {Fureby}}]{weller98tensorial}%
  \BibitemOpen
  \bibfield  {author} {\bibinfo {author} {\bibfnamefont {H.~G.}\ \bibnamefont
  {Weller}}, \bibinfo {author} {\bibfnamefont {G.}~\bibnamefont {Tabor}},
  \bibinfo {author} {\bibfnamefont {H.}~\bibnamefont {Jasak}}, \ and\ \bibinfo
  {author} {\bibfnamefont {C.}~\bibnamefont {Fureby}},\ }\bibfield  {title}
  {\enquote {\bibinfo {title} {A tensorial approach to computational continuum
  mechanics using object-oriented techniques},}\ }\href {\doibase
  http://dx.doi.org/10.1063/1.168744} {\bibfield  {journal} {\bibinfo
  {journal} {Computers in physics}\ }\textbf {\bibinfo {volume} {12}},\
  \bibinfo {pages} {620--631} (\bibinfo {year} {1998})}\BibitemShut {NoStop}%
\bibitem [{\citenamefont {Schiller}\ and\ \citenamefont
  {Naumann}(1935)}]{schillernaumann}%
  \BibitemOpen
  \bibfield  {author} {\bibinfo {author} {\bibfnamefont {L.}~\bibnamefont
  {Schiller}}\ and\ \bibinfo {author} {\bibfnamefont {Z.}~\bibnamefont
  {Naumann}},\ }\bibfield  {title} {\enquote {\bibinfo {title} {A drag
  coefficient correlation},}\ }\href@noop {} {\bibfield  {journal} {\bibinfo
  {journal} {Z. Ver. Deutsch. Ing}\ }\textbf {\bibinfo {volume} {77}},\
  \bibinfo {pages} {e323} (\bibinfo {year} {1935})}\BibitemShut {NoStop}%
\bibitem [{\citenamefont {Fureby}\ \emph {et~al.}(1996)\citenamefont {Fureby},
  \citenamefont {Tabor}, \citenamefont {Weller},\ and\ \citenamefont
  {Gosman}}]{fureby1996comparative}%
  \BibitemOpen
  \bibfield  {author} {\bibinfo {author} {\bibfnamefont {C.}~\bibnamefont
  {Fureby}}, \bibinfo {author} {\bibfnamefont {G.}~\bibnamefont {Tabor}},
  \bibinfo {author} {\bibfnamefont {H.~G.}\ \bibnamefont {Weller}}, \ and\
  \bibinfo {author} {\bibfnamefont {A.~D.}\ \bibnamefont {Gosman}},\ }\bibfield
   {title} {\enquote {\bibinfo {title} {A comparative study of subgrid scale
  models in homogeneous isotropic turbulence},}\ }\href@noop {} {\bibfield
  {journal} {\bibinfo  {journal} {Physics of Fluids}\ }\textbf {\bibinfo
  {volume} {9}} (\bibinfo {year} {1996})}\BibitemShut {NoStop}%
\bibitem [{\citenamefont {Kang}, \citenamefont {Chester},\ and\ \citenamefont
  {Meneveau}(2003)}]{kang2003decaying}%
  \BibitemOpen
  \bibfield  {author} {\bibinfo {author} {\bibfnamefont {H.~S.}\ \bibnamefont
  {Kang}}, \bibinfo {author} {\bibfnamefont {S.}~\bibnamefont {Chester}}, \
  and\ \bibinfo {author} {\bibfnamefont {C.}~\bibnamefont {Meneveau}},\
  }\bibfield  {title} {\enquote {\bibinfo {title} {Decaying turbulence in an
  active-grid-generated flow and comparisons with large-eddy simulation},}\
  }\href@noop {} {\bibfield  {journal} {\bibinfo  {journal} {Journal of Fluid
  Mechanics}\ }\textbf {\bibinfo {volume} {480}},\ \bibinfo {pages} {129--160}
  (\bibinfo {year} {2003})}\BibitemShut {NoStop}%
\bibitem [{\citenamefont {Saad}\ \emph {et~al.}(2017)\citenamefont {Saad},
  \citenamefont {Cline}, \citenamefont {Stoll},\ and\ \citenamefont
  {Sutherland}}]{saad2017turbogenpy}%
  \BibitemOpen
  \bibfield  {author} {\bibinfo {author} {\bibfnamefont {T.}~\bibnamefont
  {Saad}}, \bibinfo {author} {\bibfnamefont {D.}~\bibnamefont {Cline}},
  \bibinfo {author} {\bibfnamefont {R.}~\bibnamefont {Stoll}}, \ and\ \bibinfo
  {author} {\bibfnamefont {J.~C.}\ \bibnamefont {Sutherland}},\ }\bibfield
  {title} {\enquote {\bibinfo {title} {Scalable tools for generating synthetic
  isotropic turbulence with arbitrary spectra},}\ }\href@noop {} {\bibfield
  {journal} {\bibinfo  {journal} {AIAA journal}\ }\textbf {\bibinfo {volume}
  {55}},\ \bibinfo {pages} {327--331} (\bibinfo {year} {2017})}\BibitemShut
  {NoStop}%
\bibitem [{\citenamefont {Yeung}\ and\ \citenamefont
  {Pope}(1989)}]{yeung1989lagrangian}%
  \BibitemOpen
  \bibfield  {author} {\bibinfo {author} {\bibfnamefont {P.-K.}\ \bibnamefont
  {Yeung}}\ and\ \bibinfo {author} {\bibfnamefont {S.~B.}\ \bibnamefont
  {Pope}},\ }\bibfield  {title} {\enquote {\bibinfo {title} {Lagrangian
  statistics from direct numerical simulations of isotropic turbulence},}\
  }\href@noop {} {\bibfield  {journal} {\bibinfo  {journal} {Journal of Fluid
  Mechanics}\ }\textbf {\bibinfo {volume} {207}},\ \bibinfo {pages} {531--586}
  (\bibinfo {year} {1989})}\BibitemShut {NoStop}%
\bibitem [{\citenamefont {Mansour}\ and\ \citenamefont
  {Wray}(1994)}]{mansour1994decay}%
  \BibitemOpen
  \bibfield  {author} {\bibinfo {author} {\bibfnamefont {N.}~\bibnamefont
  {Mansour}}\ and\ \bibinfo {author} {\bibfnamefont {A.}~\bibnamefont {Wray}},\
  }\bibfield  {title} {\enquote {\bibinfo {title} {Decay of isotropic
  turbulence at low reynolds number},}\ }\href@noop {} {\bibfield  {journal}
  {\bibinfo  {journal} {Physics of Fluids}\ }\textbf {\bibinfo {volume} {6}},\
  \bibinfo {pages} {808--814} (\bibinfo {year} {1994})}\BibitemShut {NoStop}%
\bibitem [{\citenamefont {Ozel}\ \emph {et~al.}(2016)\citenamefont {Ozel},
  \citenamefont {Kolehmainen}, \citenamefont {Radl},\ and\ \citenamefont
  {Sundaresan}}]{oze16fluipart}%
  \BibitemOpen
  \bibfield  {author} {\bibinfo {author} {\bibfnamefont {A.}~\bibnamefont
  {Ozel}}, \bibinfo {author} {\bibfnamefont {J.}~\bibnamefont {Kolehmainen}},
  \bibinfo {author} {\bibfnamefont {S.}~\bibnamefont {Radl}}, \ and\ \bibinfo
  {author} {\bibfnamefont {S.}~\bibnamefont {Sundaresan}},\ }\bibfield  {title}
  {\enquote {\bibinfo {title} {Fluid and particle coarsening of drag force for
  discrete-parcel approach},}\ }\href@noop {} {\bibfield  {journal} {\bibinfo
  {journal} {Chemical engineering science}\ }\textbf {\bibinfo {volume}
  {155}},\ \bibinfo {pages} {258--267} (\bibinfo {year} {2016})}\BibitemShut
  {NoStop}%
\bibitem [{\citenamefont {Germano}\ \emph {et~al.}(1991)\citenamefont
  {Germano}, \citenamefont {Piomelli}, \citenamefont {Moin},\ and\
  \citenamefont {Cabot}}]{germano91dynamic}%
  \BibitemOpen
  \bibfield  {author} {\bibinfo {author} {\bibfnamefont {M.}~\bibnamefont
  {Germano}}, \bibinfo {author} {\bibfnamefont {U.}~\bibnamefont {Piomelli}},
  \bibinfo {author} {\bibfnamefont {P.}~\bibnamefont {Moin}}, \ and\ \bibinfo
  {author} {\bibfnamefont {W.~H.}\ \bibnamefont {Cabot}},\ }\bibfield  {title}
  {\enquote {\bibinfo {title} {A dynamic subgrid-scale eddy viscosity model},}\
  }\href@noop {} {\bibfield  {journal} {\bibinfo  {journal} {Physics of Fluids
  A: Fluid Dynamics}\ }\textbf {\bibinfo {volume} {3}},\ \bibinfo {pages}
  {1760--1765} (\bibinfo {year} {1991})}\BibitemShut {NoStop}%
\bibitem [{\citenamefont {de~Villiers}(2006)}]{devilliersphd}%
  \BibitemOpen
  \bibfield  {author} {\bibinfo {author} {\bibfnamefont {E.}~\bibnamefont
  {de~Villiers}},\ }\emph {\bibinfo {title} {The Potential of Large Eddy
  Simulation for the Modelling of Wall Bounded Flows}},\ \href@noop {} {Ph.D.
  thesis},\ \bibinfo  {school} {Imperial College London} (\bibinfo {year}
  {2006})\BibitemShut {NoStop}%
\bibitem [{\citenamefont {Nicoud}\ and\ \citenamefont
  {Ducros}(1999)}]{nicoud1999subgrid}%
  \BibitemOpen
  \bibfield  {author} {\bibinfo {author} {\bibfnamefont {F.}~\bibnamefont
  {Nicoud}}\ and\ \bibinfo {author} {\bibfnamefont {F.}~\bibnamefont
  {Ducros}},\ }\bibfield  {title} {\enquote {\bibinfo {title} {Subgrid-scale
  stress modelling based on the square of the velocity gradient tensor},}\
  }\href@noop {} {\bibfield  {journal} {\bibinfo  {journal} {Flow, turbulence
  and Combustion}\ }\textbf {\bibinfo {volume} {62}},\ \bibinfo {pages}
  {183--200} (\bibinfo {year} {1999})}\BibitemShut {NoStop}%
\bibitem [{\citenamefont {Cernick}, \citenamefont {Tullis},\ and\ \citenamefont
  {Lightstone}(2015)}]{cernick2015particlesgs}%
  \BibitemOpen
  \bibfield  {author} {\bibinfo {author} {\bibfnamefont {M.~J.}\ \bibnamefont
  {Cernick}}, \bibinfo {author} {\bibfnamefont {S.}~\bibnamefont {Tullis}}, \
  and\ \bibinfo {author} {\bibfnamefont {M.}~\bibnamefont {Lightstone}},\
  }\bibfield  {title} {\enquote {\bibinfo {title} {Particle subgrid scale
  modelling in large-eddy simulations of particle-laden turbulence},}\
  }\href@noop {} {\bibfield  {journal} {\bibinfo  {journal} {Journal of
  Turbulence}\ }\textbf {\bibinfo {volume} {16}},\ \bibinfo {pages} {101--135}
  (\bibinfo {year} {2015})}\BibitemShut {NoStop}%
\bibitem [{\citenamefont {Afkhami}\ \emph {et~al.}(2015)\citenamefont
  {Afkhami}, \citenamefont {Hassanpour}, \citenamefont {Fairweather},\ and\
  \citenamefont {Njobuenwu}}]{afkhami2015fully}%
  \BibitemOpen
  \bibfield  {author} {\bibinfo {author} {\bibfnamefont {M.}~\bibnamefont
  {Afkhami}}, \bibinfo {author} {\bibfnamefont {A.}~\bibnamefont {Hassanpour}},
  \bibinfo {author} {\bibfnamefont {M.}~\bibnamefont {Fairweather}}, \ and\
  \bibinfo {author} {\bibfnamefont {D.~O.}\ \bibnamefont {Njobuenwu}},\
  }\bibfield  {title} {\enquote {\bibinfo {title} {{Fully coupled LES-DEM of
  particle interaction and agglomeration in a turbulent channel flow}},}\
  }\href@noop {} {\bibfield  {journal} {\bibinfo  {journal} {Computers \&
  Chemical Engineering}\ }\textbf {\bibinfo {volume} {78}},\ \bibinfo {pages}
  {24--38} (\bibinfo {year} {2015})}\BibitemShut {NoStop}%
\bibitem [{\citenamefont {Wang}\ and\ \citenamefont
  {Squires}(1996)}]{wang1996LES}%
  \BibitemOpen
  \bibfield  {author} {\bibinfo {author} {\bibfnamefont {Q.}~\bibnamefont
  {Wang}}\ and\ \bibinfo {author} {\bibfnamefont {K.~D.}\ \bibnamefont
  {Squires}},\ }\bibfield  {title} {\enquote {\bibinfo {title} {Large eddy
  simulation of particle-laden turbulent channel flow},}\ }\href@noop {}
  {\bibfield  {journal} {\bibinfo  {journal} {Physics of Fluids}\ }\textbf
  {\bibinfo {volume} {8}},\ \bibinfo {pages} {1207--1223} (\bibinfo {year}
  {1996})}\BibitemShut {NoStop}%
\bibitem [{\citenamefont {Wiener}(1938)}]{wiener1938homogeneous}%
  \BibitemOpen
  \bibfield  {author} {\bibinfo {author} {\bibfnamefont {N.}~\bibnamefont
  {Wiener}},\ }\bibfield  {title} {\enquote {\bibinfo {title} {The homogeneous
  chaos},}\ }\href@noop {} {\bibfield  {journal} {\bibinfo  {journal} {American
  Journal of Mathematics}\ }\textbf {\bibinfo {volume} {60}},\ \bibinfo {pages}
  {897--936} (\bibinfo {year} {1938})}\BibitemShut {NoStop}%
\bibitem [{\citenamefont {It{\^o}}(1951)}]{ito1951multiple}%
  \BibitemOpen
  \bibfield  {author} {\bibinfo {author} {\bibfnamefont {K.}~\bibnamefont
  {It{\^o}}},\ }\bibfield  {title} {\enquote {\bibinfo {title} {Multiple wiener
  integral},}\ }\href@noop {} {\bibfield  {journal} {\bibinfo  {journal}
  {Journal of the Mathematical Society of Japan}\ }\textbf {\bibinfo {volume}
  {3}},\ \bibinfo {pages} {157--169} (\bibinfo {year} {1951})}\BibitemShut
  {NoStop}%
\bibitem [{\citenamefont {Kolmogorov}(1941)}]{kolmogorov1941dissipation}%
  \BibitemOpen
  \bibfield  {author} {\bibinfo {author} {\bibfnamefont {A.~N.}\ \bibnamefont
  {Kolmogorov}},\ }\bibfield  {title} {\enquote {\bibinfo {title} {Dissipation
  of energy in the locally isotropic turbulence},}\ }in\ \href@noop {} {\emph
  {\bibinfo {booktitle} {Dokl. Akad. Nauk SSSR A}}},\ Vol.~\bibinfo {volume}
  {32}\ (\bibinfo {year} {1941})\ pp.\ \bibinfo {pages} {16--18}\BibitemShut
  {NoStop}%
\bibitem [{\citenamefont {Minier}, \citenamefont {Peirano},\ and\ \citenamefont
  {Chibbaro}(2003)}]{minier2003schemes}%
  \BibitemOpen
  \bibfield  {author} {\bibinfo {author} {\bibfnamefont {J.-P.}\ \bibnamefont
  {Minier}}, \bibinfo {author} {\bibfnamefont {E.}~\bibnamefont {Peirano}}, \
  and\ \bibinfo {author} {\bibfnamefont {S.}~\bibnamefont {Chibbaro}},\
  }\bibfield  {title} {\enquote {\bibinfo {title} {Weak first-and second-order
  numerical schemes for stochastic differential equations appearing in
  lagrangian two-phase flow modeling},}\ }\href@noop {} {\bibfield  {journal}
  {\bibinfo  {journal} {Monte Carlo Methods and Applications}\ }\textbf
  {\bibinfo {volume} {9}},\ \bibinfo {pages} {93--133} (\bibinfo {year}
  {2003})}\BibitemShut {NoStop}%
\bibitem [{\citenamefont {Hochreiter}(1998)}]{hochreiter1998vanishing}%
  \BibitemOpen
  \bibfield  {author} {\bibinfo {author} {\bibfnamefont {S.}~\bibnamefont
  {Hochreiter}},\ }\bibfield  {title} {\enquote {\bibinfo {title} {The
  vanishing gradient problem during learning recurrent neural nets and problem
  solutions},}\ }\href@noop {} {\bibfield  {journal} {\bibinfo  {journal}
  {International Journal of Uncertainty, Fuzziness and Knowledge-Based
  Systems}\ }\textbf {\bibinfo {volume} {6}},\ \bibinfo {pages} {107--116}
  (\bibinfo {year} {1998})}\BibitemShut {NoStop}%
\bibitem [{\citenamefont {Hochreiter}\ \emph {et~al.}(2001)\citenamefont
  {Hochreiter}, \citenamefont {Bengio}, \citenamefont {Frasconi},\ and\
  \citenamefont {Schmidhuber}}]{hochreiter2001gradients}%
  \BibitemOpen
  \bibfield  {author} {\bibinfo {author} {\bibfnamefont {S.}~\bibnamefont
  {Hochreiter}}, \bibinfo {author} {\bibfnamefont {Y.}~\bibnamefont {Bengio}},
  \bibinfo {author} {\bibfnamefont {P.}~\bibnamefont {Frasconi}}, \ and\
  \bibinfo {author} {\bibfnamefont {J.}~\bibnamefont {Schmidhuber}},\
  }\bibfield  {title} {\enquote {\bibinfo {title} {Gradient flow in recurrent
  nets: the difficulty of learning long-term dependencies},}\ }in\ \href@noop
  {} {\emph {\bibinfo {booktitle} {A Field Guide to Dynamical Recurrent Neural
  Networks}}}\ (\bibinfo  {publisher} {IEEE press},\ \bibinfo {year}
  {2001})\BibitemShut {NoStop}%
\bibitem [{\citenamefont {He}\ \emph {et~al.}(2015)\citenamefont {He},
  \citenamefont {Zhang}, \citenamefont {Ren},\ and\ \citenamefont
  {Sun}}]{he2015delving}%
  \BibitemOpen
  \bibfield  {author} {\bibinfo {author} {\bibfnamefont {K.}~\bibnamefont
  {He}}, \bibinfo {author} {\bibfnamefont {X.}~\bibnamefont {Zhang}}, \bibinfo
  {author} {\bibfnamefont {S.}~\bibnamefont {Ren}}, \ and\ \bibinfo {author}
  {\bibfnamefont {J.}~\bibnamefont {Sun}},\ }\bibfield  {title} {\enquote
  {\bibinfo {title} {Delving deep into rectifiers: Surpassing human-level
  performance on imagenet classification},}\ }in\ \href@noop {} {\emph
  {\bibinfo {booktitle} {Proceedings of the IEEE international conference on
  computer vision}}}\ (\bibinfo {year} {2015})\ pp.\ \bibinfo {pages}
  {1026--1034}\BibitemShut {NoStop}%
\bibitem [{\citenamefont {Chollet}\ \emph {et~al.}(2015)\citenamefont {Chollet}
  \emph {et~al.}}]{chollet2015keras}%
  \BibitemOpen
  \bibfield  {author} {\bibinfo {author} {\bibfnamefont {F.}~\bibnamefont
  {Chollet}} \emph {et~al.},\ }\href {https://github.com/fchollet/keras}
  {\enquote {\bibinfo {title} {Keras},}\ } (\bibinfo {year} {2015})\BibitemShut
  {NoStop}%
\bibitem [{\citenamefont {Abadi}\ \emph {et~al.}(2015)\citenamefont {Abadi},
  \citenamefont {Agarwal}, \citenamefont {Barham}, \citenamefont {Brevdo},
  \citenamefont {Chen}, \citenamefont {Citro}, \citenamefont {Corrado},
  \citenamefont {Davis}, \citenamefont {Dean}, \citenamefont {Devin},
  \citenamefont {Ghemawat}, \citenamefont {Goodfellow}, \citenamefont {Harp},
  \citenamefont {Irving}, \citenamefont {Isard}, \citenamefont {Jia},
  \citenamefont {Jozefowicz}, \citenamefont {Kaiser}, \citenamefont {Kudlur},
  \citenamefont {Levenberg}, \citenamefont {Man\'{e}}, \citenamefont {Monga},
  \citenamefont {Moore}, \citenamefont {Murray}, \citenamefont {Olah},
  \citenamefont {Schuster}, \citenamefont {Shlens}, \citenamefont {Steiner},
  \citenamefont {Sutskever}, \citenamefont {Talwar}, \citenamefont {Tucker},
  \citenamefont {Vanhoucke}, \citenamefont {Vasudevan}, \citenamefont
  {Vi\'{e}gas}, \citenamefont {Vinyals}, \citenamefont {Warden}, \citenamefont
  {Wattenberg}, \citenamefont {Wicke}, \citenamefont {Yu},\ and\ \citenamefont
  {Zheng}}]{tensorflow2015}%
  \BibitemOpen
  \bibfield  {author} {\bibinfo {author} {\bibfnamefont {M.}~\bibnamefont
  {Abadi}}, \bibinfo {author} {\bibfnamefont {A.}~\bibnamefont {Agarwal}},
  \bibinfo {author} {\bibfnamefont {P.}~\bibnamefont {Barham}}, \bibinfo
  {author} {\bibfnamefont {E.}~\bibnamefont {Brevdo}}, \bibinfo {author}
  {\bibfnamefont {Z.}~\bibnamefont {Chen}}, \bibinfo {author} {\bibfnamefont
  {C.}~\bibnamefont {Citro}}, \bibinfo {author} {\bibfnamefont {G.~S.}\
  \bibnamefont {Corrado}}, \bibinfo {author} {\bibfnamefont {A.}~\bibnamefont
  {Davis}}, \bibinfo {author} {\bibfnamefont {J.}~\bibnamefont {Dean}},
  \bibinfo {author} {\bibfnamefont {M.}~\bibnamefont {Devin}}, \bibinfo
  {author} {\bibfnamefont {S.}~\bibnamefont {Ghemawat}}, \bibinfo {author}
  {\bibfnamefont {I.}~\bibnamefont {Goodfellow}}, \bibinfo {author}
  {\bibfnamefont {A.}~\bibnamefont {Harp}}, \bibinfo {author} {\bibfnamefont
  {G.}~\bibnamefont {Irving}}, \bibinfo {author} {\bibfnamefont
  {M.}~\bibnamefont {Isard}}, \bibinfo {author} {\bibfnamefont
  {Y.}~\bibnamefont {Jia}}, \bibinfo {author} {\bibfnamefont {R.}~\bibnamefont
  {Jozefowicz}}, \bibinfo {author} {\bibfnamefont {L.}~\bibnamefont {Kaiser}},
  \bibinfo {author} {\bibfnamefont {M.}~\bibnamefont {Kudlur}}, \bibinfo
  {author} {\bibfnamefont {J.}~\bibnamefont {Levenberg}}, \bibinfo {author}
  {\bibfnamefont {D.}~\bibnamefont {Man\'{e}}}, \bibinfo {author}
  {\bibfnamefont {R.}~\bibnamefont {Monga}}, \bibinfo {author} {\bibfnamefont
  {S.}~\bibnamefont {Moore}}, \bibinfo {author} {\bibfnamefont
  {D.}~\bibnamefont {Murray}}, \bibinfo {author} {\bibfnamefont
  {C.}~\bibnamefont {Olah}}, \bibinfo {author} {\bibfnamefont {M.}~\bibnamefont
  {Schuster}}, \bibinfo {author} {\bibfnamefont {J.}~\bibnamefont {Shlens}},
  \bibinfo {author} {\bibfnamefont {B.}~\bibnamefont {Steiner}}, \bibinfo
  {author} {\bibfnamefont {I.}~\bibnamefont {Sutskever}}, \bibinfo {author}
  {\bibfnamefont {K.}~\bibnamefont {Talwar}}, \bibinfo {author} {\bibfnamefont
  {P.}~\bibnamefont {Tucker}}, \bibinfo {author} {\bibfnamefont
  {V.}~\bibnamefont {Vanhoucke}}, \bibinfo {author} {\bibfnamefont
  {V.}~\bibnamefont {Vasudevan}}, \bibinfo {author} {\bibfnamefont
  {F.}~\bibnamefont {Vi\'{e}gas}}, \bibinfo {author} {\bibfnamefont
  {O.}~\bibnamefont {Vinyals}}, \bibinfo {author} {\bibfnamefont
  {P.}~\bibnamefont {Warden}}, \bibinfo {author} {\bibfnamefont
  {M.}~\bibnamefont {Wattenberg}}, \bibinfo {author} {\bibfnamefont
  {M.}~\bibnamefont {Wicke}}, \bibinfo {author} {\bibfnamefont
  {Y.}~\bibnamefont {Yu}}, \ and\ \bibinfo {author} {\bibfnamefont
  {X.}~\bibnamefont {Zheng}},\ }\href {https://www.tensorflow.org/} {\enquote
  {\bibinfo {title} {{TensorFlow}: Large-scale machine learning on
  heterogeneous systems},}\ } (\bibinfo {year} {2015}),\ \bibinfo {note}
  {software available from tensorflow.org}\BibitemShut {NoStop}%
\bibitem [{\citenamefont {Sutskever}\ \emph {et~al.}(2013)\citenamefont
  {Sutskever}, \citenamefont {Martens}, \citenamefont {Dahl},\ and\
  \citenamefont {Hinton}}]{sutskever2013importance}%
  \BibitemOpen
  \bibfield  {author} {\bibinfo {author} {\bibfnamefont {I.}~\bibnamefont
  {Sutskever}}, \bibinfo {author} {\bibfnamefont {J.}~\bibnamefont {Martens}},
  \bibinfo {author} {\bibfnamefont {G.}~\bibnamefont {Dahl}}, \ and\ \bibinfo
  {author} {\bibfnamefont {G.}~\bibnamefont {Hinton}},\ }\bibfield  {title}
  {\enquote {\bibinfo {title} {On the importance of initialization and momentum
  in deep learning},}\ }in\ \href@noop {} {\emph {\bibinfo {booktitle}
  {International conference on machine learning}}}\ (\bibinfo {organization}
  {PMLR},\ \bibinfo {year} {2013})\ pp.\ \bibinfo {pages}
  {1139--1147}\BibitemShut {NoStop}%
\bibitem [{\citenamefont {Boivin}, \citenamefont {Simonin},\ and\ \citenamefont
  {Squires}(2000)}]{boivin2000prediction}%
  \BibitemOpen
  \bibfield  {author} {\bibinfo {author} {\bibfnamefont {M.}~\bibnamefont
  {Boivin}}, \bibinfo {author} {\bibfnamefont {O.}~\bibnamefont {Simonin}}, \
  and\ \bibinfo {author} {\bibfnamefont {K.~D.}\ \bibnamefont {Squires}},\
  }\bibfield  {title} {\enquote {\bibinfo {title} {On the prediction of
  gas--solid flows with two-way coupling using large eddy simulation},}\ }\href
  {\doibase 10.1063/1.870453} {\bibfield  {journal} {\bibinfo  {journal}
  {Physics of Fluids}\ }\textbf {\bibinfo {volume} {12}},\ \bibinfo {pages}
  {2080--2090} (\bibinfo {year} {2000})}\BibitemShut {NoStop}%
\bibitem [{\citenamefont {F{\'e}vrier}, \citenamefont {Simonin},\ and\
  \citenamefont {Squires}(2005)}]{fevrier2005partitioning}%
  \BibitemOpen
  \bibfield  {author} {\bibinfo {author} {\bibfnamefont {P.}~\bibnamefont
  {F{\'e}vrier}}, \bibinfo {author} {\bibfnamefont {O.}~\bibnamefont
  {Simonin}}, \ and\ \bibinfo {author} {\bibfnamefont {K.~D.}\ \bibnamefont
  {Squires}},\ }\bibfield  {title} {\enquote {\bibinfo {title} {Partitioning of
  particle velocities in gas--solid turbulent flows into a continuous field and
  a spatially uncorrelated random distribution: theoretical formalism and
  numerical study},}\ }\href@noop {} {\bibfield  {journal} {\bibinfo  {journal}
  {Journal of Fluid Mechanics}\ }\textbf {\bibinfo {volume} {533}},\ \bibinfo
  {pages} {1--46} (\bibinfo {year} {2005})}\BibitemShut {NoStop}%
\bibitem [{\citenamefont {Capecelatro}, \citenamefont {Desjardins},\ and\
  \citenamefont {Fox}(2014)}]{capecelatro2014numerical}%
  \BibitemOpen
  \bibfield  {author} {\bibinfo {author} {\bibfnamefont {J.}~\bibnamefont
  {Capecelatro}}, \bibinfo {author} {\bibfnamefont {O.}~\bibnamefont
  {Desjardins}}, \ and\ \bibinfo {author} {\bibfnamefont {R.~O.}\ \bibnamefont
  {Fox}},\ }\bibfield  {title} {\enquote {\bibinfo {title} {Numerical study of
  collisional particle dynamics in cluster-induced turbulence},}\ }\href@noop
  {} {\bibfield  {journal} {\bibinfo  {journal} {Journal of Fluid Mechanics}\
  }\textbf {\bibinfo {volume} {747}} (\bibinfo {year} {2014})}\BibitemShut
  {NoStop}%
\bibitem [{\citenamefont {Capecelatro}, \citenamefont {Desjardins},\ and\
  \citenamefont {Fox}(2015)}]{capecelatro2015fluid}%
  \BibitemOpen
  \bibfield  {author} {\bibinfo {author} {\bibfnamefont {J.}~\bibnamefont
  {Capecelatro}}, \bibinfo {author} {\bibfnamefont {O.}~\bibnamefont
  {Desjardins}}, \ and\ \bibinfo {author} {\bibfnamefont {R.~O.}\ \bibnamefont
  {Fox}},\ }\bibfield  {title} {\enquote {\bibinfo {title} {On fluid--particle
  dynamics in fully developed cluster-induced turbulence},}\ }\href@noop {}
  {\bibfield  {journal} {\bibinfo  {journal} {Journal of Fluid Mechanics}\
  }\textbf {\bibinfo {volume} {780}},\ \bibinfo {pages} {578--635} (\bibinfo
  {year} {2015})}\BibitemShut {NoStop}%
\bibitem [{\citenamefont {Yeung}\ and\ \citenamefont
  {Pope}(1988)}]{yeungpope1988algorithm}%
  \BibitemOpen
  \bibfield  {author} {\bibinfo {author} {\bibfnamefont {P.}~\bibnamefont
  {Yeung}}\ and\ \bibinfo {author} {\bibfnamefont {S.}~\bibnamefont {Pope}},\
  }\bibfield  {title} {\enquote {\bibinfo {title} {An algorithm for tracking
  fluid particles in numerical simulations of homogeneous turbulence},}\
  }\href@noop {} {\bibfield  {journal} {\bibinfo  {journal} {Journal of
  computational physics}\ }\textbf {\bibinfo {volume} {79}},\ \bibinfo {pages}
  {373--416} (\bibinfo {year} {1988})}\BibitemShut {NoStop}%
\bibitem [{\citenamefont {Balachandar}\ and\ \citenamefont
  {Maxey}(1989)}]{balachandarmaxey1989}%
  \BibitemOpen
  \bibfield  {author} {\bibinfo {author} {\bibfnamefont {S.}~\bibnamefont
  {Balachandar}}\ and\ \bibinfo {author} {\bibfnamefont {M.}~\bibnamefont
  {Maxey}},\ }\bibfield  {title} {\enquote {\bibinfo {title} {Methods for
  evaluating fluid velocities in spectral simulations of turbulence},}\
  }\href@noop {} {\bibfield  {journal} {\bibinfo  {journal} {Journal of
  Computational Physics}\ }\textbf {\bibinfo {volume} {83}},\ \bibinfo {pages}
  {96--125} (\bibinfo {year} {1989})}\BibitemShut {NoStop}%
\bibitem [{\citenamefont {Rovelstad}, \citenamefont {Handler},\ and\
  \citenamefont {Bernard}(1994)}]{rovelstad1994interpolation}%
  \BibitemOpen
  \bibfield  {author} {\bibinfo {author} {\bibfnamefont {A.~L.}\ \bibnamefont
  {Rovelstad}}, \bibinfo {author} {\bibfnamefont {R.~A.}\ \bibnamefont
  {Handler}}, \ and\ \bibinfo {author} {\bibfnamefont {P.~S.}\ \bibnamefont
  {Bernard}},\ }\bibfield  {title} {\enquote {\bibinfo {title} {The effect of
  interpolation errors on the lagrangian analysis of simulated turbulent
  channel flow},}\ }\href@noop {} {\bibfield  {journal} {\bibinfo  {journal}
  {Journal of Computational Physics}\ }\textbf {\bibinfo {volume} {110}},\
  \bibinfo {pages} {190--195} (\bibinfo {year} {1994})}\BibitemShut {NoStop}%
\bibitem [{\citenamefont {Garg}\ \emph {et~al.}(2007)\citenamefont {Garg},
  \citenamefont {Narayanan}, \citenamefont {Lakehal},\ and\ \citenamefont
  {Subramaniam}}]{garg2007accurate}%
  \BibitemOpen
  \bibfield  {author} {\bibinfo {author} {\bibfnamefont {R.}~\bibnamefont
  {Garg}}, \bibinfo {author} {\bibfnamefont {C.}~\bibnamefont {Narayanan}},
  \bibinfo {author} {\bibfnamefont {D.}~\bibnamefont {Lakehal}}, \ and\
  \bibinfo {author} {\bibfnamefont {S.}~\bibnamefont {Subramaniam}},\
  }\bibfield  {title} {\enquote {\bibinfo {title} {Accurate numerical
  estimation of interphase momentum transfer in lagrangian--eulerian
  simulations of dispersed two-phase flows},}\ }\href@noop {} {\bibfield
  {journal} {\bibinfo  {journal} {International Journal of Multiphase Flow}\
  }\textbf {\bibinfo {volume} {33}},\ \bibinfo {pages} {1337--1364} (\bibinfo
  {year} {2007})}\BibitemShut {NoStop}%
\bibitem [{\citenamefont {Marchioli}\ \emph {et~al.}(2008)\citenamefont
  {Marchioli}, \citenamefont {Soldati}, \citenamefont {Kuerten}, \citenamefont
  {Arcen}, \citenamefont {Taniere}, \citenamefont {Goldensoph}, \citenamefont
  {Squires}, \citenamefont {Cargnelutti},\ and\ \citenamefont
  {Portela}}]{marchioli2008benchmark}%
  \BibitemOpen
  \bibfield  {author} {\bibinfo {author} {\bibfnamefont {C.}~\bibnamefont
  {Marchioli}}, \bibinfo {author} {\bibfnamefont {A.}~\bibnamefont {Soldati}},
  \bibinfo {author} {\bibfnamefont {J.}~\bibnamefont {Kuerten}}, \bibinfo
  {author} {\bibfnamefont {B.}~\bibnamefont {Arcen}}, \bibinfo {author}
  {\bibfnamefont {A.}~\bibnamefont {Taniere}}, \bibinfo {author} {\bibfnamefont
  {G.}~\bibnamefont {Goldensoph}}, \bibinfo {author} {\bibfnamefont
  {K.}~\bibnamefont {Squires}}, \bibinfo {author} {\bibfnamefont
  {M.}~\bibnamefont {Cargnelutti}}, \ and\ \bibinfo {author} {\bibfnamefont
  {L.}~\bibnamefont {Portela}},\ }\bibfield  {title} {\enquote {\bibinfo
  {title} {Statistics of particle dispersion in direct numerical simulations of
  wall-bounded turbulence: Results of an international collaborative benchmark
  test},}\ }\href@noop {} {\bibfield  {journal} {\bibinfo  {journal}
  {International Journal of Multiphase Flow}\ }\textbf {\bibinfo {volume}
  {34}},\ \bibinfo {pages} {879--893} (\bibinfo {year} {2008})}\BibitemShut
  {NoStop}%
\bibitem [{\citenamefont {Soldati}\ and\ \citenamefont
  {Marchioli}(2009)}]{soldati2009physics}%
  \BibitemOpen
  \bibfield  {author} {\bibinfo {author} {\bibfnamefont {A.}~\bibnamefont
  {Soldati}}\ and\ \bibinfo {author} {\bibfnamefont {C.}~\bibnamefont
  {Marchioli}},\ }\bibfield  {title} {\enquote {\bibinfo {title} {Physics and
  modelling of turbulent particle deposition and entrainment: Review of a
  systematic study},}\ }\href@noop {} {\bibfield  {journal} {\bibinfo
  {journal} {International Journal of Multiphase Flow}\ }\textbf {\bibinfo
  {volume} {35}},\ \bibinfo {pages} {827--839} (\bibinfo {year}
  {2009})}\BibitemShut {NoStop}%
\end{thebibliography}%

\end{document}


\maketitle

\section{Interpolation error analysis} \label{sec:appendixA}
To determine the optimal scheme for interpolation of Eulerian properties to the particle position, we performed simulations of one particle flowing through a domain where the fluid velocity can be described analytically \citep{marchioli2007simple}. \citet{marchioli2007simple} proposed a case where the fluid velocity followed:
\begin{eqnarray}
    \boldsymbol{u}_{fx} &=& u_0, \\
    \boldsymbol{u}_{fy} &=& 0, \\
    \boldsymbol{u}_{fz} &=& u_0 \sin (B \pi \boldsymbol{x}_x).
\end{eqnarray}
where $\boldsymbol{u}_{fx}$ is the x component of fluid velocity, $\boldsymbol{x}_x$ is the x coordinate of particle position, the parameters describing the flow are $u_0 = 0.025 \unit{m/s}$ and $B = 4$. The particle position at time, $t$, can be obtained analytically by
\begin{eqnarray}
    \boldsymbol{x}_{x} (t) &=& \boldsymbol{x}_{x}^0 + u_0 \, t, \\
    \boldsymbol{x}_{y} (t) &=& 0, \\
    \boldsymbol{x}_{z} (t) &=& \boldsymbol{x}_{z}^0 + \frac{1}{B \pi}\left[ \cos(B\pi\boldsymbol{x}_{x}^0) - \cos(B\pi\boldsymbol{x}_{x}^0 +\pi u_0 t) \right] \label{eq:zposAnalytical}.
\end{eqnarray}

We evaluated the absolute relative error of particle position in the z-direction ($\boldsymbol{x}_z$) obtained from interpolating the fluid velocity and integrating the equation, compared to the analytical solution \eqref{eq:zposAnalytical}. We tested interpolating the fluid velocity using trilinear interpolation (`cellPoint') and using the cell value (`cell'). The computational domain was a cuboid extending from $\boldsymbol{x}=(-2, -1, -1) \unit{m}$ to $\boldsymbol{x} = (2, 1, 1) \unit{m}$. The domain was discretised into $16^3$, $32^3$, $64^3$ and $128^3$ cells.

\begin{figure}
    \centering
    \begin{tabular}{c c}
        (a) cell, $N_{cell}=16^3$ & (b) cellPoint, $N_{cell}=16^3$  \\
        \includegraphics{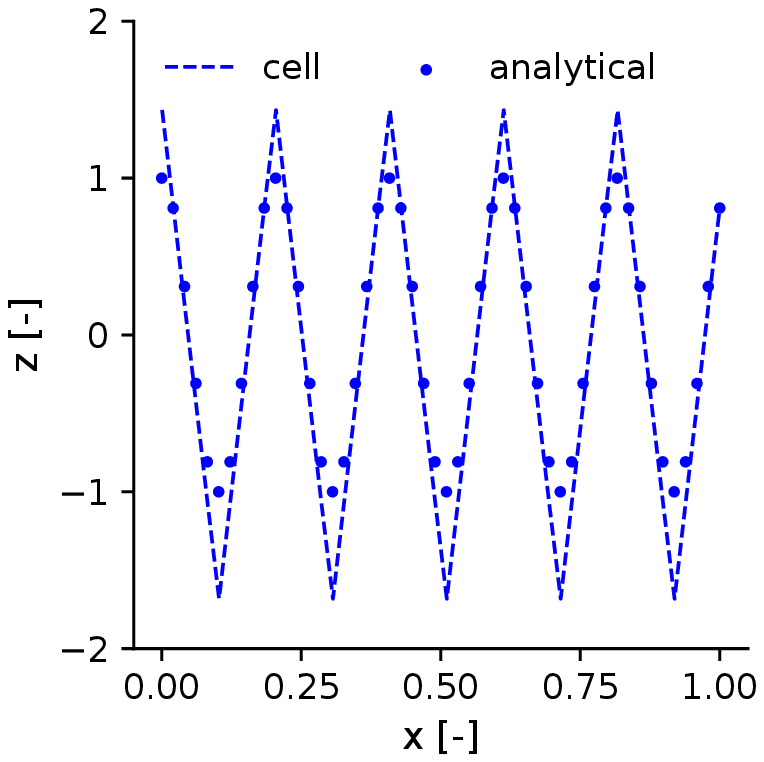} & 
        \includegraphics{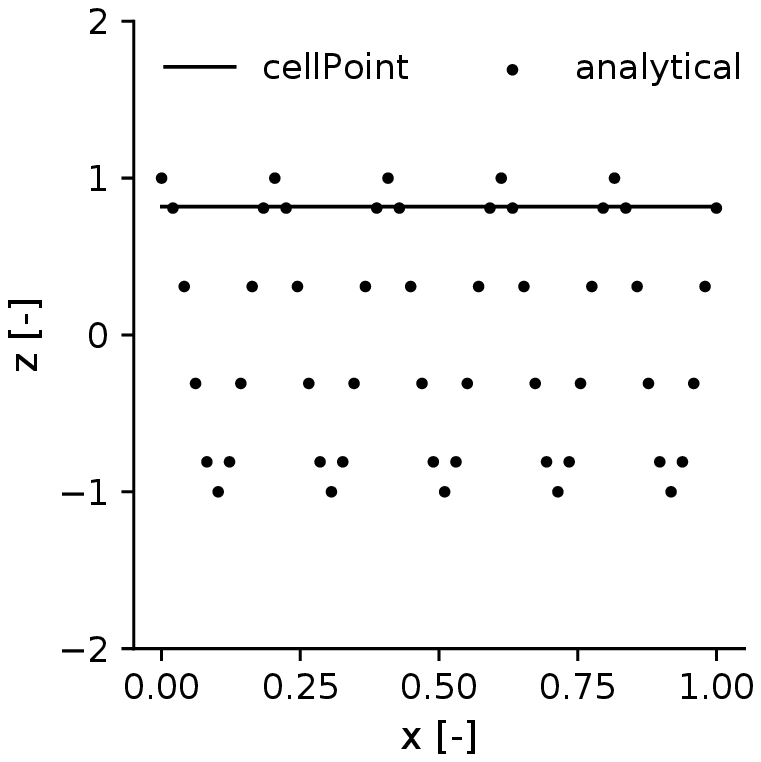} \\
        (c) cell, $N_{cell}=64^3$ & (d) cellPoint, $N_{cell}=64^3$  \\
        \includegraphics{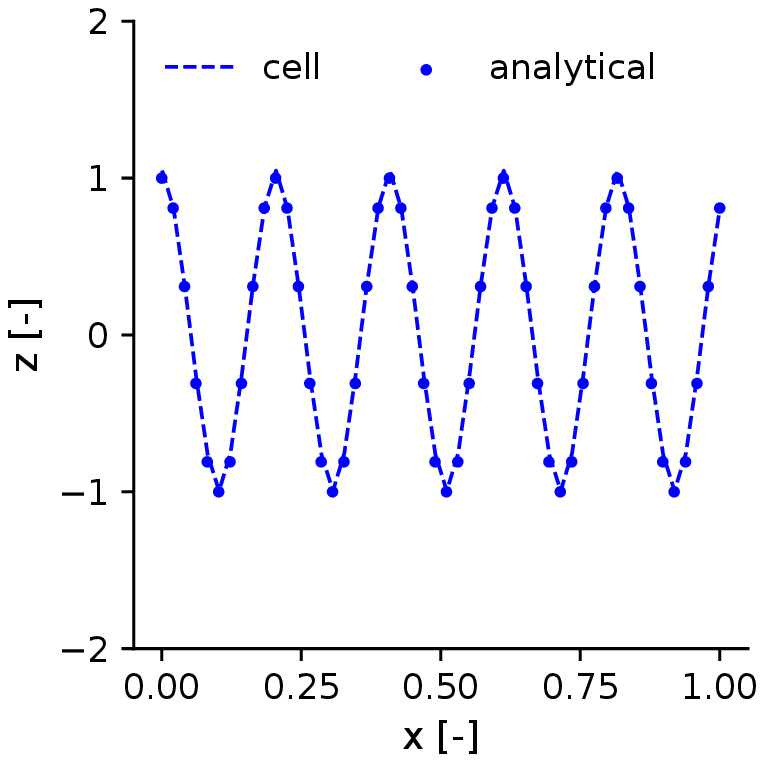} &
        \includegraphics{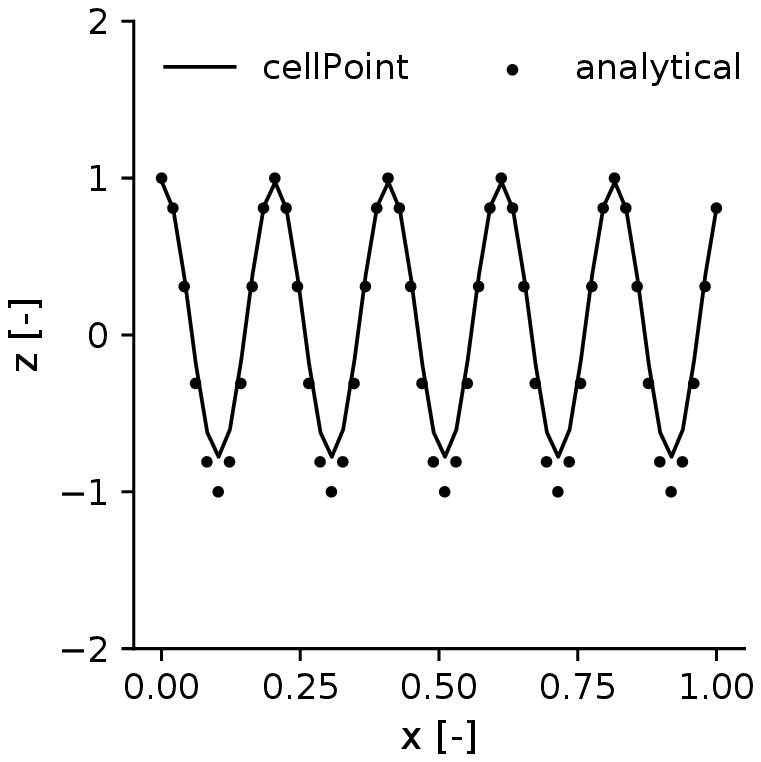}
    \end{tabular}
    \caption{Evolution particle position in $x$ and $z$ directions for varying interpolation scheme and mesh resolution. Columns show (a,c) cell interpolation, and (b,d) cellPoint interpolation. Rows show (a,b) $N_{cell}=16^3$ and (c,d) $N_{cell}=64^3$.}
    \label{fig:posEvol}
\end{figure}

By examining particle trajectories, we observed that cellPoint interpolation scheme produces zero motion in the transverse direction on a coarse mesh and gave a maximum relative error of 165\% (\fref{fig:posEvol}b). In contrast, when using the cell value for calculating particle motion (\fref{fig:posEvol}a), the particle shows qualitatively correct behaviour, with some over-prediction due to the coarse mesh (maximum error 43\%). When $N_{cell}=64^3$, both interpolation schemes show an improved agreement with the analytical solution (\fref{fig:posEvol}c,d). However, the maximum error was 25\% for cellPoint and 6.4\% for cell interpolation. The poor performance of cellPoint interpolation is surprising, as previous studies have shown that using linear interpolation of fluid properties produces lower error than using the cell value \citep{peirano2006meanfield, innocenti2016}. However, some studies have discussed the limitations of trilinear interpolation compared to more advanced methods such as high-order Lagrange polynomials, cubic spline or Hermite interpolation \citep{yeungpope1988algorithm, rovelstad1994interpolation, garg2007accurate}. Due to the lower error of cell interpolation and an apparent implementation issue in cellPoint interpolation, we chose to use cell interpolation in this study.

\bibliography{refs_appraisal}